\documentclass[onecolumn,pra,aps,superscriptaddress,floatfix,notitlepage,groupedaddress]{revtex4-1}

\usepackage{graphicx}
\usepackage{dcolumn}
\usepackage{bm}
\usepackage{amsmath}
\usepackage{amsfonts}
\usepackage{caption}
\usepackage{subcaption}
\usepackage[section]{placeins}
\usepackage{hyperref}

\DeclareMathOperator{\sech}{sech}

\begin{document}

\title{Quantum Tunneling and Information Entropy in a Double Square Well Potential: Ammonia Molecule}

\author{S.T. Tserkis}  \email{stserkis@physics.auth.gr}

\author{Ch.C. Moustakidis}

\author{S.E. Massen}

\author{C.P. Panos}

\affiliation{Department of Theoretical Physics, Aristotle University of Thessaloniki, 54124 Thessaloniki, Greece}

\date{\today}

\begin{abstract}

Quantum tunneling is the quantum-mechanical effect where a particle tunnels through a classically forbidden region. Double Square Well Potential (DSWP) is a system where this phenomenon is feasible. Numerous phenomena can be illustrated by considering motion in a pair of wells that are separated by a barrier of finite height and width. The energy level splitting, resulting from barrier penetration, is the reason of the so-called inversion spectrum, which is an example of quantum tunneling. Out of several molecules ($NH_3$, $PH_3$, $AsH_3$, $NH_2CN$) where this inversion phenomenon occurs, ammonia molecule $NH_3$ provides a nice physical realization of a vibrational system with a DSWP. The main goal of the present work is to examine the implications of quantum tunneling on information entropy measures (Shannon's and Fisher's) and statistical complexity.

\end{abstract}

\maketitle


\section{Introduction}

Classically if a particle is located in one of the wells of a DSWP and does not have sufficient energy to surmount the barrier in the potential, it will be forever confined to that well. Quantum mechanics shows that due to the wave-like nature of the particle, after a certain length of time, there is a non-zero probability that it will be located in the other well. It is therefore quantum mechanically possible for a particle to pass through a barrier that it cannot classically overcome. This phenomenon is known as quantum tunneling effect.

The quantum tunneling results in the splitting of the low-lying energy levels which occur in pairs with slightly different energy values. The transition frequency between the energy levels of each pair is associated with the emission or absorption of electromagnetic radiation. Particularly in the ammonia molecule this transition frequency for the ground-state has been measured at about $24\ \rm{GHz}$ \cite{Cleeton,Herzberg,Basdevant}. This phenomenon, namely the inversion spectrum of the ammonia molecule, has been observed through infrared spectroscopy and plays a fundamental role in the principle of operation of the ammonia MASER\cite{Major}. Although inversion effect occurs in other molecules as well ($PH_3$, $AsH_3$, $NH_2CN$) \cite{Peacock-Lopez,Basdevant}, $NH_3$ provides a tractable vibrational system for experimental observation and exploitation, since inversion frequency falls in the microwave region.

Until recently, oscillation of probability density in position space $\rho(x,t)$ between the wells has been the usual way to approach the quantum tunneling phenomenon and consequently the inversion spectrum. Heisenberg uncertainty relation reflects this phenomenon, but information entropy offers a more sensitive approach, to study a particle moving in a non-classical way through the barrier.

Information-theoretic tools, initially applied to communication systems, have been employed extensively to investigate various classical and quantum systems e.g. in physics \cite{Bialynicki-Birula}, chemistry \cite{Sears,KP}, biology \cite{Adami} and many other scientific branches as well. Specifically, the well-known information measures defined by Shannon \cite{Shannon} and Fisher \cite{Fisher}, have been applied with considerable success in quantum systems e.g. atoms \cite{GSCB,Gadre}. Shannon information entropy has been correlated fairly well with experimental data for atomic ionization potentials and dipole polarizabilities \cite{SPCM}. A comprehensive account of applications to molecules can be found in \cite{Nalewajski1} and \cite{Bonchev}. Another example is an information-theoretic treatment of a molecule ($\pi$-system) described in \cite{KP}.

To begin with, one needs a probabilistic treatment of a system, which in fact is especially suitable and relevant for quantum systems, and then use the corresponding probability densities, $\rho(\bf{r})$ in position space and $n(\bf{k})$ in momentum space as input to the definitions of Shannon information entropy and Fisher information. Thus, one proceeds to the calculation of the information content of the system and investigate its related properties. An additional merit of the probabilistic treatment is that one can calculate quantitatively, in a systematic manner, a measure of complexity of the quantum system, the so-called LMC statistical complexity \cite{LMC}. The LMC complexity of the $H_2^{+}$ ion was studied using a simple wavefunction of Coulson type, leading to a promising relation of complexity with chemical bonding \cite{MS}. Atomic complexity has been calculated for the first time in the literature in \cite{CMP}, where another definition of complexity was employed, namely the SDL measure \cite{SDL}. Last but not least, calculations of molecular information entropies were carried out in \cite{HSPSE,HCSWGSE}.

In the present paper we apply the above methods in order to study the ammonia molecule as a test bed, and in particular to assess the effect of tunneling on the Shannon information entropy, the Fisher information and the LMC complexity, together with Heisenberg's uncertainty. We employ a simple model for the ammonia molecule, i.e. the Double Square Well Potential (DSWP), which captures its essential properties required for a probabilistic treatment via the Schr\"{o}dinger Equation.

This paper is organized as follows. In section \ref{DSWP}, we derive the time-dependent wavefunction for the ammonia molecule (DSWP), and we also plot the time dependence of probability density in both position and momentum spaces. In section \ref{ISWP}, we consider the Infinite Square Well Potential (ISWP), in order to compare the corresponding results of the DSWP with a system, where quantum tunneling is absent. In section \ref{Statistical Measures} we define the relevant statistical measures and present the intrinsic relation among them. In section \ref{Results} we illustrate and comment the results i.e. the time evolution of statistical measures for both ammonia molecule (DSWP) and ISWP. The conclusions of the analysis are drawn in section \ref{Summary and Conclusions}.

\section{Ammonia Molecule $NH_3$ (DSWP)}\label{DSWP}

The ammonia molecule has the shape of a pyramid where the nitrogen atom is at the apex and the three hydrogen atoms form the base in the shape of an equilateral triangle \cite{Basdevant}. The position of the nitrogen atom is chosen as the origin of the x axis. Manning Potential \cite{Manning} constitutes a good approximation of the ammonia molecule potential, which is defined by the function

\begin{equation}
V(x) =-C\sech^2(x/2 \rho)+D\sech^4(x/2 \rho).
\end{equation}

\begin{figure}[!htb]
\centering
  \includegraphics[height=3.7cm,width=5.6cm]{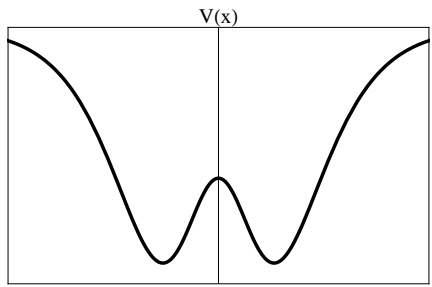}
   \caption{\small Manning Potential}
  \label{manning}
\end{figure}

\begin{figure}[!htb]
\minipage{0.32\textwidth}
  \includegraphics[height=3.7cm,width=5.6cm]{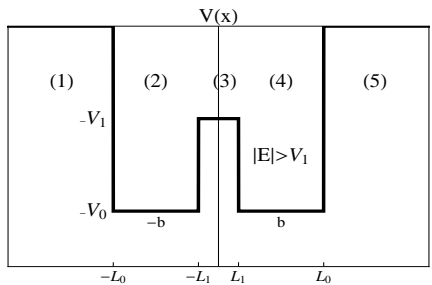}
   \caption{\small DSWP of $NH_3$}
  \label{DSWP-potential}
\endminipage\hfill
\end{figure}

In the following analysis we replace the Manning potential (Fig. \ref{manning}) by the simplified DSWP (Fig. \ref{DSWP-potential}). DSWP in general, is a tractable quantum system which can be solved analytically, and therefore it will be used to illustrate the main features of the actual potential. 

We study the quantum motion of a ``particle" representing the collective motion of the three hydrogen atoms. The mass of this hydrogen plane is equal to $m=3m_h$, where $m_h$ is the mass of a hydrogen atom. The depth is assumed to be $V_0=0.5\ \rm{eV}$, the height of the central barrier $V_1=0.25\ \rm{eV}$ and the maximum distance that the hydrogen plane can take from nitrogen is $b=0.4$ \AA \, where $b=\left( \frac{L_0+L_1}{2} \right)$ \cite{Peacock-Lopez,Basdevant,Herzberg}. Adjustments to the length of the wells $L_0$ and $L_1$ have been made, so that the calculated frequencies are in accordance with the ones that have been found experimentally ($\nu=23.98\ \rm{GHz}$) \cite{Cleeton}. 

DSWP in the ammonia molecule is defined by the following piecewise function, 

\begin{equation}
V(x) = \left\{ \begin{array}{rr}
  0, & \quad |x| > L_0  \\
  -V_0, &\quad L_1 < |x| < L_0 \\
  -V_1, &\quad |x| < L_1 
  \end{array}
  \right. ,
\end{equation}

\noindent where $V_0=0.5\ \rm{eV}$, $V_1=0.25\ \rm{eV}$, $L_0=0.672$ \AA \, and $L_1=0.128$ \AA. The one-dimensional Schr\"{o}dinger Equation (SE) is:

\begin{equation}
-\frac{\hbar^2}{2m}\frac{d^2u(x)}{dx^2 }+ V(x)u(x) = Eu(x).
\end{equation}

By dividing the $x$ axis into five spatial regions, and following the standard procedure \cite{Basdevant, Massen} we solve analytically the corresponding second order differential equations, taking into account continuity and smoothness conditions for the wavefunction at $x=L_0$ and $x=L_1$.

The potential is an even function of position, i.e., $V (x) = V (-x)$. Due to this symmetry, the solutions of the SE are either odd or even functions of position. We solve these equations only for $|E|>V_1$, because only in that case the energy is less than the barrier's height, and so quantum tunneling is possible.

The symmetric (even) solution is:

\begin{equation}
u_{S}(x) = \left\{ \begin{array}{lr}
   A_1 e^{\gamma x}, & x \leq -L_0 \\
   A_2 \cos (\varphi - k_0x), & -L_0\leq x \leq -L_1 \\
   A_3 \cosh ( |k_1|x ), & |x|\leq L_1\\
   A_2 \cos (\varphi + k_0x), & L_1 \leq x \leq L_0\\
   A_1e^{-\gamma x}, & x \geq L_0
  \end{array}
  \right. ,
\end{equation}

\noindent while the antisymmetric one (odd) is:

\begin{equation}
u_{A}(x) = \left\{ \begin{array}{lr}
  A_1 e^{\gamma x}, & x\leq -L_0 \\
  A_2 \cos (\varphi - k_0x), & -L_0 \leq x \leq -L_1 \\
  B_3 \sinh ( |k_1|x), & |x|\leq L_1 \\
  -A_2 \cos (\varphi + k_0x), & L_1 \leq x \leq L_0 \\
  -A_1 e^{-\gamma x}, & x\geq L_0
  \end{array}
  \right. ,
\end{equation}

\noindent where
  
\begin{equation}
\gamma^2 = -\frac{2m}{\hbar^2}E, \quad
k_0^2 = \frac{2m}{\hbar^2}(E+V_0),  \quad
k_1^2 = \frac{2m}{\hbar^2}(E+V_1).
\end{equation} 

In the even case we set $A_1=1$, and the values of the constants $A_2, A_3$ are obtained by solving the resulting transcendental equation numerically. Working similarly in the odd case, we set again $A_1=1$ and find $A_2$ and $B_3$. The energies and phases found numerically are:

\begin{table}[ht]
\caption{Energies and Phases in ammonia molecule $NH_3$} 
\centering 
\begin{tabular}{c c c } 
\hline\hline 
Indices \footnote{The indices show the number of nodes of the wavefunction and the parity of the symmetric S and antisymmetric A eigenstates respectively.} & \qquad $eV$ &  $\qquad  \varphi$  \\ [0.5ex] 
\hline 
0S & \quad -0.4831090 &  \qquad 1.20559  \\ 
1A & \quad  -0.4830108  & \qquad 1.19540  \\
2S & \quad  -0.4331151 &  \qquad 0.86912 \\
3A & \quad  -0.4325328 & \qquad 0.83870  \\ [1ex] 
\hline 
\end{tabular}
\label{table:nonlin} 
\end{table}

The eigenfunctions have well-defined parities: a symmetric one $(u_S)$, and an antisymmetric one $(u_A)$. We observe that the energy levels which lie below the potential barrier occur in pairs. For each parity, the two smallest values of energy correspond to the ground-state, while the rest to the excited ones, thus we find

\begin{equation*}
\begin{array}{l}
\Delta E_{10}=9.82\cdot 10^{-5}\ \rm{eV},\\
\Delta E_{32}=58.23\cdot 10^{-5}\ \rm{eV},
\end{array} 
\end{equation*}

In the excited states we obtain lower accuracy, since the DSWP has been adjusted only for the ground-state, in which we are more interested in. We can form linear superpositions of the energy eigenstates,

\begin{equation}
 u_L=\frac{u_S+u_A}{\sqrt{2}} ,
\label{superposition}
\end{equation}
 
\noindent and

\begin{equation}
 u_R=\frac{u_S-u_A}{\sqrt{2}},
\end{equation}

\noindent which are not eigenfunctions of the system, and correspond to states for which the probability density in position space $\rho(x)$ is concentrated on the left and on the right well respectively. The time-dependent wavefunction of the particle, assumed to be in the left well at t = 0, evolves in the ground-state according to

\begin{eqnarray}
\psi(x,t) &=& \frac{1}{\sqrt{2}}[u_{0S}(x)e^{-i{E_{0S}}t/\hbar}+u_{1A}(x)e^{-i{E_{1A}}t/\hbar}]      \nonumber \\
   &=& \frac{1}{\sqrt{2}}[u_{0S}(x)+u_{1A}(x)e^{-i\Delta E_{10}t/\hbar}]e^{-i{E_{0S}}t/\hbar},
\end{eqnarray}

\noindent and for the first excited state

\begin{equation}
\psi(x,t) =\frac{1}{\sqrt{2}}[u_{2S}(x)+u_{3A}(x)e^{-i\Delta E_{32}t/\hbar}]e^{-i{E_{2S}}t/\hbar}.
\end{equation}

The normalized wavefunction in momentum space is given by the the Fourier transform

\begin{equation}
\phi(k,t)=\frac{1}{\sqrt{2\pi}} \int \psi(x,t)e^{-ikx} \,dx.
\end{equation}

\begin{figure}[!htb]
        \centering
        \begin{subfigure}[b]{0.3\textwidth}
                \centering
                \includegraphics[width=\textwidth]{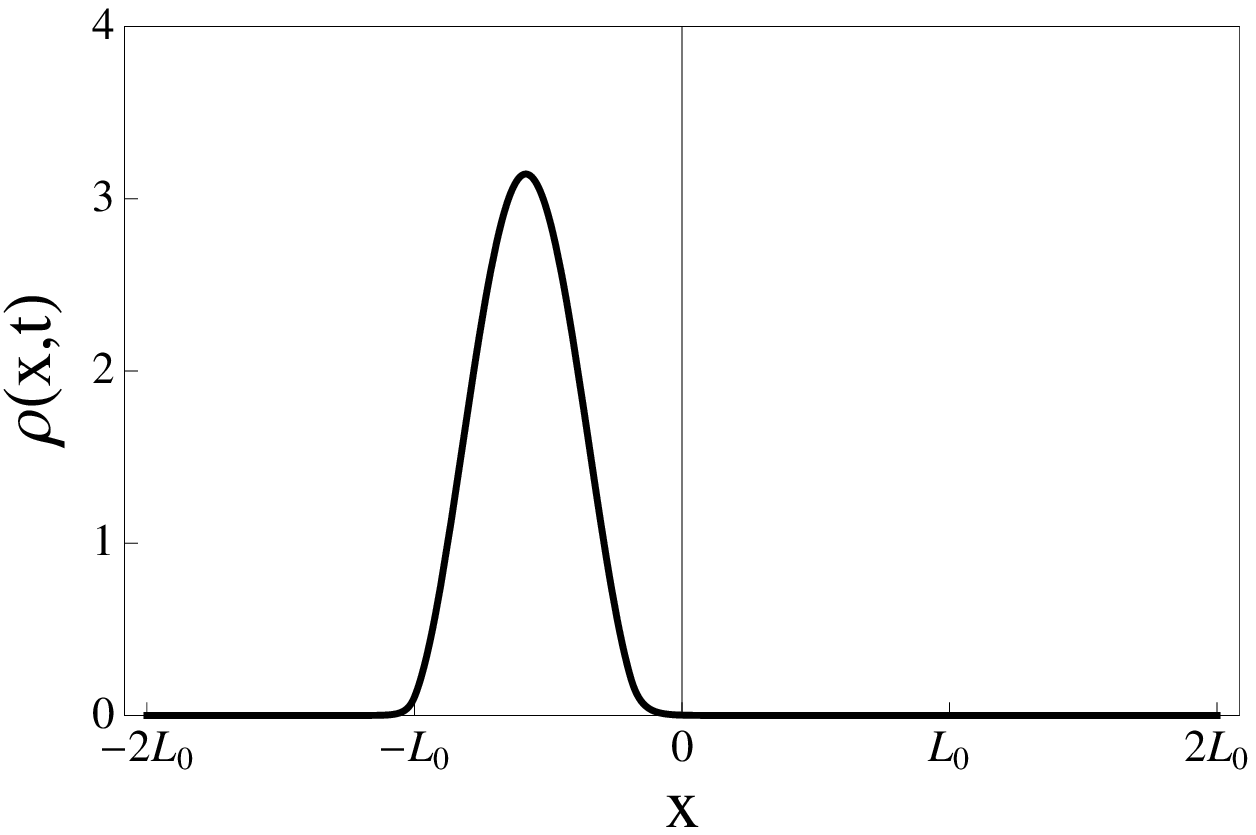}
                \caption{\small t=0, T}
             
        \end{subfigure} 
         \begin{subfigure}[b]{0.3\textwidth}
                \centering
                \includegraphics[width=\textwidth]{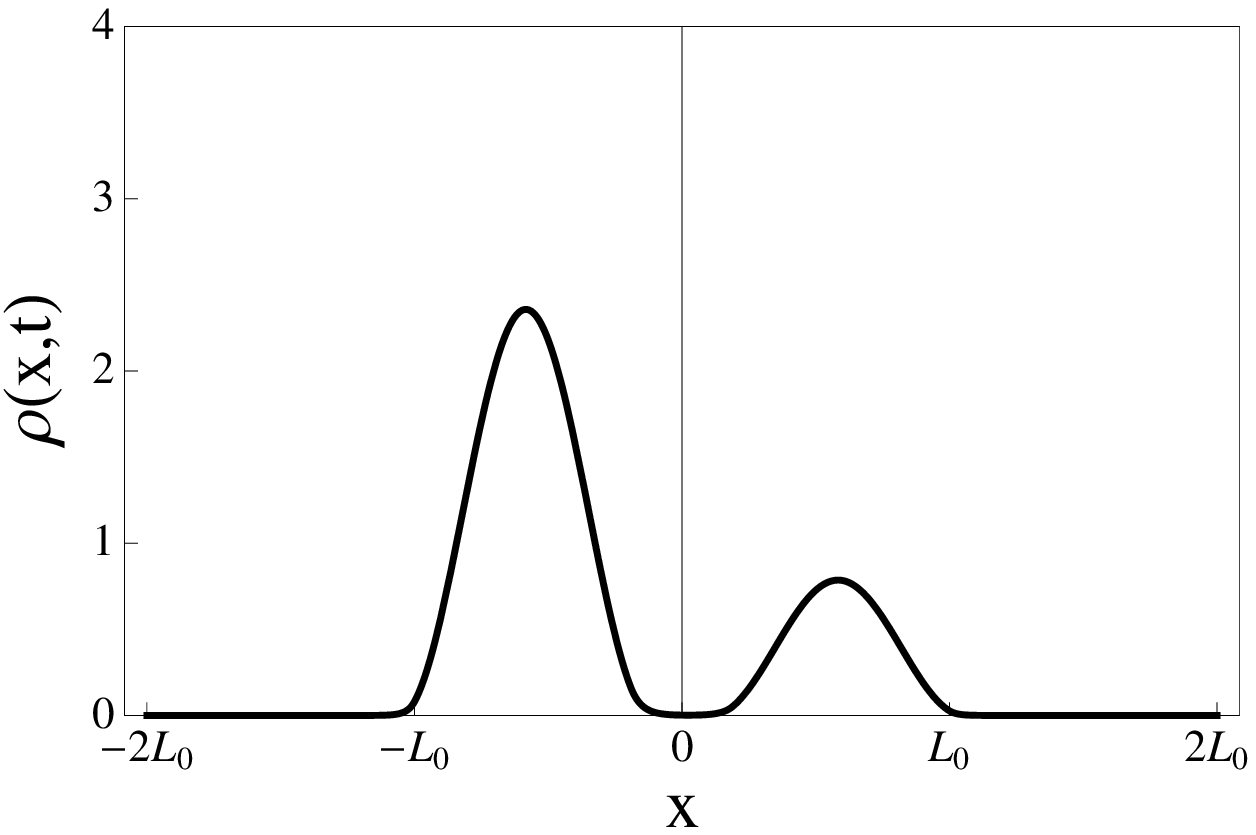}
                \caption{\small t=T/6, 5T/6}
          
        \end{subfigure}
       \begin{subfigure}[b]{0.3\textwidth}
                \centering
                \includegraphics[width=\textwidth]{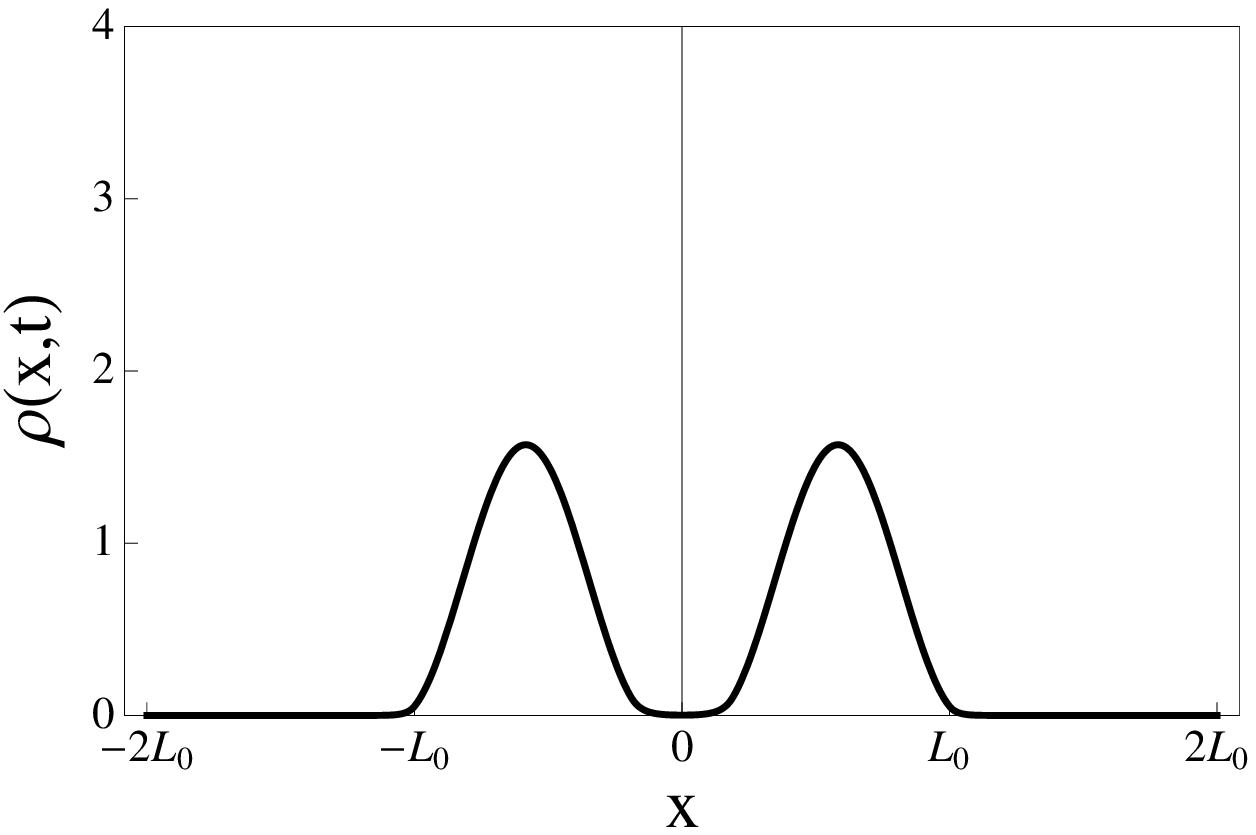}
                \caption{\small t=T/4, 3T/4}
         
        \end{subfigure}
         \begin{subfigure}[b]{0.3\textwidth}
                \centering
                \includegraphics[width=\textwidth]{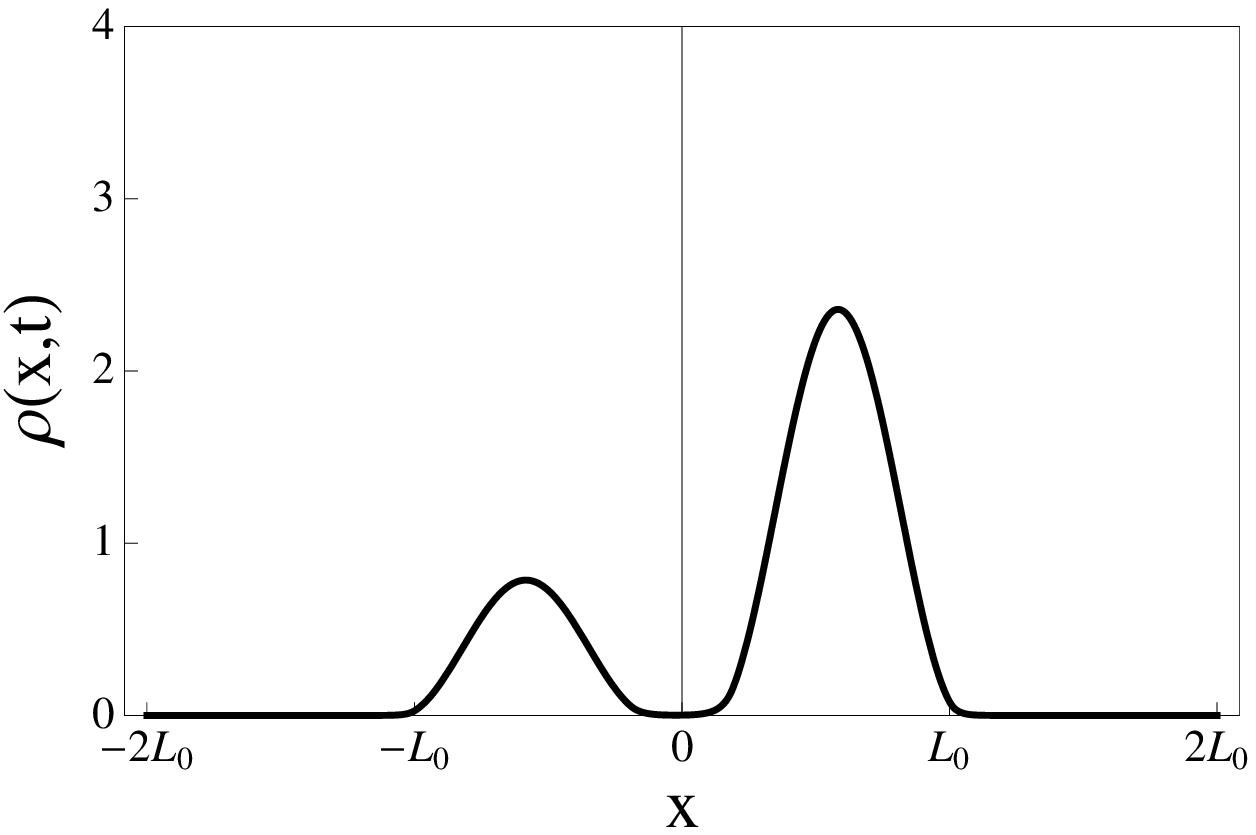}
                \caption{\small t=T/3, 2T/3}
            
        \end{subfigure} 
         \begin{subfigure}[b]{0.3\textwidth}
                \centering
                \includegraphics[width=\textwidth]{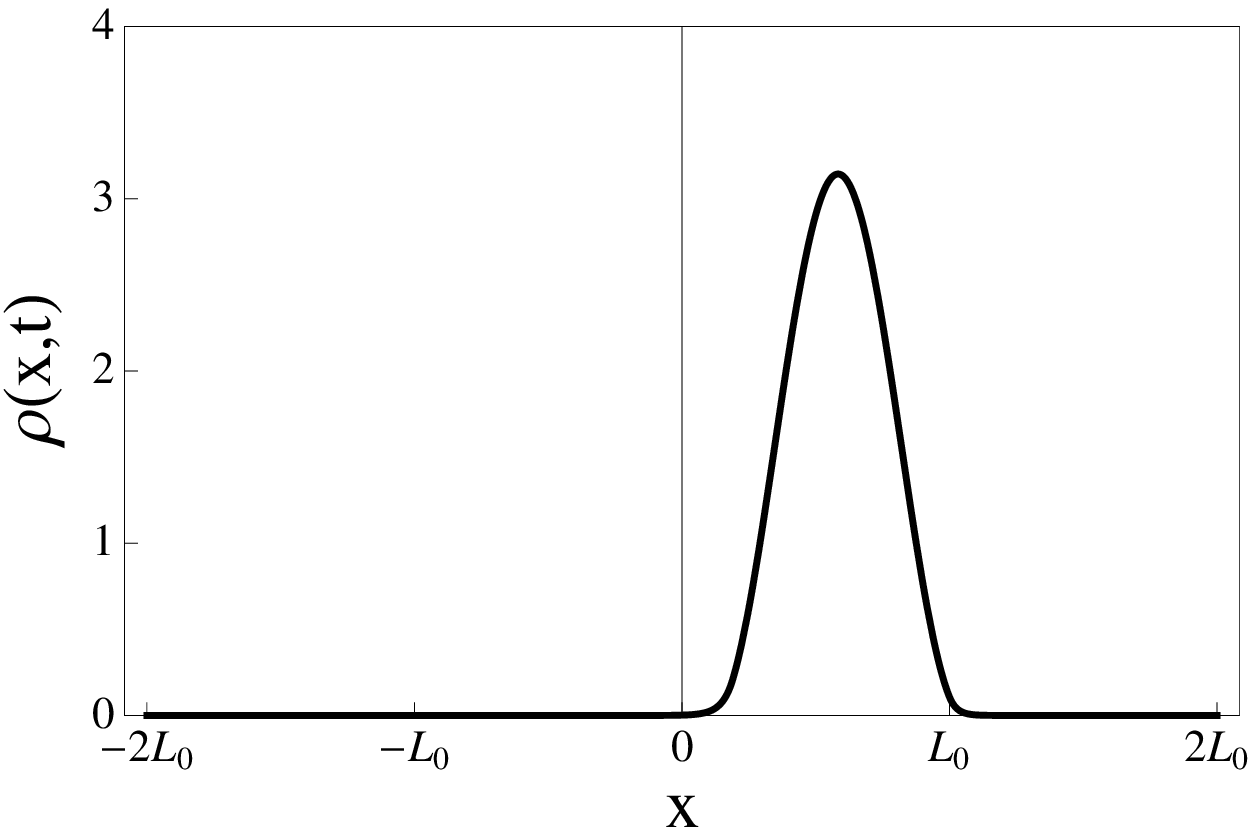}
                \caption{\small t=T/2}
                
        \end{subfigure}
   \caption{\small Probability density $\rho(x,t)$ in position space (DSWP)}\label{fig:DSWP-Oscillation-Position}
\end{figure}

\begin{figure}[!htb]
        \centering
        \begin{subfigure}[b]{0.3\textwidth}
                \centering
                \includegraphics[width=\textwidth]{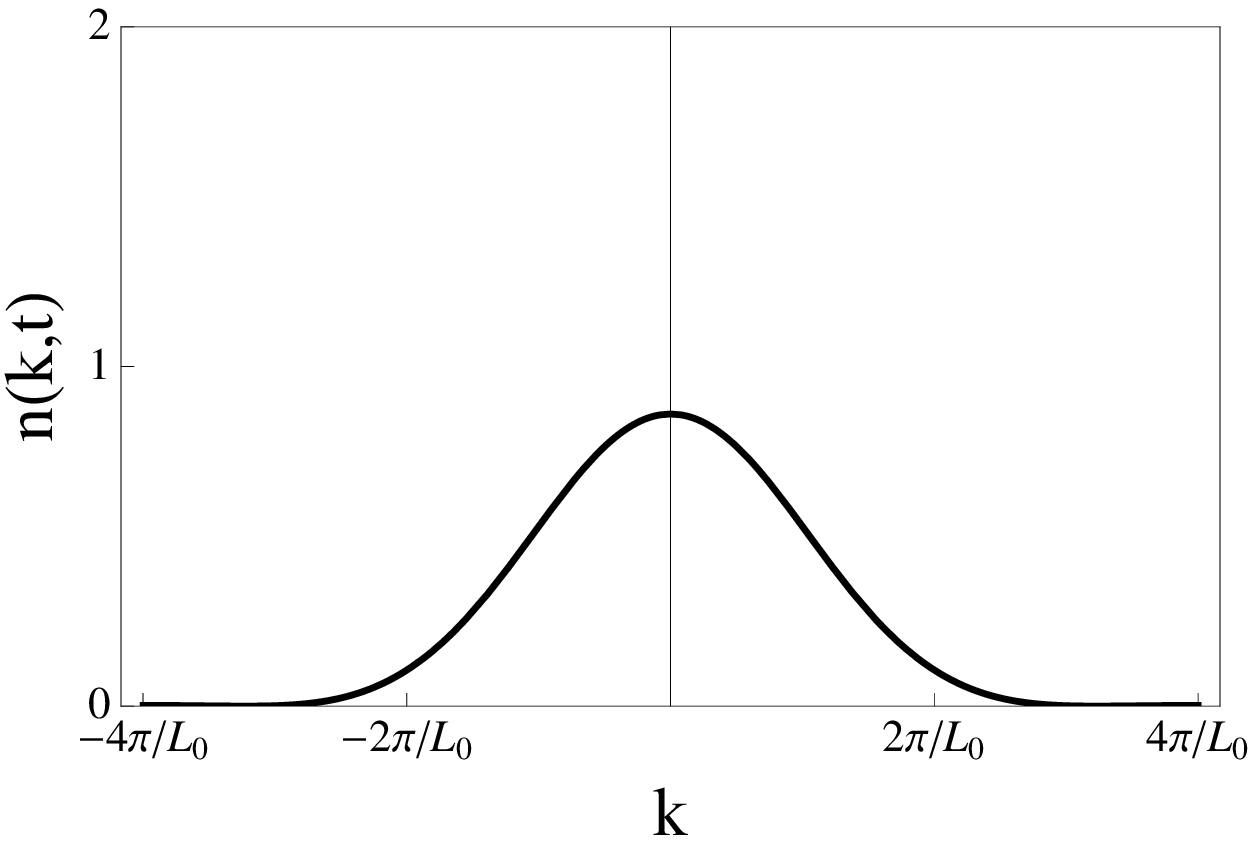}
                \caption{\small t=0}
          
        \end{subfigure} 
         \begin{subfigure}[b]{0.3\textwidth}
                \centering
                \includegraphics[width=\textwidth]{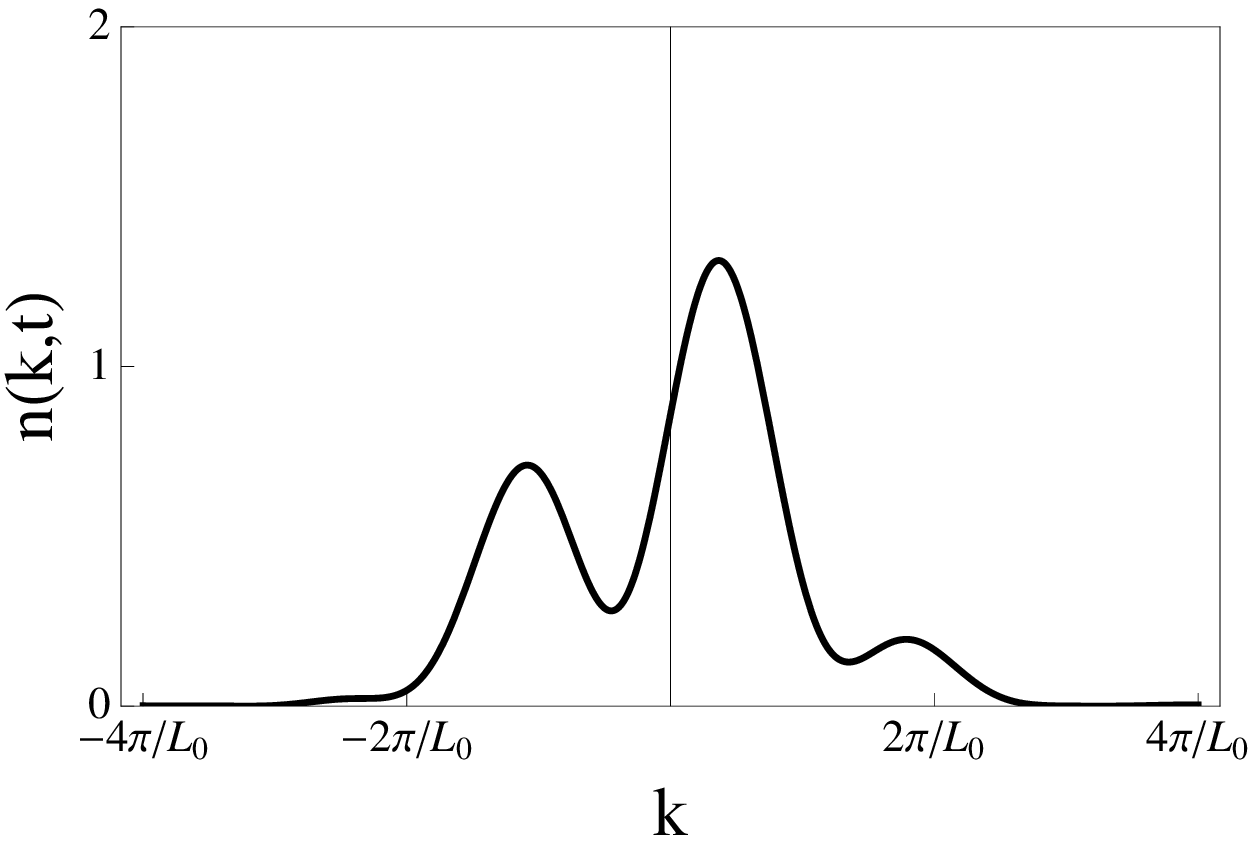}
                \caption{\small t=T/9}
          
        \end{subfigure}
       \begin{subfigure}[b]{0.3\textwidth}
                \centering
                \includegraphics[width=\textwidth]{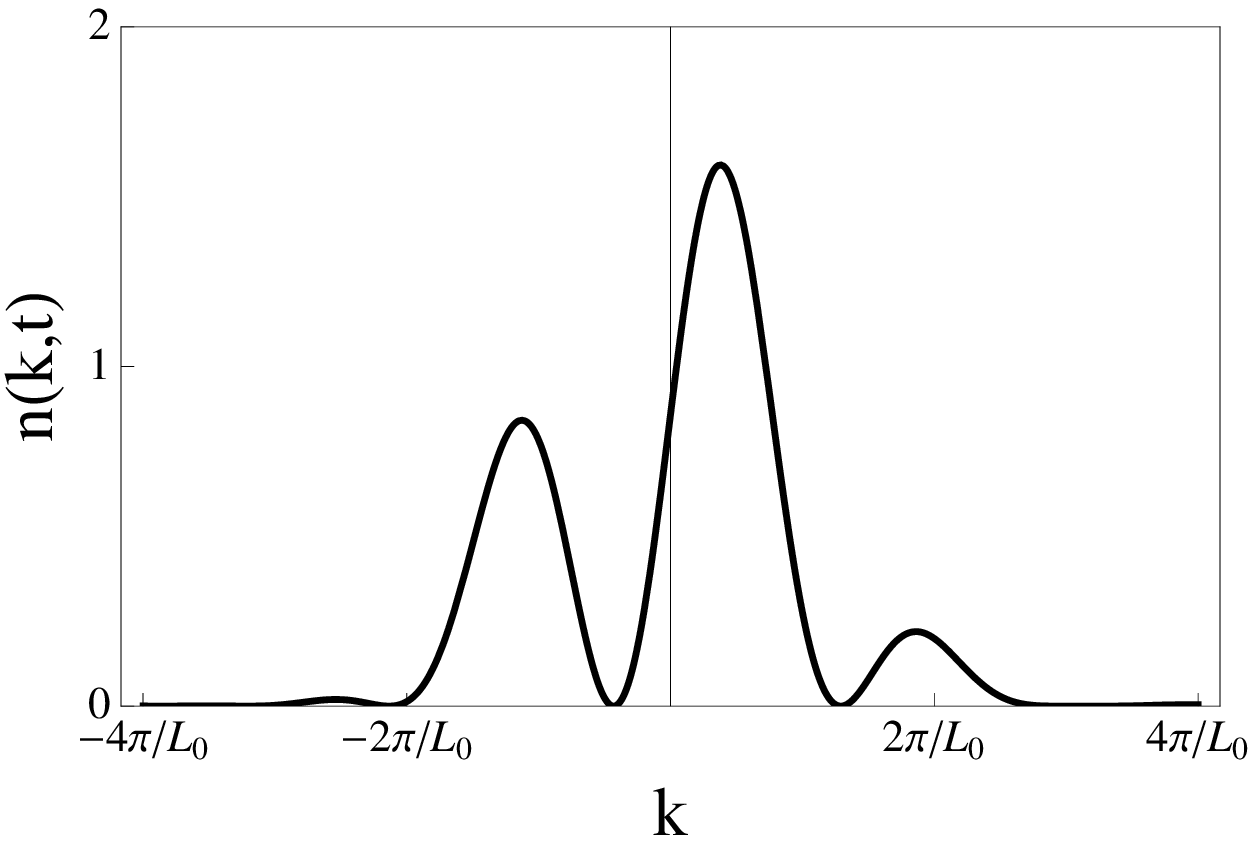}
                \caption{\small t=T/4}
            
        \end{subfigure}
         \begin{subfigure}[b]{0.3\textwidth}
                \centering
                \includegraphics[width=\textwidth]{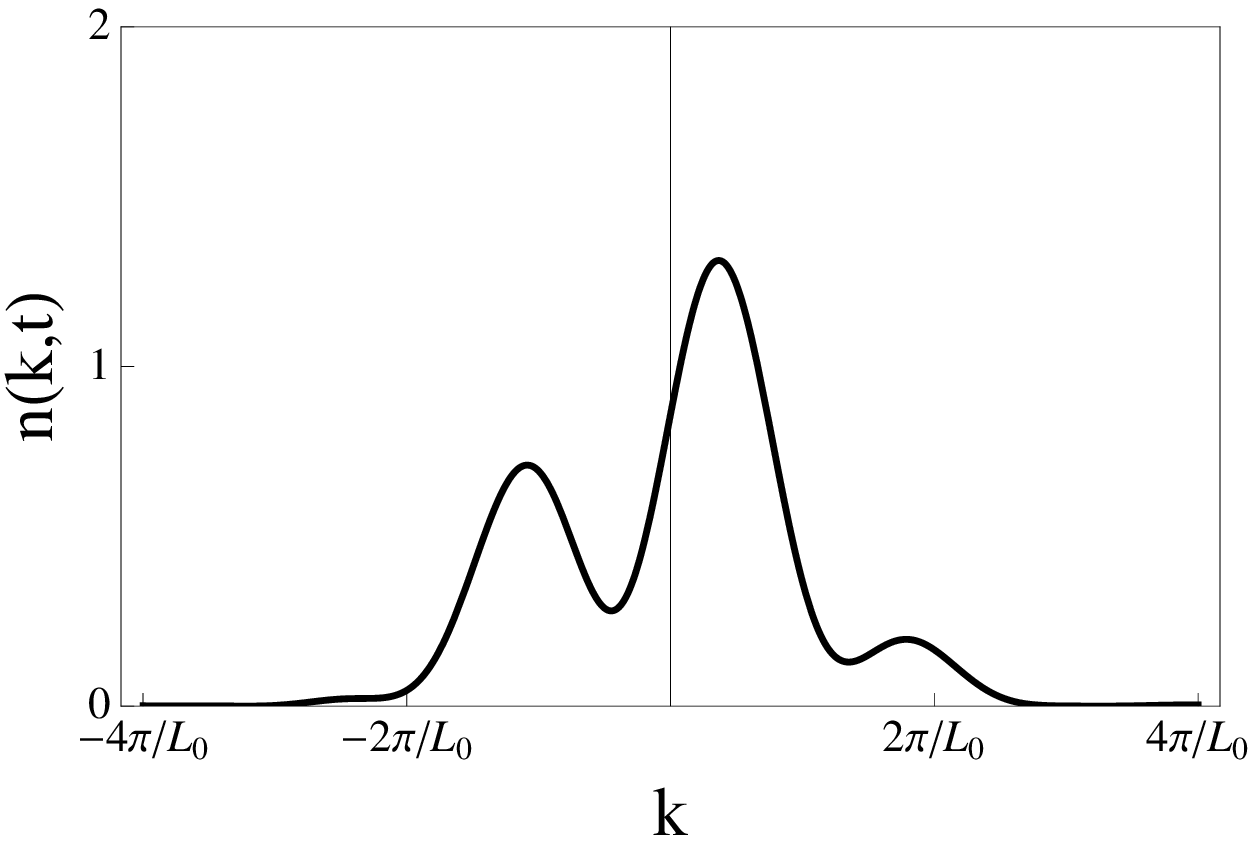}
                \caption{\small t=7T/18}
         
        \end{subfigure} 
         \begin{subfigure}[b]{0.3\textwidth}
                \centering
                \includegraphics[width=\textwidth]{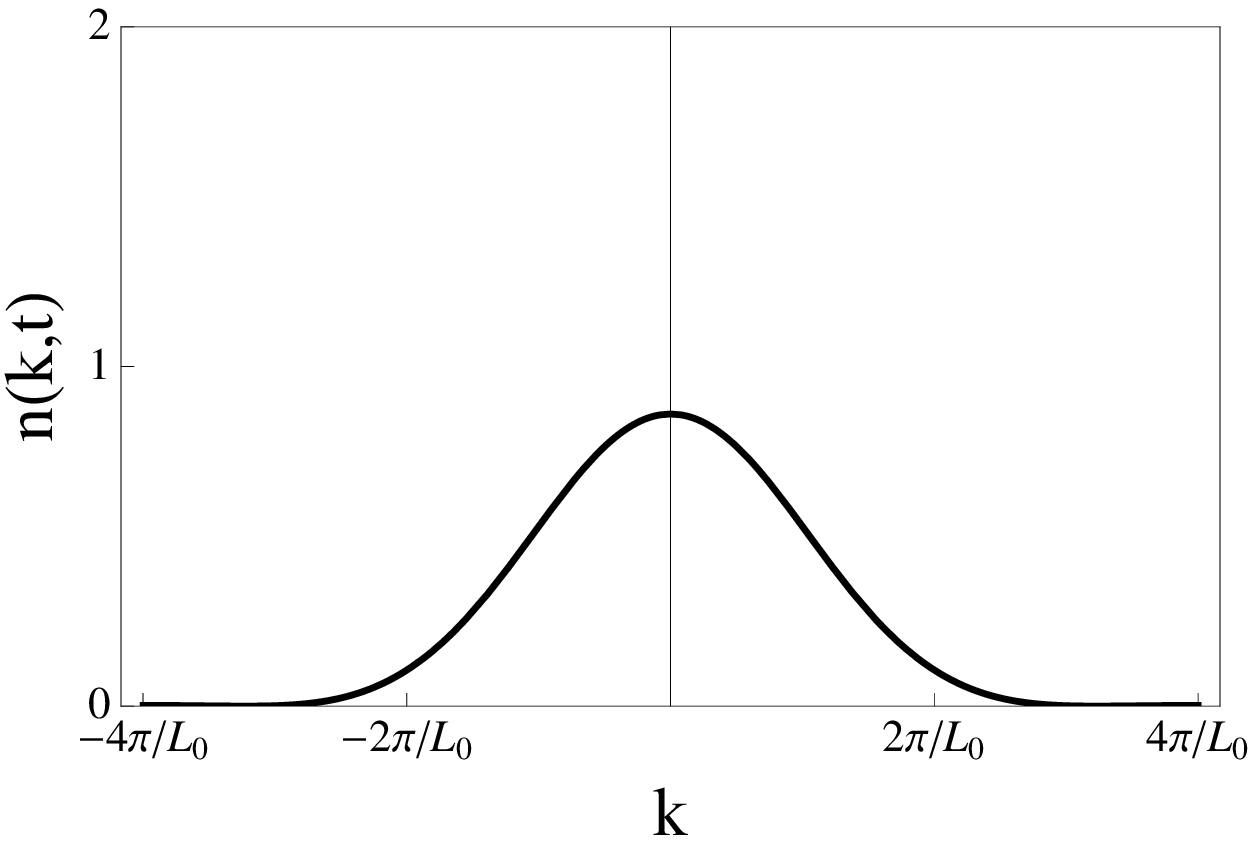}
                \caption{\small t=T/2}
         
        \end{subfigure}
     \begin{subfigure}[b]{0.3\textwidth}
                \centering
                \includegraphics[width=\textwidth]{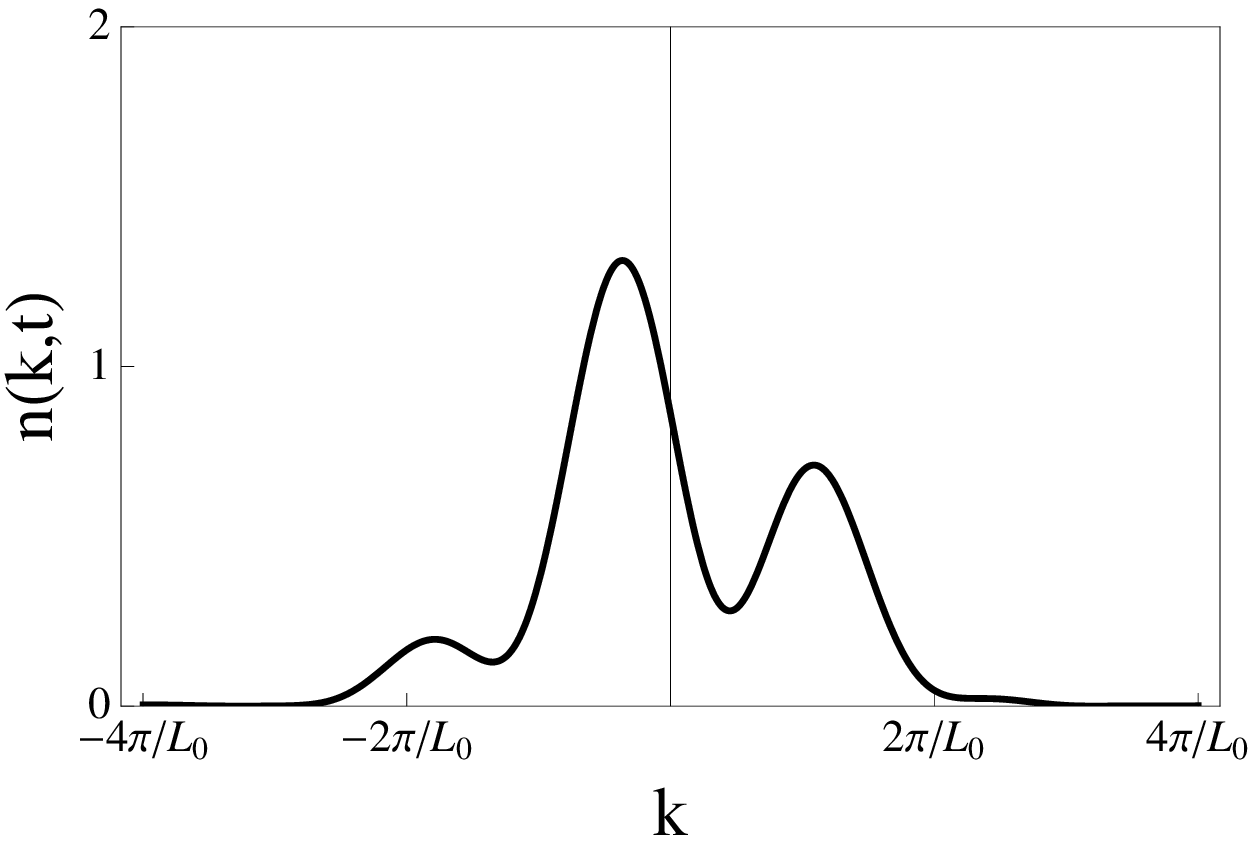}
                \caption{\small t=11T/18}
            
        \end{subfigure}
         \begin{subfigure}[b]{0.3\textwidth}
                \centering
                \includegraphics[width=\textwidth]{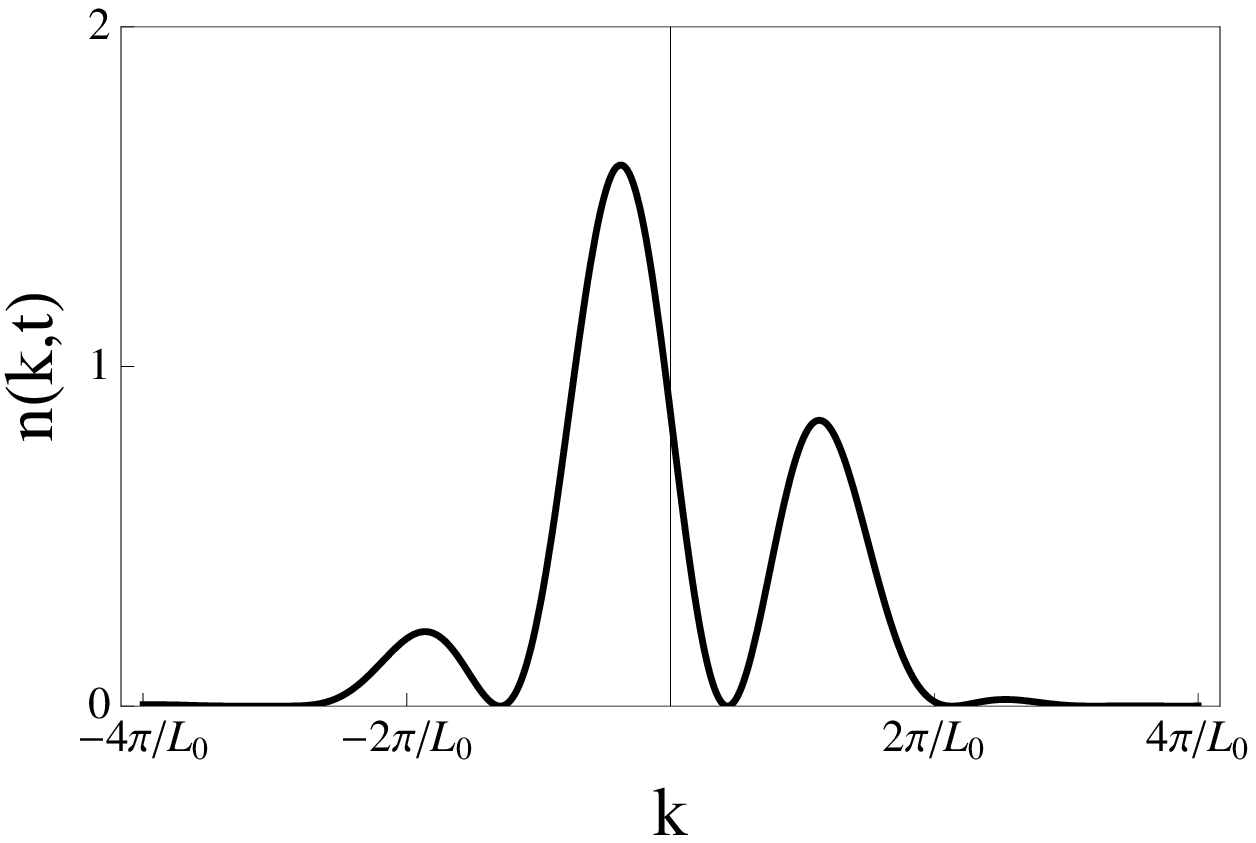}
                \caption{\small t=3T/4}
            
        \end{subfigure} 
         \begin{subfigure}[b]{0.3\textwidth}
                \centering
                \includegraphics[width=\textwidth]{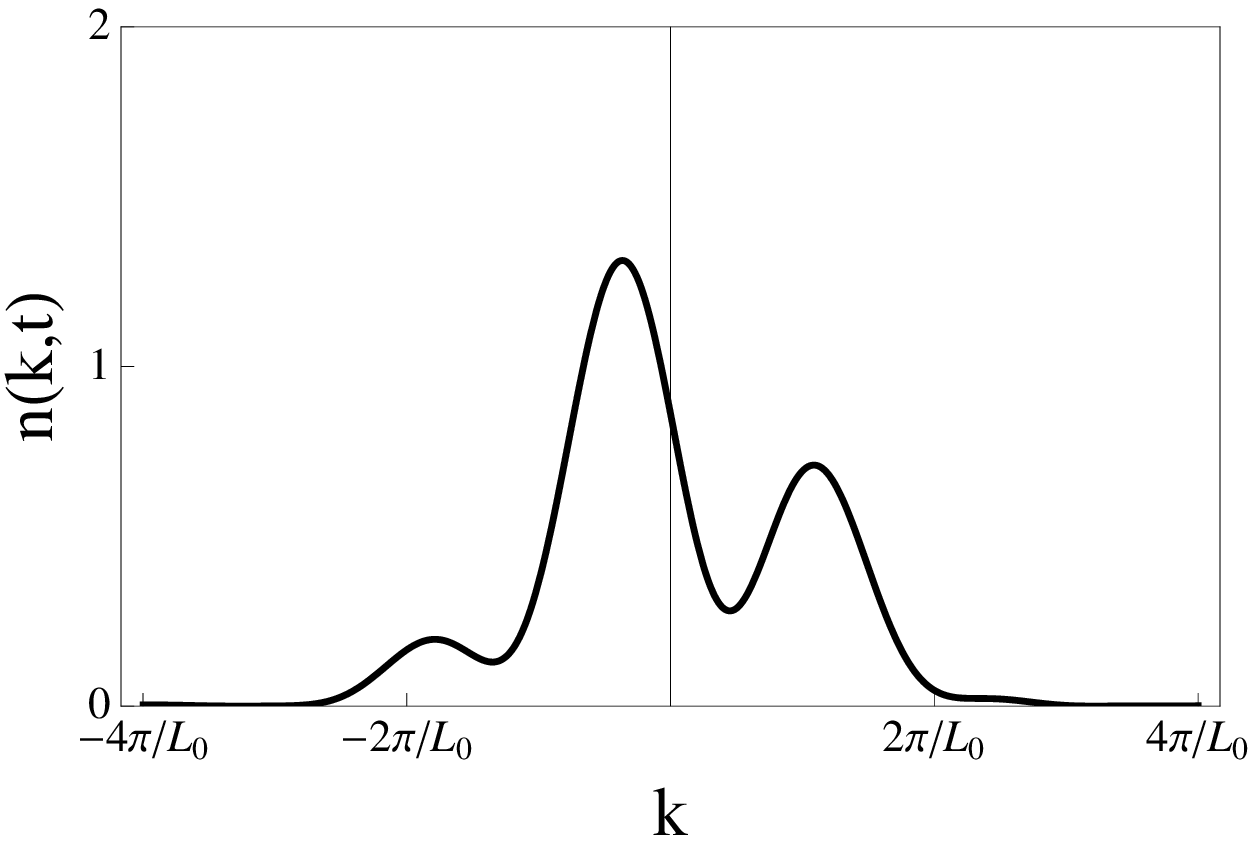}
                \caption{\small t=8T/9}
             
        \end{subfigure}
        \begin{subfigure}[b]{0.3\textwidth}
                \centering
                \includegraphics[width=\textwidth]{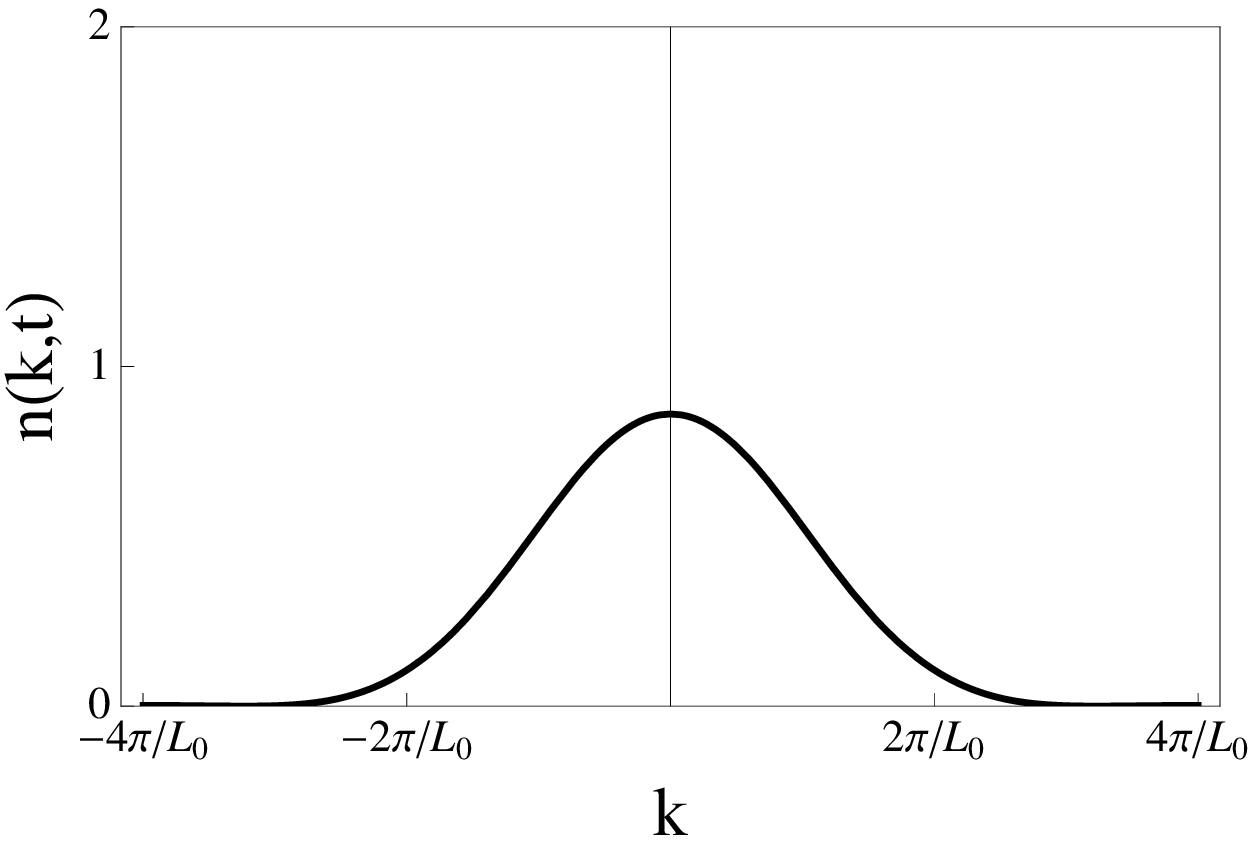}
                \caption{\small t=T}
                
        \end{subfigure}
   \caption{\small Probability density $n(k,t)$ in momentum space (DSWP)}\label{fig:DSWP-Oscillation-Momentum}
\end{figure}

Probability densities in position and momentum space are given by

\begin{equation}
\rho(x,t)=|\psi(x,t)|^2=\psi(x,t)^*\psi(x,t),
\label{pdr}
\end{equation}

\noindent and

\begin{equation}
n(k,t)=|\phi(k,t)|^2=\phi(k,t)^*\phi(k,t),
\label{pdn}
\end{equation}

\noindent respectively. In Figs. \ref{fig:DSWP-Oscillation-Position} and \ref{fig:DSWP-Oscillation-Momentum} we present how probability density develops over time for the ground-state. Instead of $\rho$ and $n$, we put the values $\rho \cdot 10^{-10}$ and $n \cdot 10^{11}$ respectively, in order to avoid extremely large or small magnitudes. It is noted that the Schr\"{o}dinger equation was solved in the S.I. system of units.

We notice that after time $t=\pi \hbar/\Delta E$, the wavefunction $\psi(x,t)$ is equal to $u_R$, and the particle is in the right configuration. At $t=2\pi \hbar/\Delta E$ the wavefunction $\psi(x,t)$ is equal to $u_L$ and the particle is back at the left well. Therefore, the particle in the ground-state oscillates from the left well to the right one at the Bohr frequency $\nu=\Delta E/2\pi \hbar=23.76\ \rm{GHz}$, while the experimental measured frequency is $\nu=23.98\ \rm{GHz}$ \cite{Cleeton}, and the corresponding value calculated through Manning Potential is $\nu=24.88\ \rm{GHz}$ \cite{Manning}. In other words, the hydrogen plane can be moved to a position diametrically opposite to where it was, which leads to the inversion of the molecule. The resulting molecule is indistinguishable from the initial one and this phenomenon is called inversion spectrum of ammonia molecule.

\clearpage

\section{Infinite Square Well Potential (ISWP)}\label{ISWP}

It is obvious that the Bohr frequency plays a fundamental role in quantum tunneling, since the time-dependent wavefunction of the particle is the result of a superposition state. Similar superpositions of eigenstates can also be formed in other systems, where quantum tunneling is absent. In this section, we apply an analysis similar to the DSWP case, and form the corresponding superpositions, which evolve with time in a well without a barrier, and in particular in the Infinite Square Well Potential (ISWP) \cite{LSHR}. Since ISWP is discussed in every undergraduate Quantum Mechanics textbook \cite{Basdevant,Grif}, we skip the mathematical analysis and directly present the eigenfunctions and the energy levels of the system. 

The ISWP is defined by the function

\begin{equation}
V(x) = \left\{ \begin{array}{rr}
  0, & \quad 0 < x < L \\
  \infty, & \quad x < 0 ,\, x >L \\
  \end{array}
  \right. .
\end{equation}

The normalized eigenfunctions $\psi_n(x)$ are

\begin{eqnarray}
\psi_n(x)=\sqrt {\frac{2}{L}} \sin\left(\frac{n\pi x}{L}\right) \quad, \quad n=1,2,3 \ldots
\end{eqnarray}

The corresponding quantized energy levels of the ISWP are given by

\begin{equation}
E_n=\frac{\hbar^2\pi^2n^2}{2mL^2} \quad, \quad n=1,2,3 \ldots
\end{equation}

We set $L=2 L_0$ and $m=3m_h$, where $2L_0$ is the width of the DSWP used in section \ref{DSWP}, and form again linear superpositions (as we did for DSWP case) of the two lowest energy eigenstates $\psi_1$ and $\psi_2$. 

\begin{equation}
\psi_L=\frac{\psi_1+\psi_2}{\sqrt{2}}, 
\end{equation}
 
\noindent and

\begin{equation}
\psi_R=\frac{\psi_1-\psi_2}{\sqrt{2}}, 
\end{equation}

\noindent corresponding to wavefunctions in which the particle is located predominantly in the left or right side of the well respectively. The time-dependent wavefunction for a particle initially located at the left side of the well at $t = 0$, is given by

\begin{equation}
\psi(x,t) = \frac{1}{\sqrt{2}}[\psi_1(x)-\psi_2(x)e^{-i\Delta E_{21}t/\hbar}]e^{-iE_1t/\hbar}.
\end{equation}

\begin{figure}[!htb]
        \centering
        \begin{subfigure}[b]{0.3\textwidth}
                \centering
                \includegraphics[width=\textwidth]{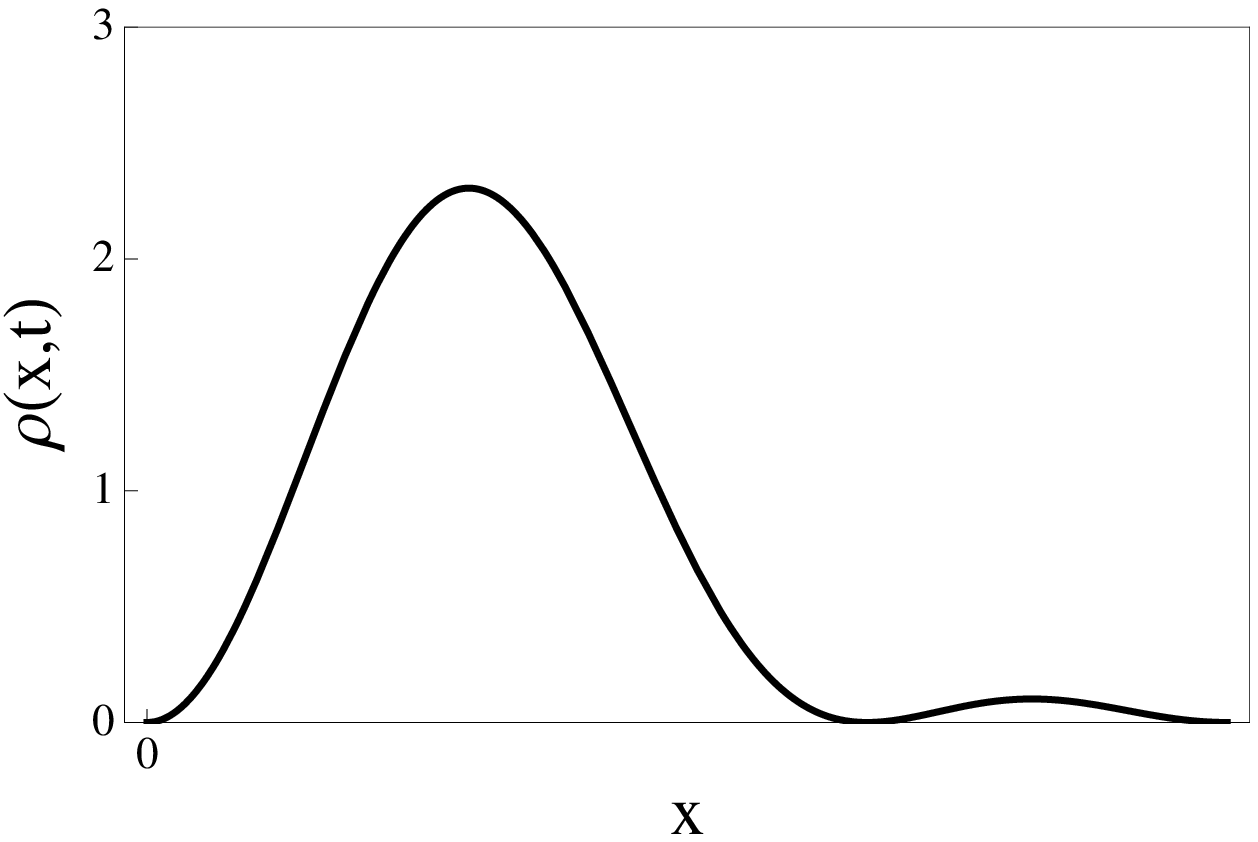}
                \caption{\small t=0, T}
             
        \end{subfigure} 
         \begin{subfigure}[b]{0.3\textwidth}
                \centering
                \includegraphics[width=\textwidth]{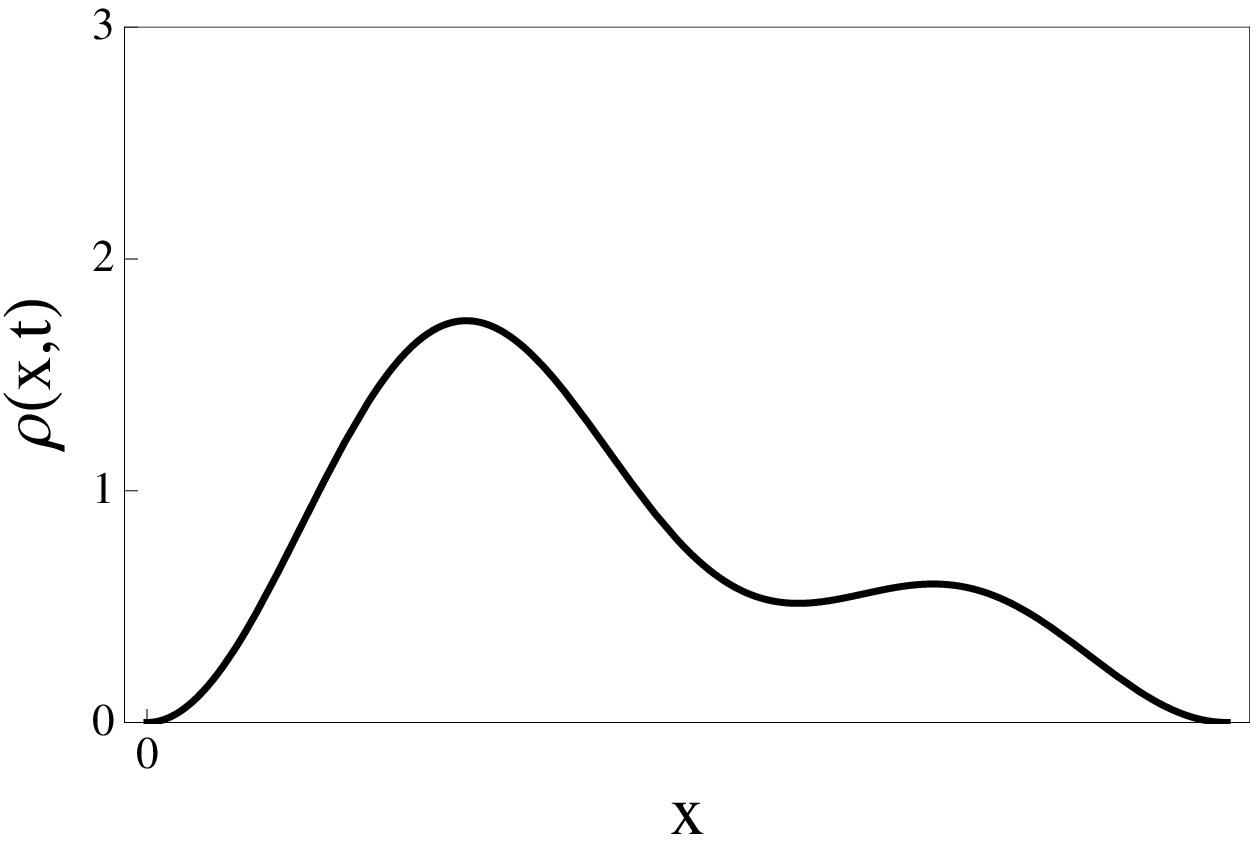}
                \caption{\small t=T/6, 5T/6}
            
        \end{subfigure}
       \begin{subfigure}[b]{0.3\textwidth}
                \centering
                \includegraphics[width=\textwidth]{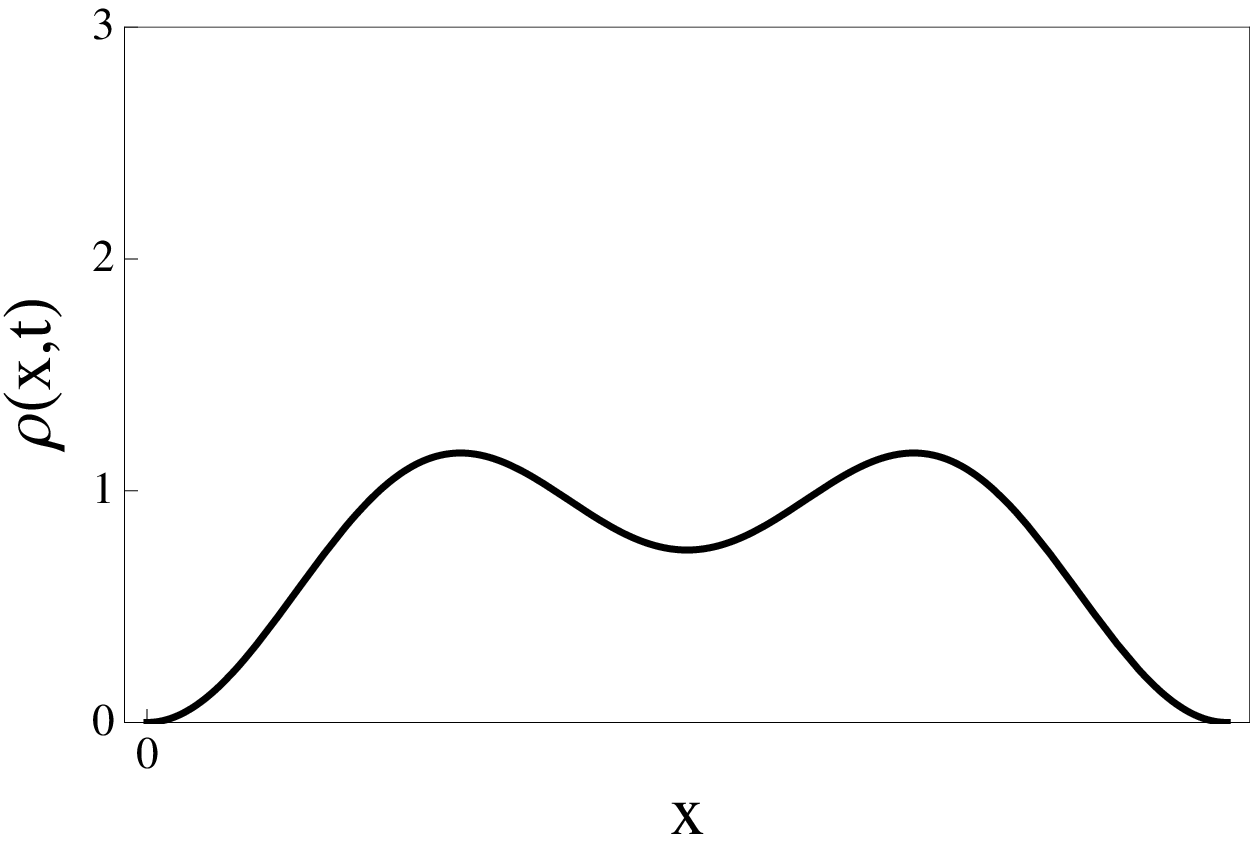}
                \caption{\small t=T/4, 3T/4}
      
        \end{subfigure}
         \begin{subfigure}[b]{0.3\textwidth}
                \centering
                \includegraphics[width=\textwidth]{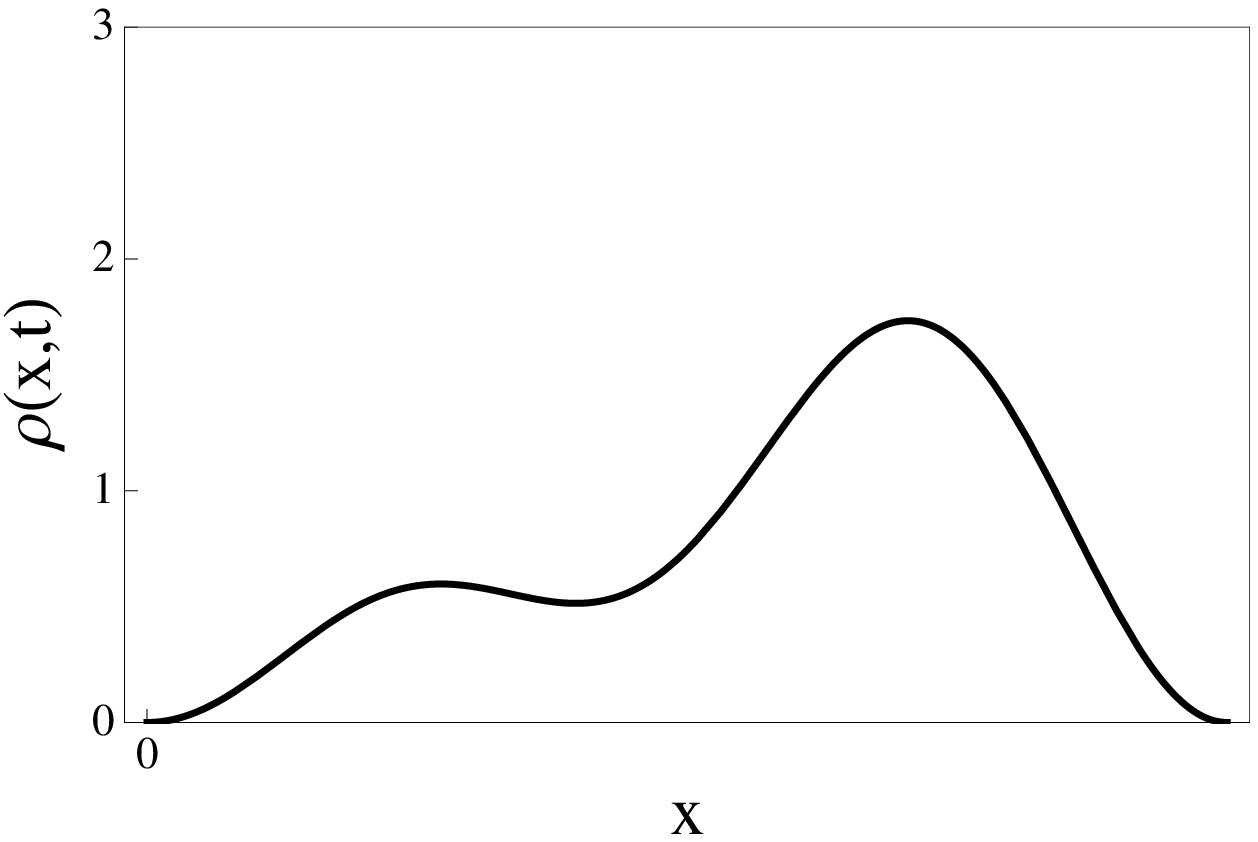}
                \caption{\small t=T/3, 2T/3}
           
        \end{subfigure} 
         \begin{subfigure}[b]{0.3\textwidth}
                \centering
                \includegraphics[width=\textwidth]{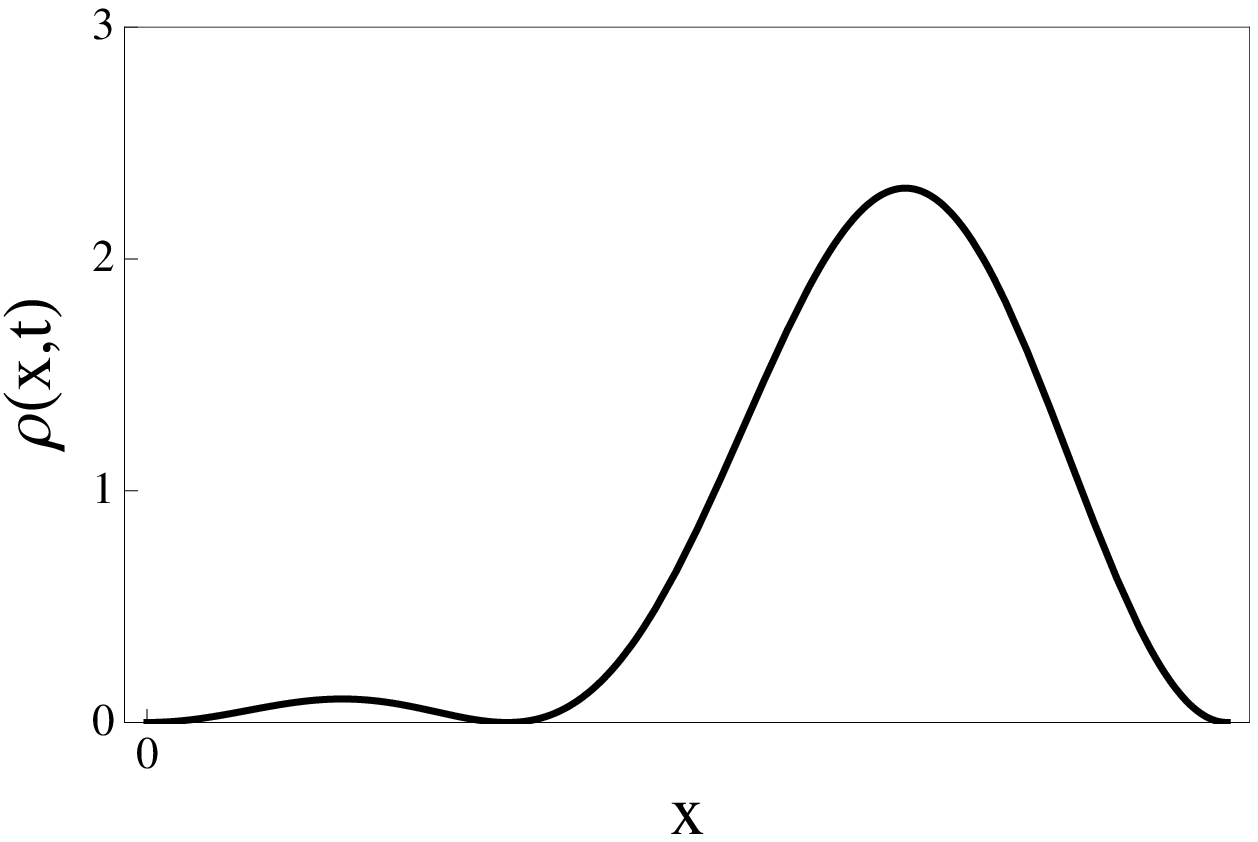}
                \caption{\small t=T/2}
              
        \end{subfigure}
\caption{\small Probability density $\rho(x,t)$ in position space (ISWP)}\label{fig:ISWP-Oscillation-Position}
\end{figure}

\begin{figure}[!htb]
        \centering
        \begin{subfigure}[b]{0.3\textwidth}
                \centering
                \includegraphics[width=\textwidth]{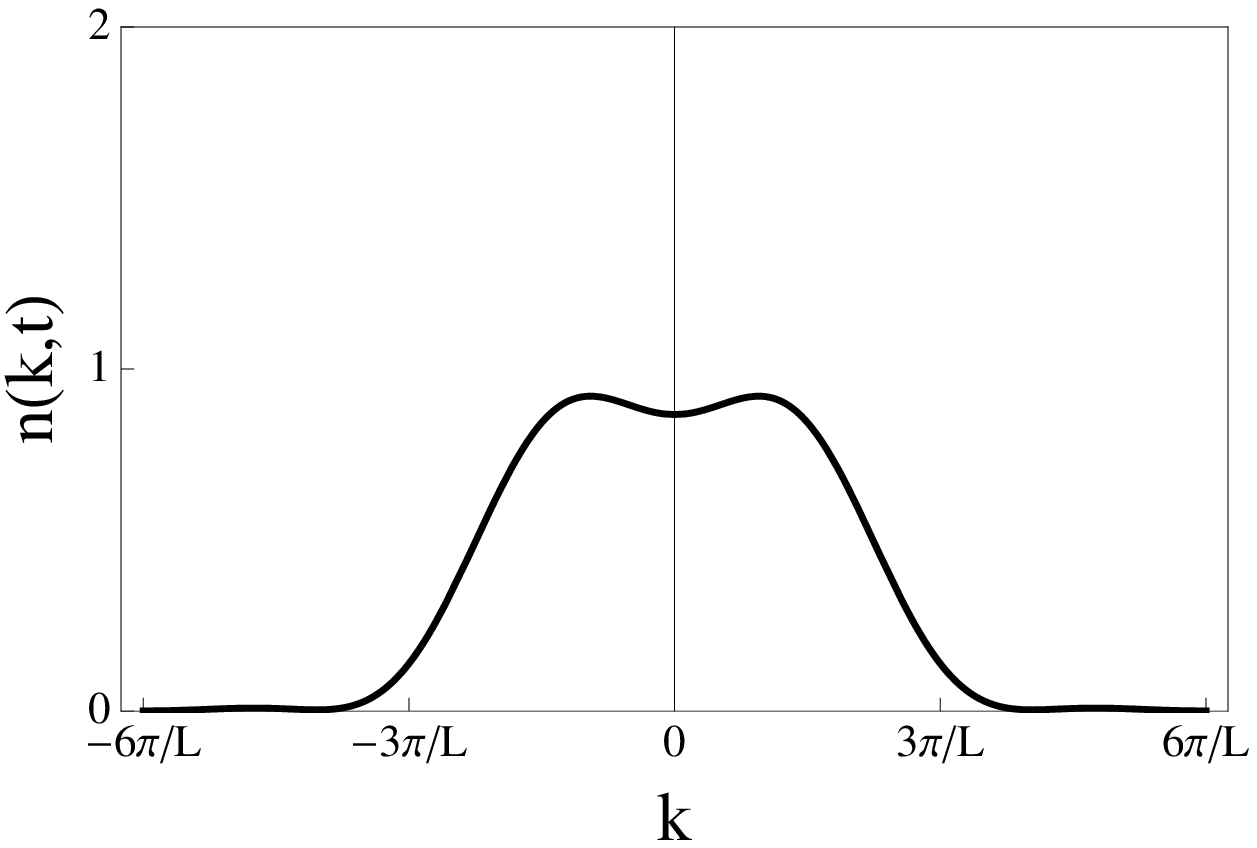}
                \caption{\small t=0}
               
        \end{subfigure} 
         \begin{subfigure}[b]{0.3\textwidth}
                \centering
                \includegraphics[width=\textwidth]{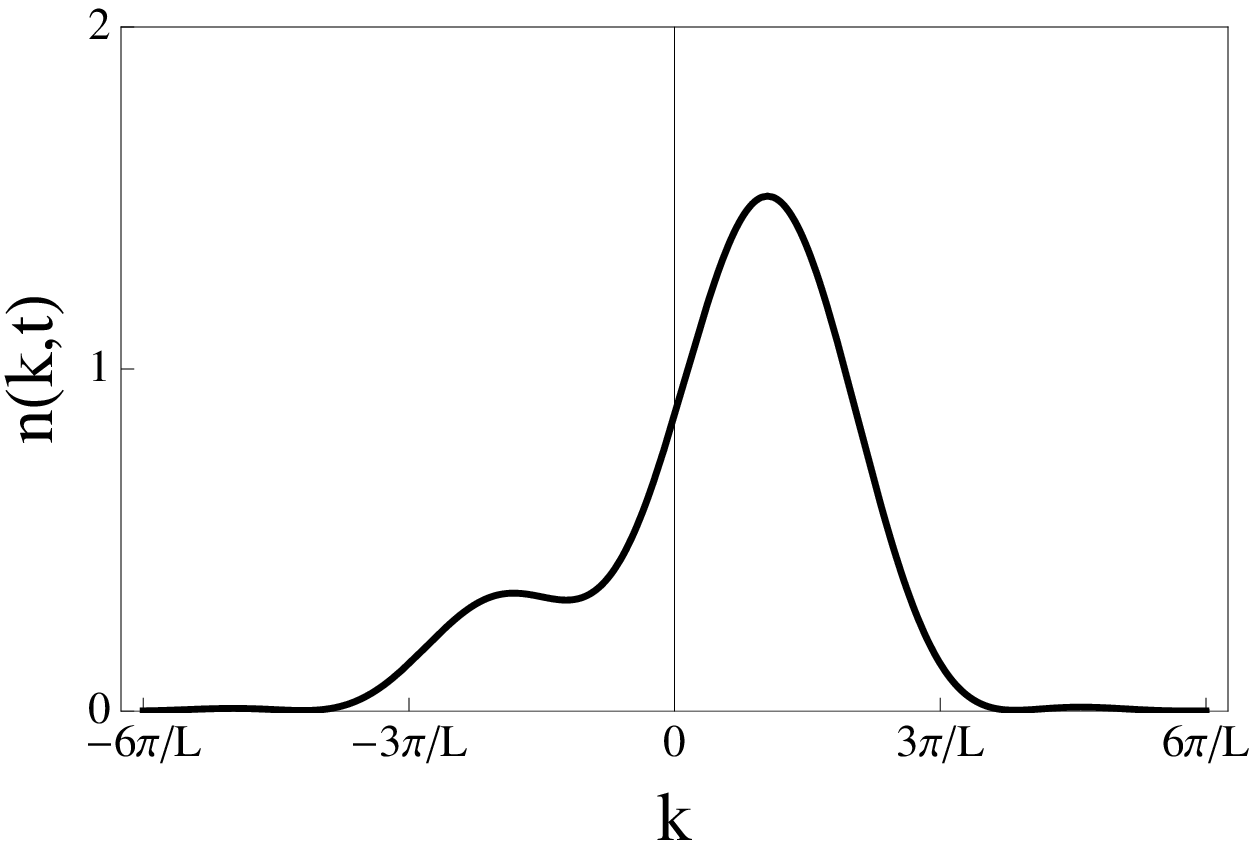}
                \caption{\small t=T/9}
            
        \end{subfigure}
       \begin{subfigure}[b]{0.3\textwidth}
                \centering
                \includegraphics[width=\textwidth]{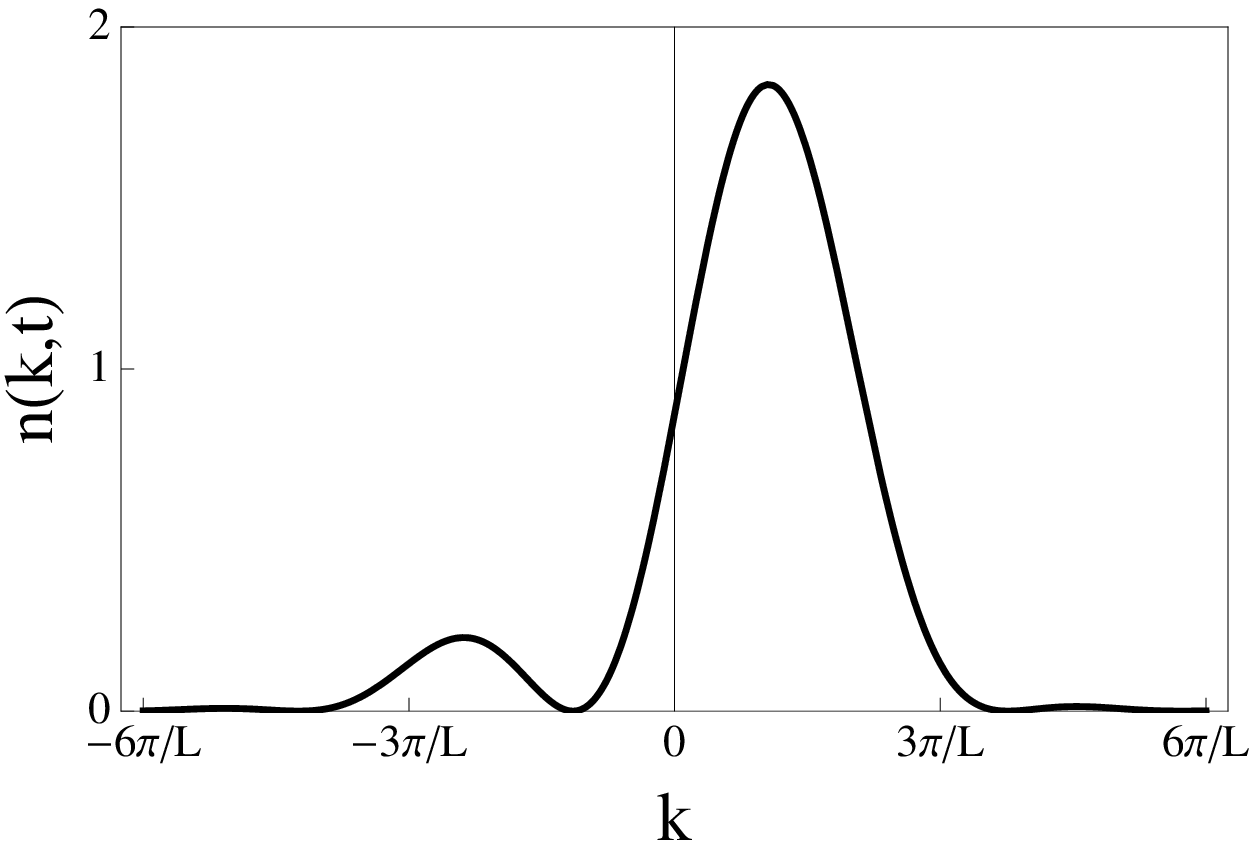}
                \caption{\small t=T/4}
        
        \end{subfigure}
         \begin{subfigure}[b]{0.3\textwidth}
                \centering
                \includegraphics[width=\textwidth]{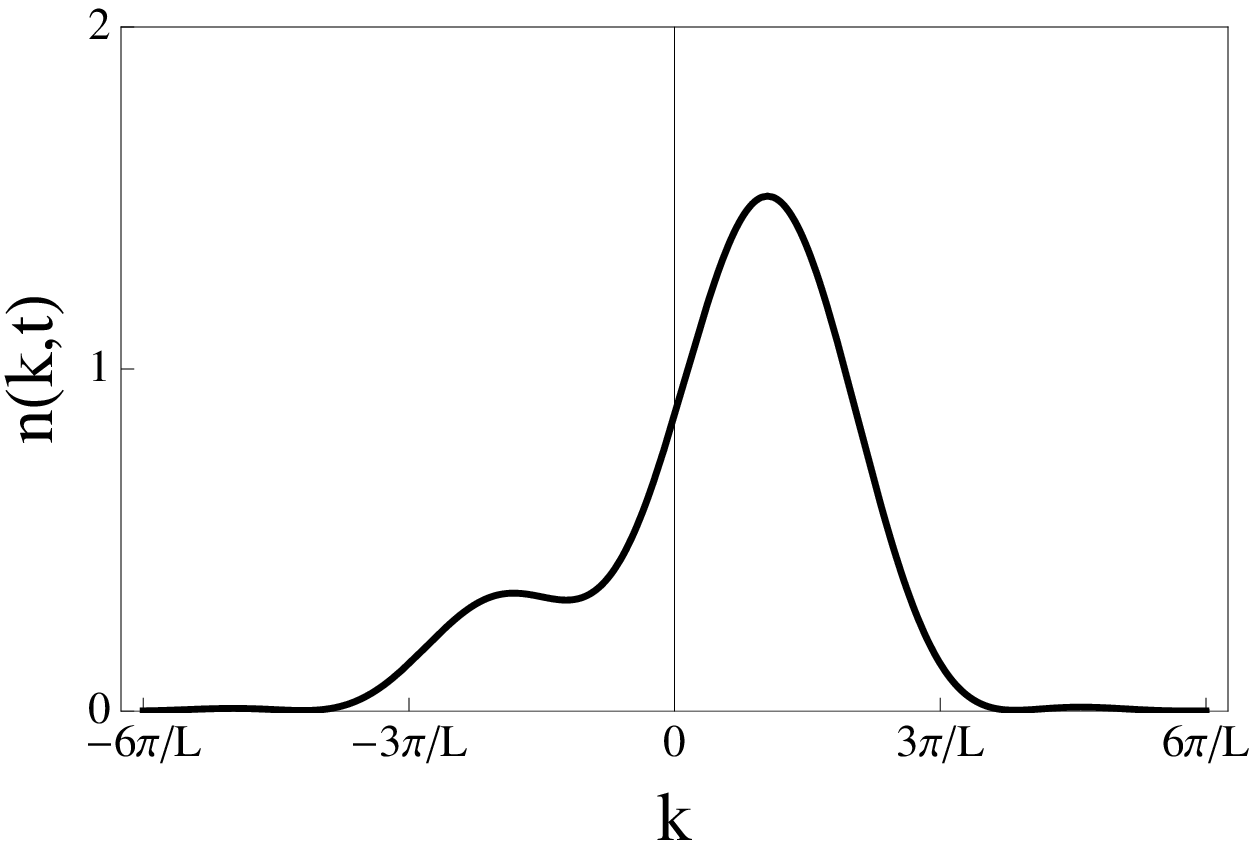}
                \caption{\small t=7T/18}
          
        \end{subfigure} 
         \begin{subfigure}[b]{0.3\textwidth}
                \centering
                \includegraphics[width=\textwidth]{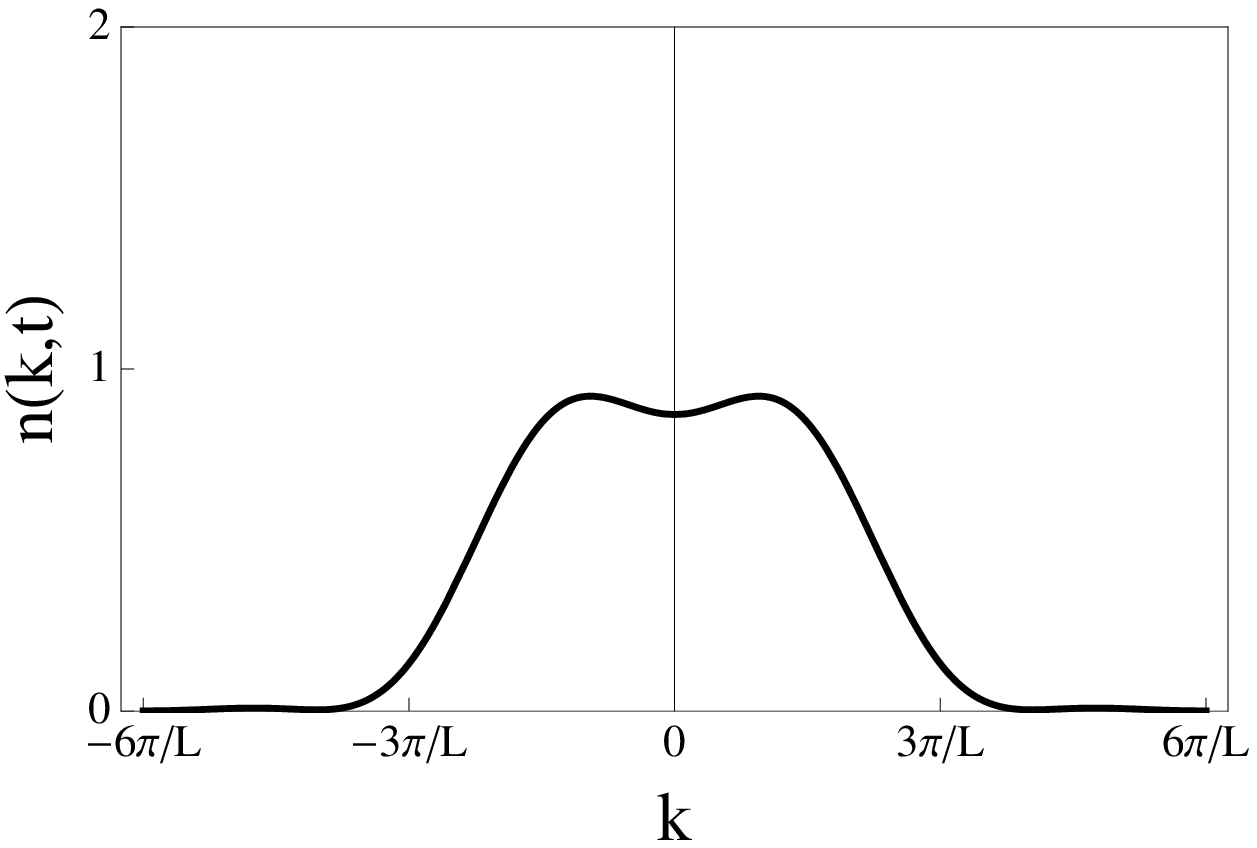}
                \caption{\small t=T/2}
      
        \end{subfigure}
     \begin{subfigure}[b]{0.3\textwidth}
                \centering
                \includegraphics[width=\textwidth]{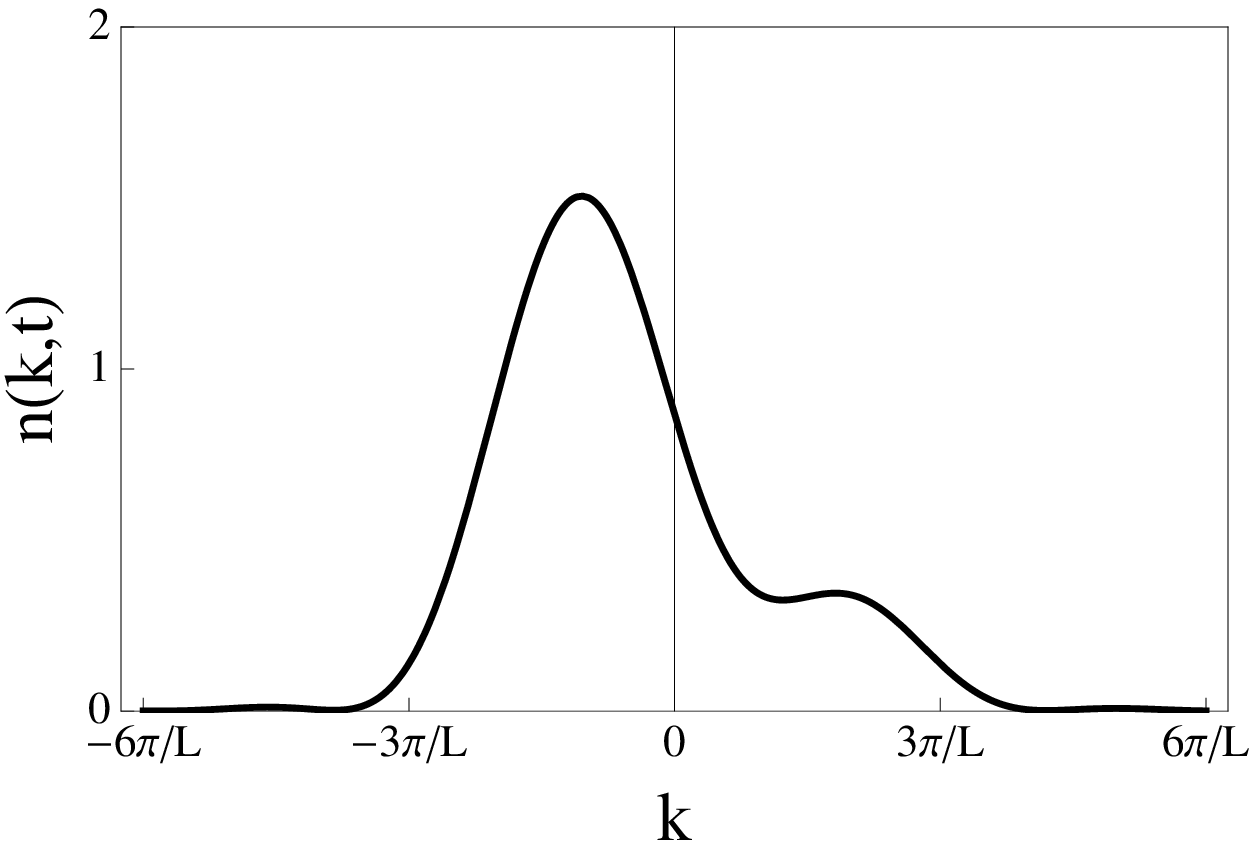}
                \caption{\small t=11T/18}
             
        \end{subfigure}
         \begin{subfigure}[b]{0.3\textwidth}
                \centering
                \includegraphics[width=\textwidth]{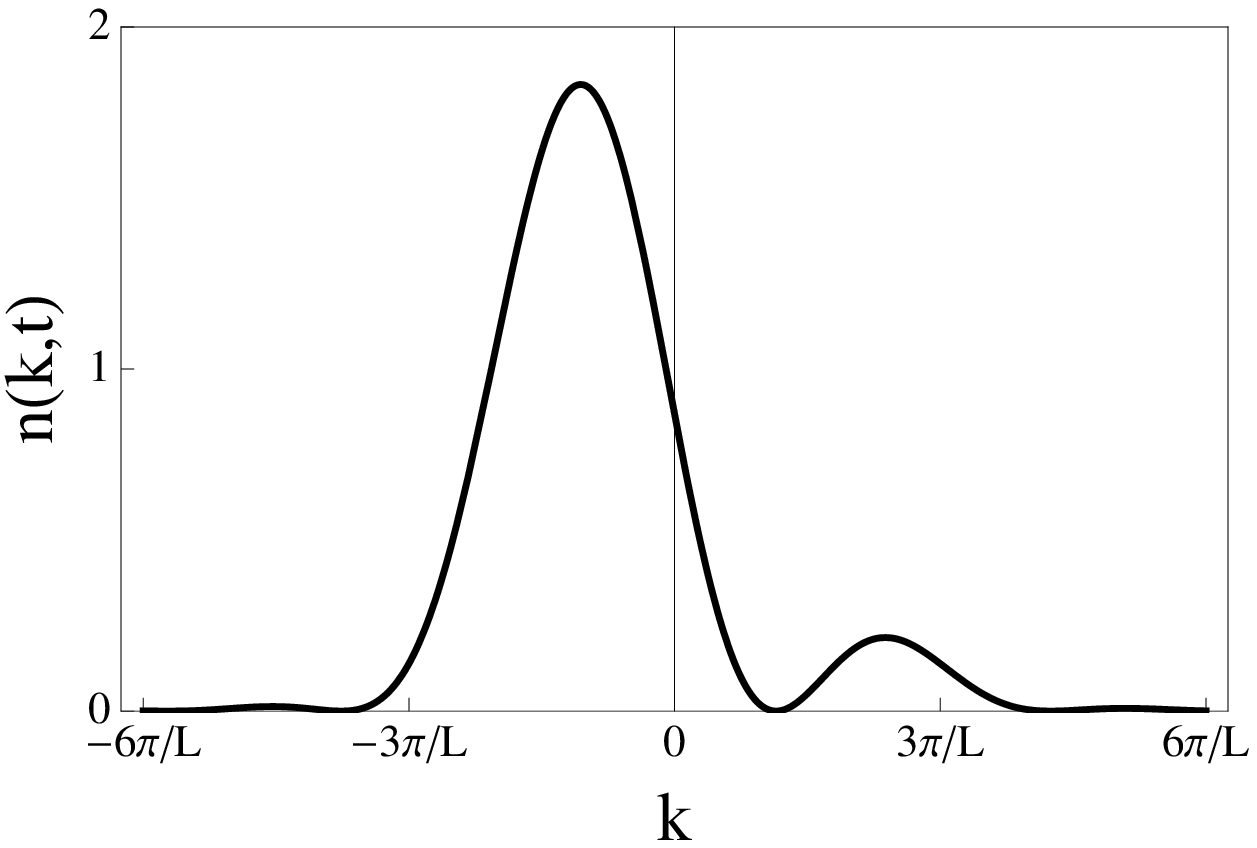}
                \caption{\small t=3T/4}
            
        \end{subfigure} 
         \begin{subfigure}[b]{0.3\textwidth}
                \centering
                \includegraphics[width=\textwidth]{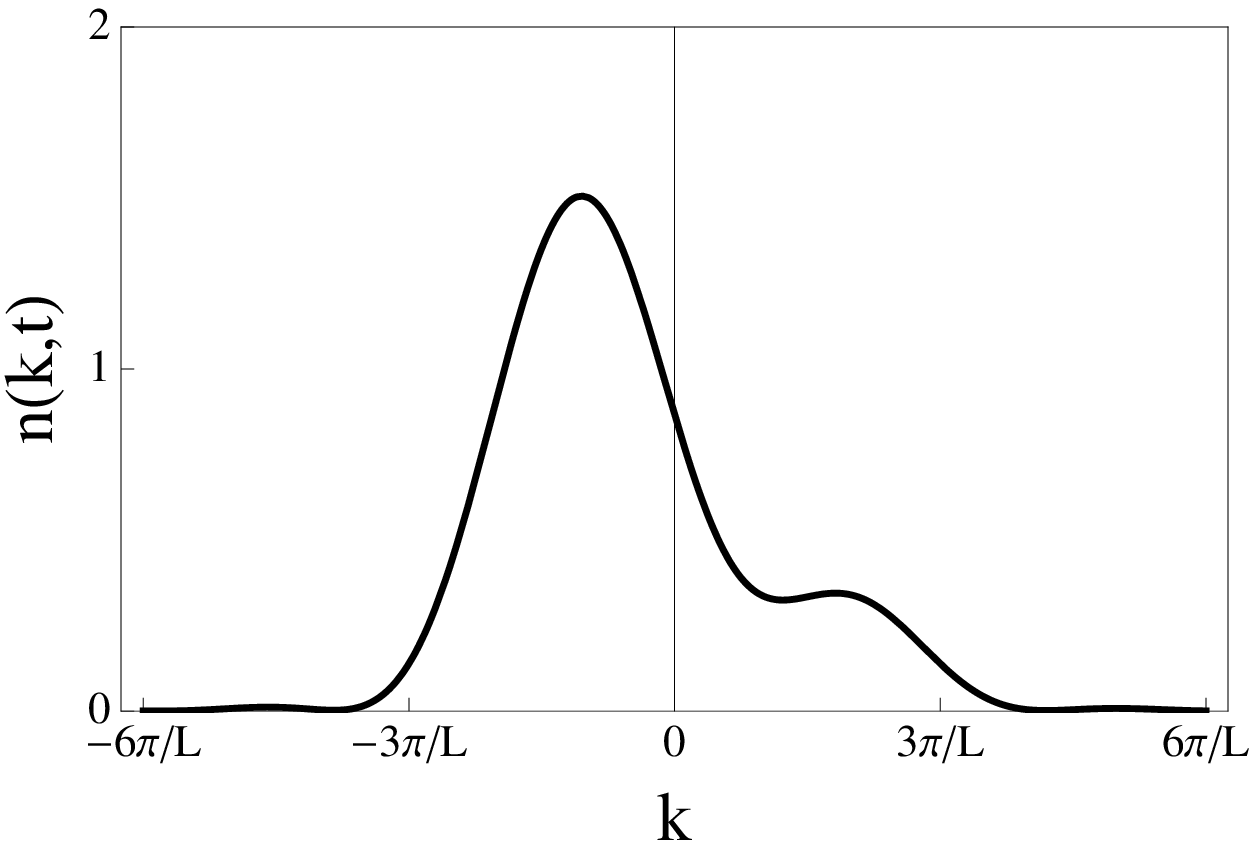}
                \caption{\small t=8T/9}
          
        \end{subfigure}
        \begin{subfigure}[b]{0.3\textwidth}
                \centering
                \includegraphics[width=\textwidth]{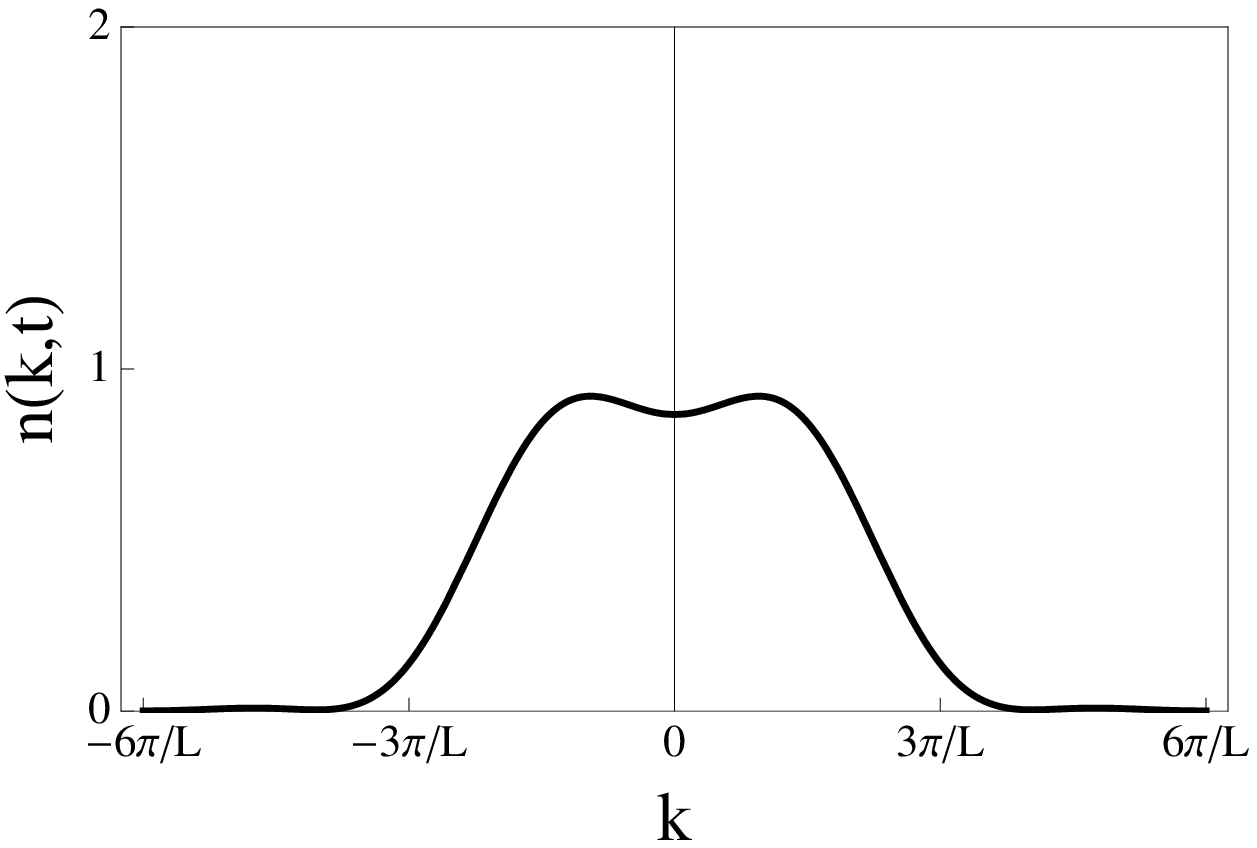}
                \caption{\small t=T}
            
        \end{subfigure}
   \caption{\small Probability density $n(k,t)$ in momentum space (ISWP)}\label{fig:ISWP-Oscillation-Momentum}
\end{figure}

Probability densities in position and momentum space are given by equations (\ref{pdr}) and (\ref{pdn}), and its time dependence is illustrated in Figs. \ref{fig:ISWP-Oscillation-Position} and \ref{fig:ISWP-Oscillation-Momentum}

\section{Statistical Measures}\label{Statistical Measures}

We shall begin our presentation of statistical measures with the generalized uncertainty principle. For any pair of observables $A$ and $B$ whose operators do not commute (incompatible observables), their corresponding standard deviations obey the lower bound \cite{Grif}

\begin{equation}
\sigma_A ^2 \sigma_B ^2 \geq \left( \frac{1}{2i}\langle[\hat{A},\hat{B}]\rangle  \right)^2.
\end{equation}

Supposing that the first observable is position $x$ and the second one is momentum $k=p/\hbar$, their product represents the so-called Heisenberg uncertainty principle, which defines a lower bound that depends not only on the operators, but also on the quantum mechanical state

\begin{equation}
\Delta x \Delta k \geq \frac{1}{2},
\label{hp2}
\end{equation}

\noindent where $\Delta x= \left( \langle x^2 \rangle - {\langle x \rangle}^2  \right)^{1/2}$ and $\Delta k= \left( \langle k^2 \rangle - {\langle k \rangle}^2  \right)^{1/2}$. 

The Shannon information entropy \cite{Shannon,Cover} for a discrete probability distribution $p_i$ with $N$ accessible states, is defined as

\begin{equation}
S=-\sum\limits_{i=1}^N p_i\log{p_i}, 
\end{equation}

\noindent while for a continuous probability density $f(x)$ is usually called ``differential entropy" \cite{Cover} and is defined as 

\begin{equation}
S=-\int  f(x)\log{f(x)} \,dx.
\end{equation}

In quantum mechanics, for a continuous distribution representing the probability density in position space $\rho({\bf r})$, takes the form

\begin{equation}
S_r=-\int \rho({\bf r})\ln{\rho({\bf r})}  \,d{\bf r},
\end{equation}

\noindent and the corresponding momentum space entropy $S_k$ is given by

\begin{equation}
S_k=-\int n({\bf k})\ln{n({\bf k})}  \,d{\bf k},
\end{equation}

\noindent where $n({\bf k})$ denotes the momentum probability density \cite{MP}. The densities $\rho({\bf r})$ and $n({\bf k})$ are respectively normalized to one. The information entropy sum in conjugate spaces $S_T=S_r+S_k$, contains the net information of the system and is typically measured in nats. Individual entropies $S_r$ and $S_k$ depend on the units used to measure $r$ and $k$ respectively, but their sum $S_T$ does not i.e. it is invariant to uniform scaling of coordinates.

The net Shannon information entropy, in D-dimensions, obeys the following lower bound, also known as the entropic uncertainty relation (EUR)

\begin{equation}
S_T=S_r+S_k\geq D(1+\ln{\pi}),
\label{EUR}
\end{equation}

\noindent which represents a stronger version of the Heisenberg uncertainty principle of quantum mechanics, in the sense that the EUR leads to Heisenberg relation, while the inverse is not true. Additionally, the right-hand side of Heisenberg relation depends on the quantum state of the system, while EUR does not \cite{Bialynicki-Birula}.

Shannon's information entropy (``uncertainty") provides a global measure of smoothness \cite{Frieden} and reflects the indeterminacy (``spread") of a distribution, since a highly localized $\rho({\bf r})$ is associated with a diffuse $n({\bf k})$, leading to low $S_r$ and high $S_k$ and vice-versa. In other words, Shannon information entropy measures the average amount of the information received, when this uncertainty is removed by an appropriate ``localization" experiment \cite{Nalewajski2}.

The Fisher information measure $I_{\theta}$ \cite{Fisher,Frieden,Sanchez-Moreno}, also called the ``intrinsic accuracy", corresponding to a family of probability densities $f(x;\theta)$ and depending on a parameter $\theta$ is given by

\begin{equation}
I_{\theta}=\int \frac{1}{f(x;\theta)} \left( \frac{\partial f(x;\theta)}{\partial \theta}   \right)^2 \,dx,
\end{equation}

\noindent while for a discrete distribution \cite{Frieden,PNCT} is defined as

\begin{equation}
I=\sum\limits_{i=1}^N \frac{(p_{i+1}-p_i)^2}{p_i}.
\end{equation}

In quantum mechanics, Fisher information in position space takes the form

\begin{equation}
I_r=\int \frac{{|\bigtriangledown \rho({\bf r}) |}^2}{\rho({\bf r})} \,d{\bf r},
\end{equation}

\noindent and the corresponding momentum space measure is given by

\begin{equation}
I_k=\int \frac{{|\bigtriangledown n({\bf k}) |}^2}{n({\bf k})} \,d{\bf k}.
\end{equation}

The individual Fisher measures are bounded through the Cramer-Rao inequality according to $I_r\geq \frac{1}{V_r}$ and $I_k\geq \frac{1}{V_k}$, where $V_r$ and $V_k$ denote the corresponding spatial and momentum variances respectively \cite{Rao,Stam}. 

In contrast to Shannon's information entropy which provides a global way of characterizing ``uncertainty", Fisher's information provides a local measure of smoothness and reflects the ``narrowness" of the probability distribution \cite{Nalewajski2,Frieden}. Furthermore, Fisher's information is strongly sensitive to the local oscillatory character of probability density, due to the fact that it depends on its gradient \cite{Sanchez-Moreno}.

In position space, the Fisher information measures the ``sharpness" of probability density i.e. a strongly localized probability density gives rise to a larger value of Fisher information and vice-versa. In this sense, Fisher information is complementary to Shannon information entropy and their reciprocal relation is, in fact, utilized in this work. 

If either the momentum space wavefunction $\phi(k)$ or the position space wavefunction $\psi(x)$ is real, it has been shown \cite{Sanchez-Moreno} that the net Fisher information ($I_T=I_r I_k$), in D-dimensions, obeys the following lower bound

\begin{equation}
I_T=I_r I_k\geq 4 D^2.
\label{EURF}
\end{equation}

The lower bounds of both Shannon sum $(S_r + S_k )$ and Fisher product $(I_r I_k)$ get saturated for the Gaussian distributions \cite{Frieden}.

Fisher's information is also intimately related to the Shannon information entropy via de Bruijn identity \cite{Cover,Stam}

\begin{equation}
\frac{\partial}{\partial t}S(x+\sqrt{t}z)\bigg|_{t=0} =\frac{1}{2}I(x),
\end{equation}

\noindent where $x$ is a random variable with a finite variance with a density $f(x)$, and $z$ an independent normally distributed random variable with zero mean and unit variance.

In a statistical analysis we are usually interested in knowing how far the system deviates from equilibrium. An isolated system in equilibrium is characterized by equiprobability ($p_i=1/N$), a case where Shannon information entropy takes its maximum value

\begin{equation}
S_{max}=\log N.
\end{equation}

In the neighborhood of equilibrium, we can expand Shannon entropy $S$ around its maximum value $S_{max}$ 

\begin{equation}
S=S_{max}-\frac{N}{2} \sum\limits_{i=1}^N  \left( p_i -\frac{1}{N}\right)^2 + \cdots 
\label{expand}
\end{equation}

\noindent where the quantity $D=\sum\limits_{i=1}^N  \left( p_i -\frac{1}{N}\right)^2$ is called disequilibrium, and represents the distance from equilibrium. Multiplying both parts of equation (\ref{expand}) with $S$ and setting $C=S \cdot D$ we get

\begin{equation}
C=\frac{2}{N}\cdot S(S_{max}-S),
\label{ds}
\end{equation}

\noindent where $C$ is another statistical measure called LMC complexity, due to L\' {o}pez-Ruiz, Mancini and Calbet, who first defined it and gave the above analysis \cite{LMC}. Ideal gas and crystal are two systems which help us to illustrate the basic properties of those measures. An isolated ideal gas is completely disordered, while each accessible state has the same probability ($p_i=1/N$). Equiprobable distribution results in maximizing Shannon's information entropy ($S=S_{max}$) and minimizing disequilibrium ($D\rightarrow0$), so deductively, LMC complexity is also minimized ($C\rightarrow0$). On the other hand, in a crystal the probability distribution is centered around a prevailing state of perfect symmetry, something that imposes disequilibrium to take its maximum value ($D=D_{max}$), and Shannon's information entropy its minimum ($S\rightarrow0$) respectively. Finally, LMC complexity is again minimized ($C\rightarrow0$).

The irreversibility property of $S$ implies that $\frac{dS}{dt}\geq0$, so another important property of LMC complexity is that while the system reaches equilibrium, $C$ is always decreasing \cite{LMC}

\begin{equation}
\frac{dC}{dt}\leq0,
\end{equation}

\noindent however, this does not forbid complexity to increase when the system is far from equilibrium.

In order to generalize these notions for the continuous case, we redefine disequilibrium as

\begin{equation}
D=\int p^2(x) \,dx,
\end{equation}

\noindent and complexity as

\begin{equation}
C=S \cdot D = - \left( \int p(x)\log p(x) \,dx\right) \cdot \left( \int p^2(x) \,dx \right).
\end{equation}

Another way of extending LMC complexity for a continuous system is to use the exponential of Shannon information entropy \cite{Catalan}

\begin{equation}
C=e^S  D.
\end{equation}

The reason of this extension is that in the continuous case, $S$ and consequently $C$ can become negative. Since in quantum mechanics we have to deal with two spaces i.e. position and momentum ones, in order to calculate the net complexity $C_T$, we use the formula

\begin{equation}
C_T=e^{S_T}  D_T,
\end{equation}

\noindent where $S_T=S_r+S_k$, while for the net disequilibrium $D_T$ we employ the definition

\begin{equation}
D_T=D_r D_k,
\end{equation}

\noindent where $D_r=\int \rho^2({\bf r}) \,d{\bf r}$ and $D_k=\int n^2({\bf k}) \,d{\bf k}$. $S_T$ and $D_T$ are chosen in such a way that they are dimensionless quantities, characterizing the system.

R\' {e}nyi entropy is another information measure \cite{Renyi} defined as 

\begin{equation}
H_a=\frac{1}{1-a}\log \left(\sum\limits_{i=1}^N p_i^a  \right),
\end{equation}

\noindent where $a>0$ and $a \neq 1$. R\' {e}nyi entropy is a generalized measure of information, which converges to Shannon information entropy as $a \rightarrow 1$. It can be easily proved \cite{LMC} that for a discrete distribution, $C$ takes the form

\begin{equation}
C=S \cdot D = H_1 \cdot \left( e^{-H_2}-\frac{1}{N} \right),
\end{equation}

\noindent while for a continuous distribution 

\begin{equation}
C=S \cdot D = H_1 \cdot  e^{-H_2}.
\end{equation}

\section{Results}\label{Results}

\subsection{Ammonia Molecule (DSWP)}
We are interested in studying the time evolution of information measures during the tunneling effect for one period. Below, instead of investigating the transitions associated with both ground and first excited states, we focus only on the ground-state for two reasons. At first, each energy state (which lie below the potential barrier) shows qualitatively similar results. Secondly, transitions that correspond to the ground-state fall in the microwave region, for which there is strong coupling to the radiation field, making it easier to observe and exploit in building a MASER \cite{Major}.

\begin{figure}[!htb]
        \centering
        \begin{subfigure}[b]{0.32\textwidth}
                \centering
                \includegraphics[height=3.7cm,width=5.6cm]{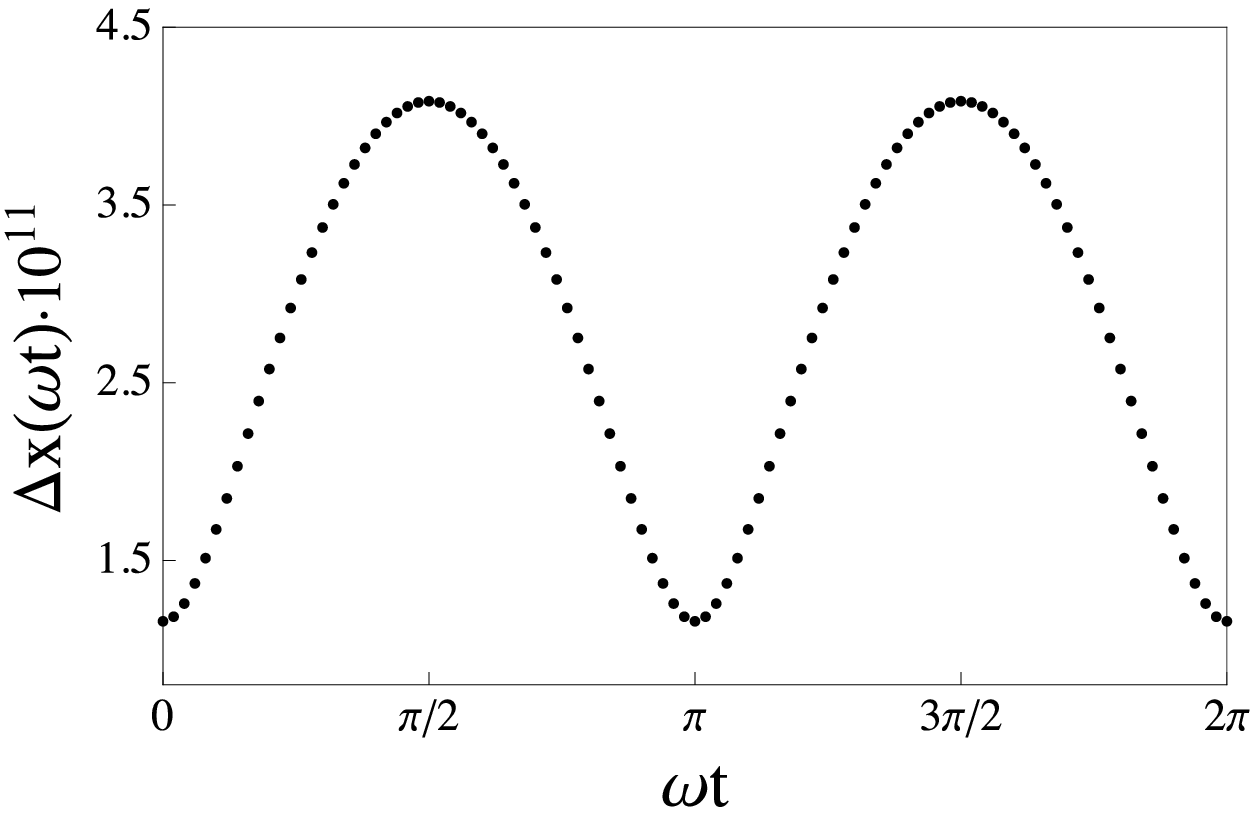}
                \caption{}
                \label{fig:NH3-HPx}
        \end{subfigure}
        \begin{subfigure}[b]{0.32\textwidth}
                \centering
                \includegraphics[height=3.7cm,width=5.9cm]{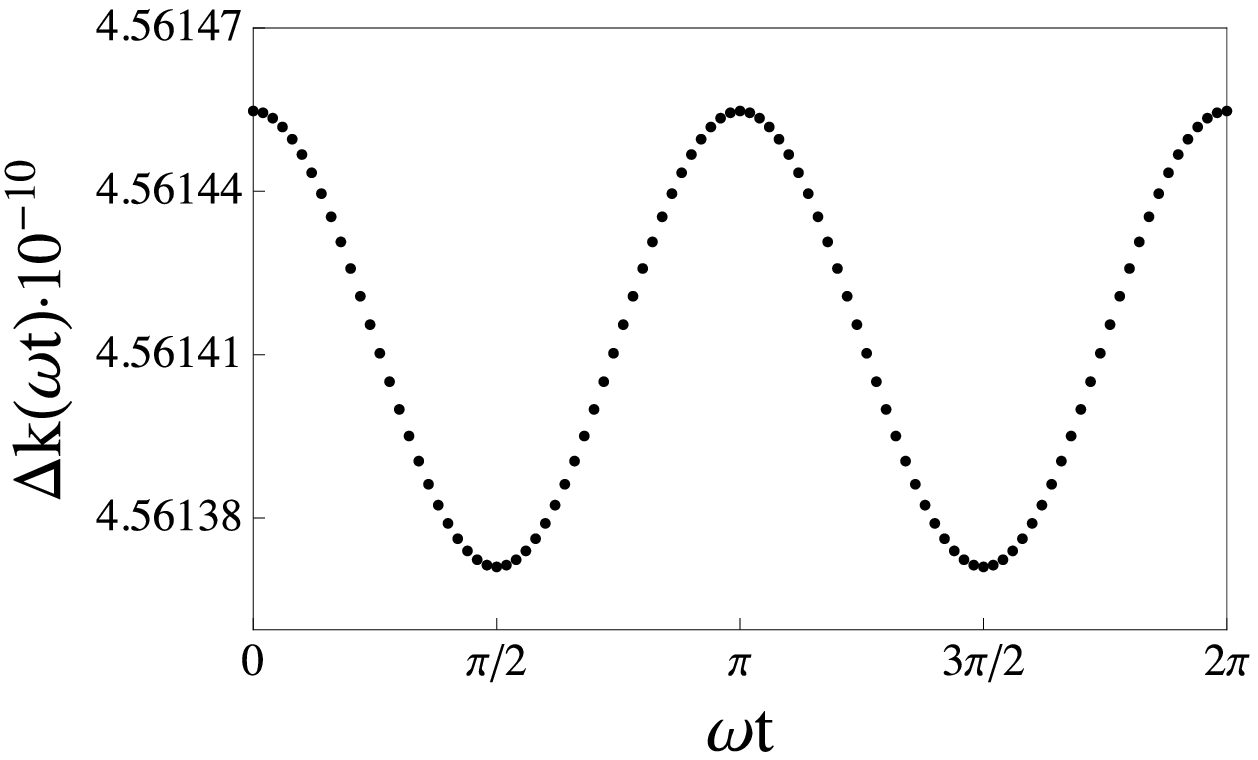}
                \caption{}
                \label{fig:NH3-HPk}
        \end{subfigure}
        \begin{subfigure}[b]{0.32\textwidth}
                \centering
                \includegraphics[height=3.7cm,width=5.6cm]{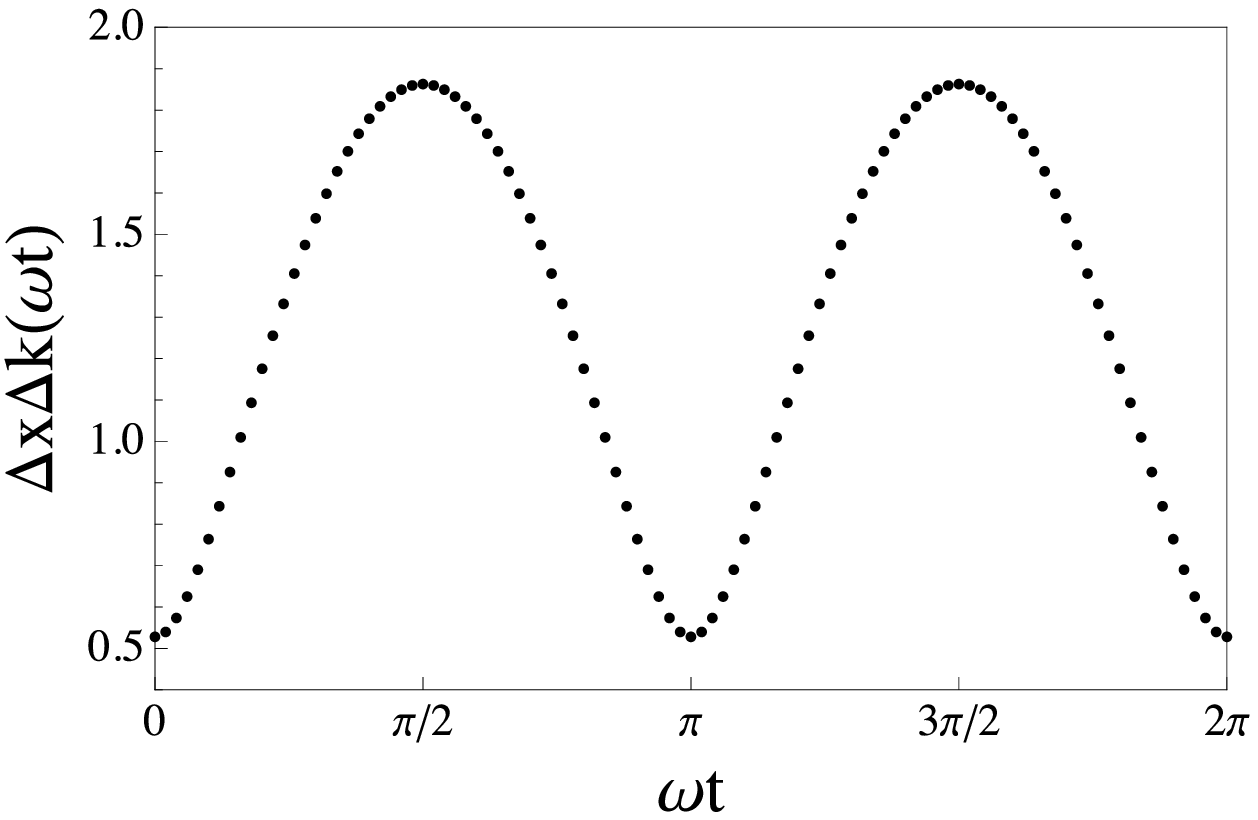}
                \caption{}
                \label{fig:NH3-HP}
        \end{subfigure}
        \caption{\small Time evolution of Heisenberg relation in $NH_3$ \\ a)in position space $\Delta x \cdot 10^{11}$, b) in momentum space $\Delta k \cdot 10^{-10}$ and c)$\Delta x \Delta k$}\label{fig:NH3-HP-ALL}
\end{figure}

In Fig. \ref{fig:NH3-HP-ALL} we plot the time evolution of Heisenberg uncertainties for one oscillation of the particle between the wells. The values of $\Delta x$ are multiplied by $10^{11}$ and $\Delta k$ by $10^{-10}$ to improve the presentation. It is observed that they oscillate between two extreme values. The minimum value is obtained when the particle is located in one of the wells ($t=0,T/2,T$) and the maximum when the particle penetrates the barrier ($t=T/4,3T/4$).

\begin{figure}[!htb]
        \centering
        \begin{subfigure}[b]{0.32\textwidth}
                \centering
                \includegraphics[height=3.7cm,width=5.6cm]{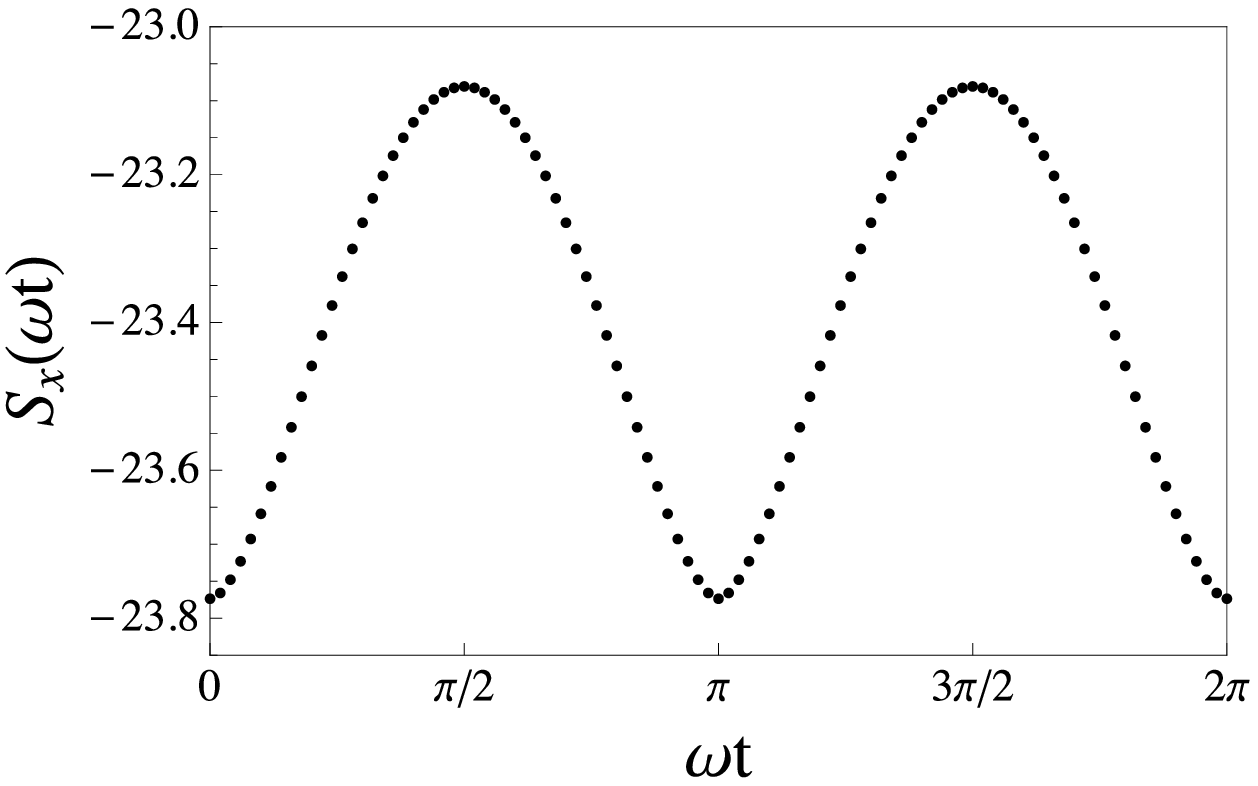}
                \caption{}
                \label{fig:NH3-Shannon-Position-1}
        \end{subfigure}
        \begin{subfigure}[b]{0.32\textwidth}
                \centering
                \includegraphics[height=3.7cm,width=5.6cm]{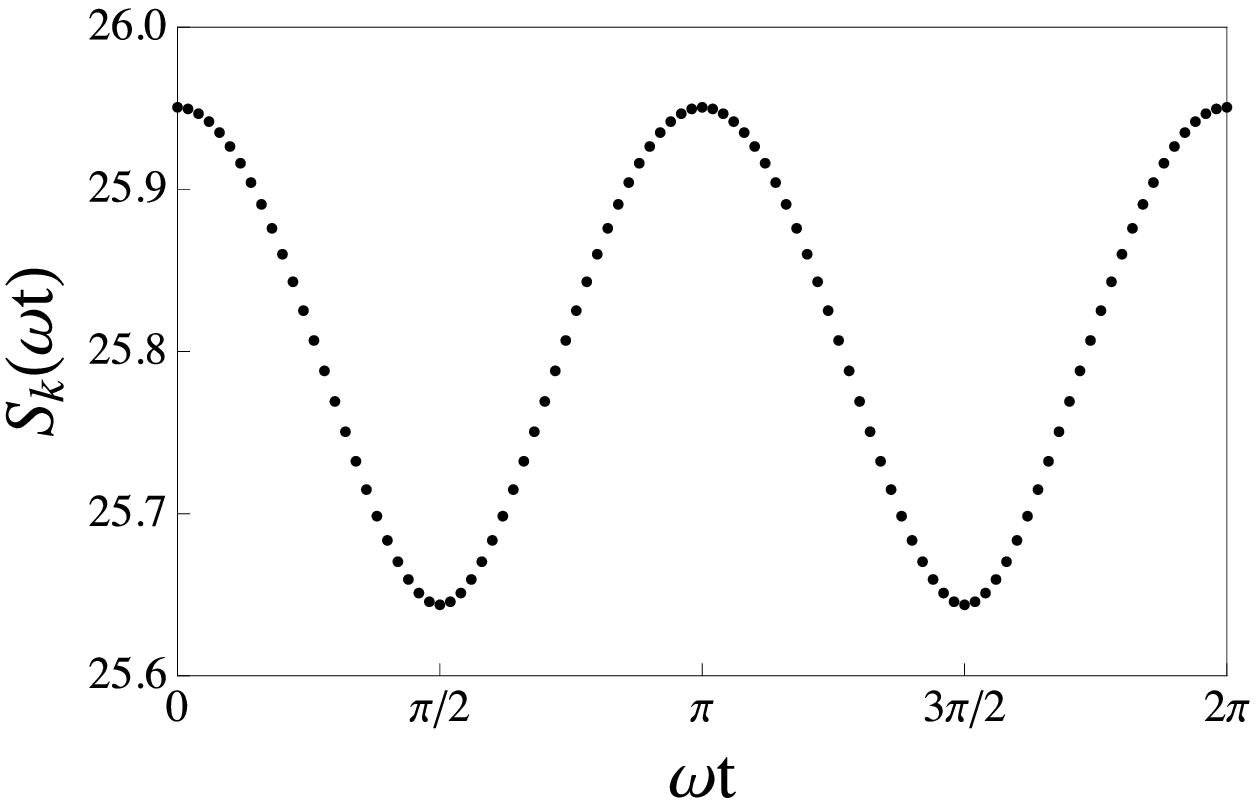}
                \caption{}
                \label{fig:NH3-Shannon-Momentum-1}
        \end{subfigure}
        \begin{subfigure}[b]{0.32\textwidth}
                \centering
                \includegraphics[height=3.7cm,width=5.6cm]{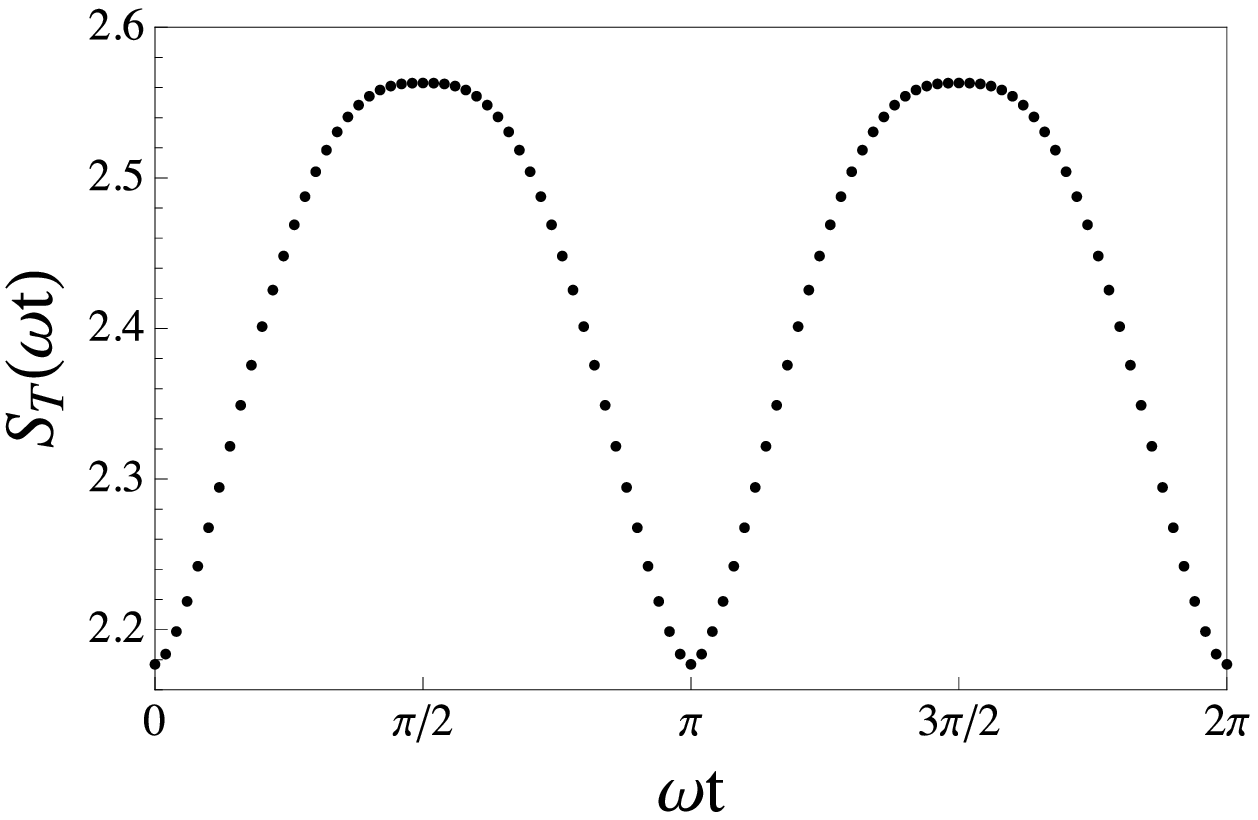}
                \caption{}
                \label{fig:NH3-Shannon-Net-1}
        \end{subfigure}
        \caption{\small Time evolution of Shannon information entropy in $NH_3$ \\ a)in position space $S_x$, b) in momentum space $S_k$ and c)$S_T=S_x+S_k$}\label{fig:NH3-Shannon}
\end{figure}

Shannon information entropy versus time is plotted in Fig. \ref{fig:NH3-Shannon}. As we can see in Fig. \ref{fig:NH3-Shannon-Position-1}, $S_x$ is minimized when the particle is exactly in one well ($t=0,T/2,T$), and maximized when the particle is penetrating the barrier. This behavior reflects the localization characteristics of the probability distribution $\rho(x,t)$ plotted in Fig. \ref{fig:DSWP-Oscillation-Position}. On the other hand, in momentum space, we observe the opposite behavior (Fig. \ref{fig:NH3-Shannon-Momentum-1}), since at time $t=T/4$ and $t=3T/4$ probability density $n(k,t)$ is clearly more localized than at time $t=0$ (or $t=T/2$ and $T$) when the particle is only in one well (Fig. \ref{fig:DSWP-Oscillation-Momentum}). 

Shannon's information entropy is a measure of uncertainty, thus its behavior observed is similar to Heisenberg relation (\ref{fig:NH3-HP}). This behavior also verifies the property stated in section \ref{Statistical Measures}, that smaller values of Shannon information entropy correspond to more localized distributions. It is also observed in Fig. \ref{fig:NH3-Shannon-Net-1}, that the net Shannon information ($S_T=S_x+S_k$) presents a similar trend compared with its component in position space $S_x$. Sometimes, in one of the spaces (position and momentum) the information entropy is negative, but the net information content $S_T=S_x+S_k$ is always positive for normalized probability densities $\rho(x,t)$ and $n(k,t)$

\begin{figure}[!htb]
        \centering
        \begin{subfigure}[b]{0.32\textwidth}
                \centering
                \includegraphics[height=3.7cm,width=5.6cm]{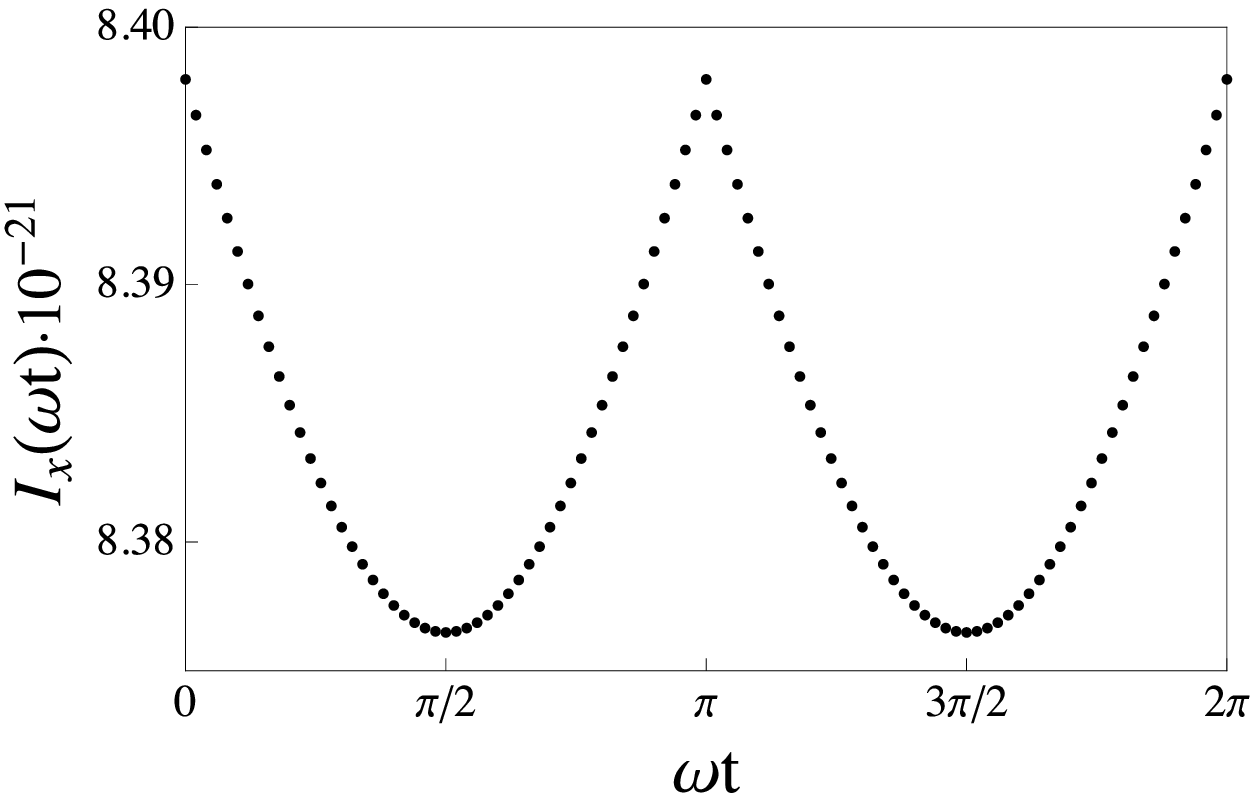}
                \caption{}
                \label{fig:NH3-Fisher-Position-1}
        \end{subfigure}
        \begin{subfigure}[b]{0.32\textwidth}
                \centering
                \includegraphics[height=3.7cm,width=5.6cm]{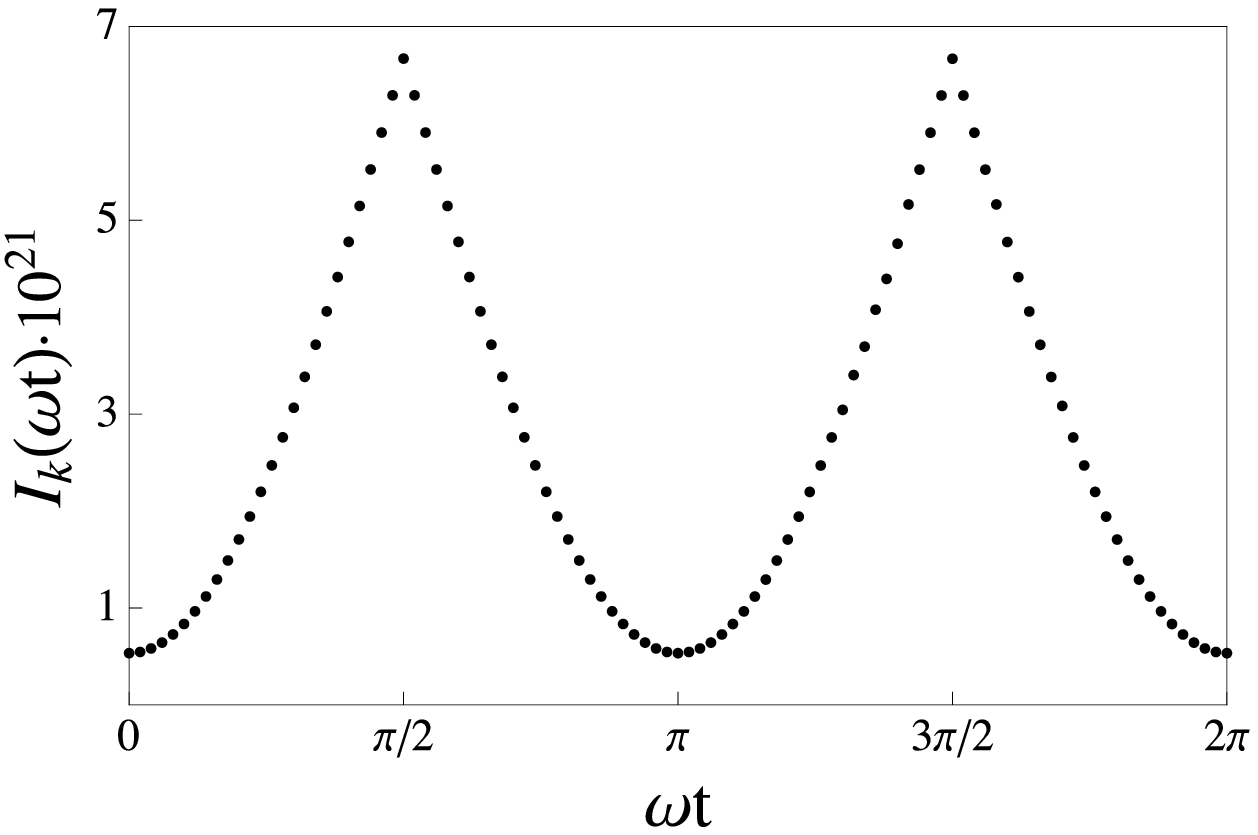}
                \caption{}
                \label{fig:NH3-Fisher-Momentum-1}
        \end{subfigure}
        \begin{subfigure}[b]{0.32\textwidth}
                \centering
                \includegraphics[height=3.7cm,width=5.6cm]{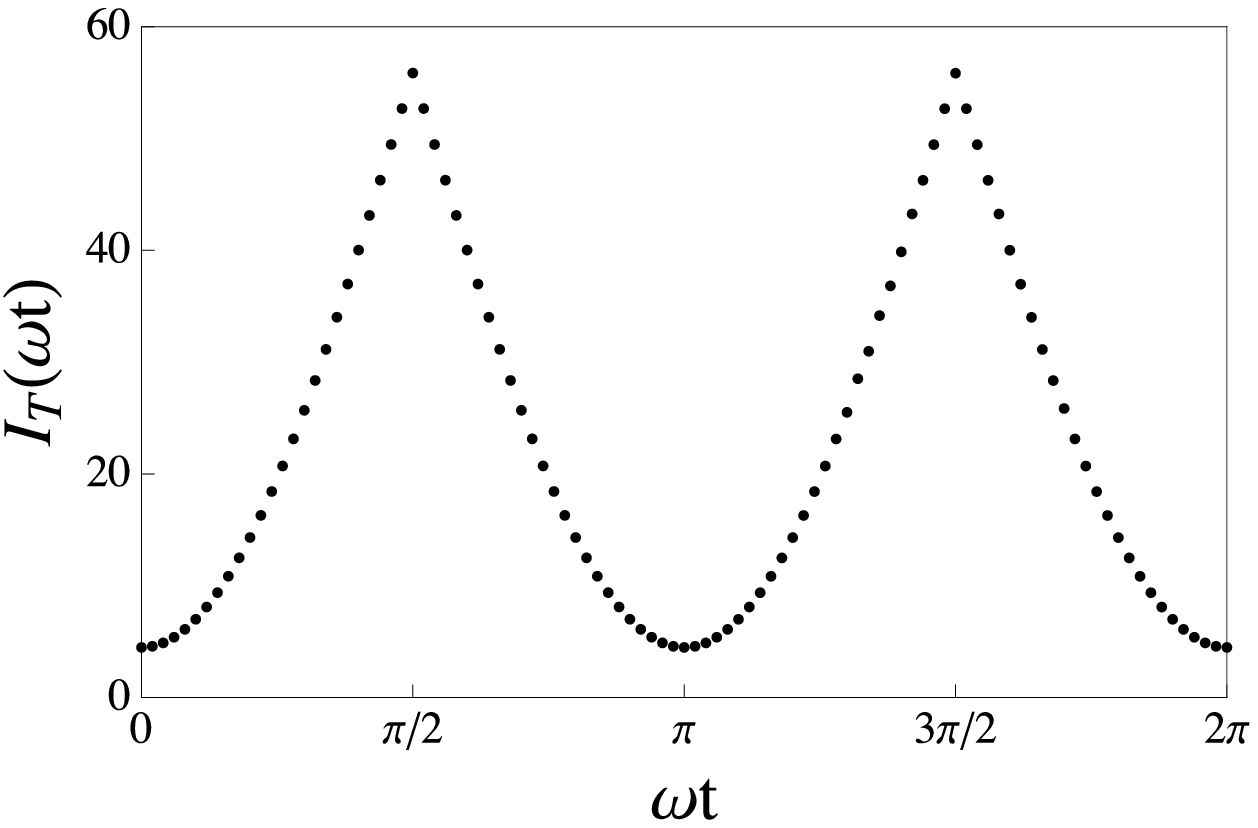}
                \caption{}
                \label{fig:NH3-Fisher-Net-1}
        \end{subfigure}
        \caption{\small Time evolution of Fisher information in $NH_3$ \\ a)in position space $I_x \cdot 10^{-21}$, b) in momentum space $I_k \cdot 10^{21}$ and c)$I_T=I_x I_k$}\label{fig:NH3-Fisher}
\end{figure}

In Fig. \ref{fig:NH3-Fisher} we plot the time dependence of Fisher information  $I_x$, $I_k$ and $I_T$. It is noted that the values of $I_x$ are multiplied by $10^{-21}$, while those of $I_k$ by $10^{21}$. In Fig. \ref{fig:NH3-Fisher-Position-1} Fisher information in position space $I_x$ is peaked when the particle is located in one of the wells, while at time $t=0,T/2$ and $T$, $I_k$ is minimized (Fig. \ref{fig:NH3-Fisher-Momentum-1}). As we stated in section 4, a localized probability density gives rise to a larger value of Fisher information, as expected. Furthermore, compared to Fig. \ref{fig:NH3-Shannon-Net-1}, the curve of $I_T$ simulates the tunneling phenomenon in a more sensitive way. Fisher's information depends not only on the probability density, but also on its gradient. Thus it is reasonable to provide a more sensitive measure of information, reflecting more accurately the oscillation that takes place. We also note that the lower limits for $S_T =1+\ln{\pi}$ and $I_T=4$, obtained from relations (\ref{EUR}) and (\ref{EURF}) putting $D=1$ (for our one dimensional problem), are fulfilled throughout our numerical calculations. 

\begin{figure}[!htb]
        \centering
       \begin{subfigure}[b]{0.32\textwidth}
                \centering
                \includegraphics[height=3.7cm,width=5.6cm]{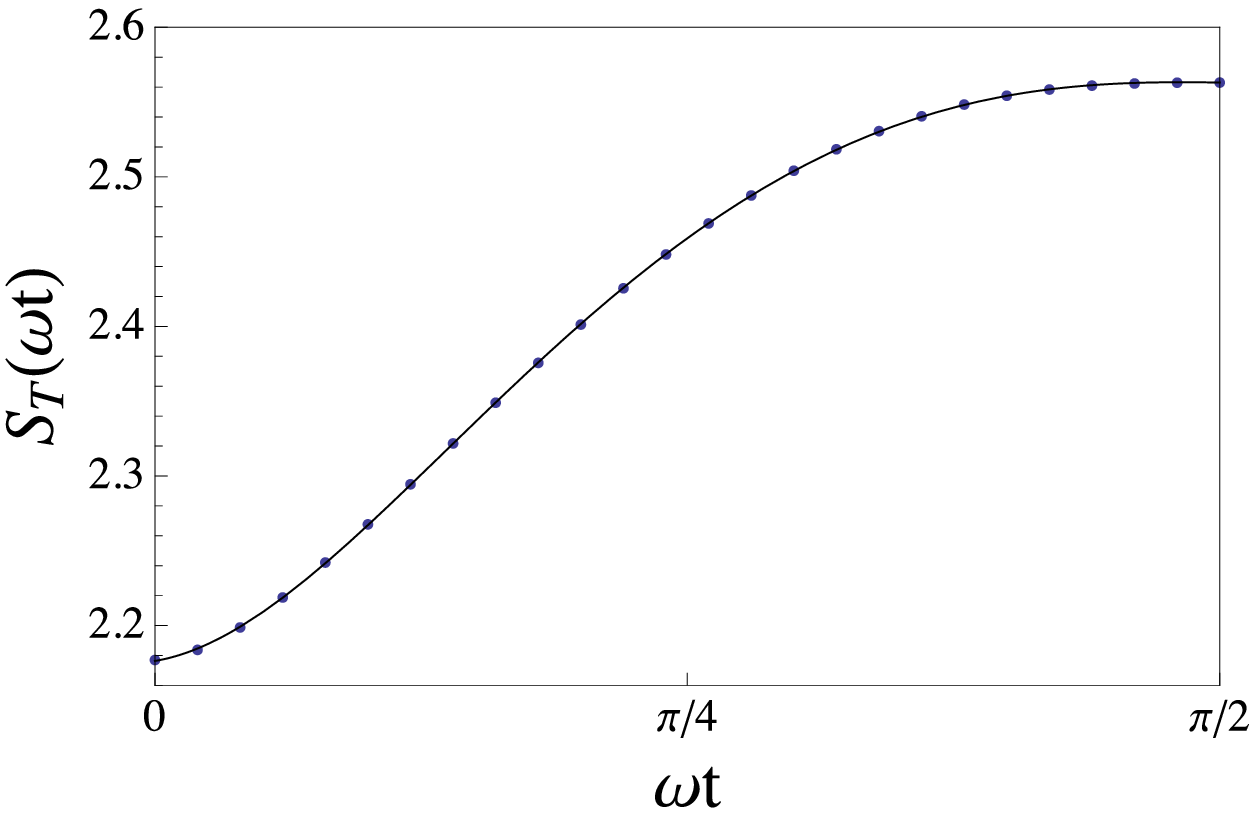}
                \caption{\small $S_T(\omega t)$}
             \label{fig:NH3-Shannon-Fitting}
        \end{subfigure} 
         \begin{subfigure}[b]{0.32\textwidth}
                \centering
                \includegraphics[height=3.7cm,width=5.6cm]{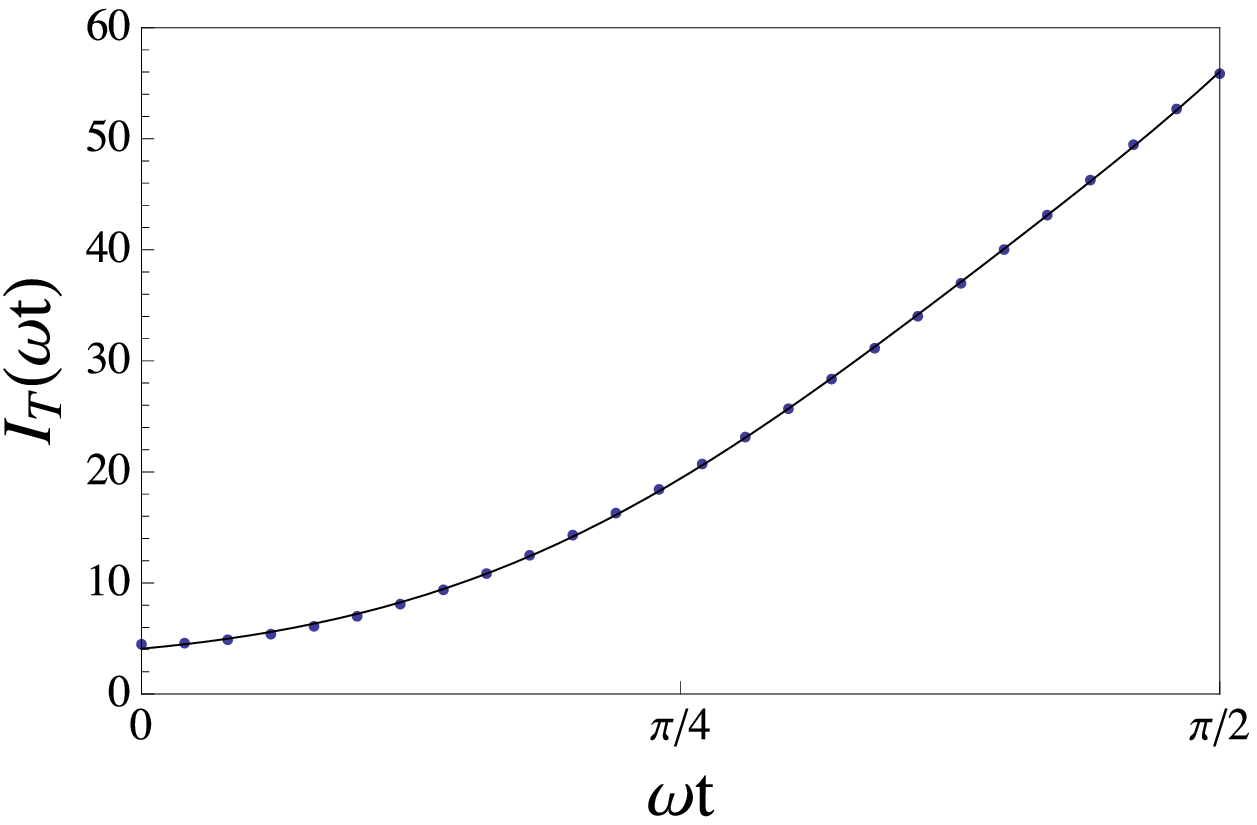}
                \caption{\small $I_T(\omega t)$}
              \label{fig:NH3-Fisher-Fitting}
        \end{subfigure}
   \caption{\small Fitted functions for Shannon and Fisher measures ($0\leq \omega t\leq \pi/2$)}
\end{figure}

More specifically, a fitting curve obtained for $S_T$ as function of time, shows that $S_T$ is logarithmically maximized at $t=T/4$ (Fig. \ref{fig:NH3-Shannon-Fitting})

\begin{equation}
S_T(\omega t)=\ln \left( \alpha_0+\alpha_1 \omega t+\alpha_2 (\omega t)^{2}+\alpha_3  (\omega t)^{3}+\alpha_4  (\omega t)^{4} \right),
\end{equation}

\noindent while Fisher's information takes its maximum value at $t=T/4$ exponentially (Fig. \ref{fig:NH3-Fisher-Fitting})

\begin{equation}
I_T(\omega t)= e^{\left( \alpha_0+\alpha_1 \omega t+\alpha_2  (\omega t)^{2}+\alpha_3  (\omega t)^{3}+\alpha_4  (\omega t)^{4} \right) }.
\end{equation}

The evaluated fitting parameters are 

\begin{table}[ht]
\caption{The values of the fitting parameters $\alpha$} 
\centering 
\begin{tabular}{c c c } 
\hline\hline 
Parameter &\qquad $S_T$ & \qquad $I_T$  \\ [0.5ex] 
\hline 
$\alpha_0$ & \qquad $8.81379$ & \qquad $1.40713$  \\ 
$\alpha_1$ & \qquad $0.66905$ & \qquad $1.35802$  \\
$\alpha_2$ & \qquad $8.47152$ & \qquad $1.99491$ \\
$\alpha_3$ & \qquad $-7.26716$ & \qquad $-1.90413$ \\
$\alpha_4$ & \qquad $1.70397$  & \qquad $0.48349$  \\ [1ex]
\hline 
\end{tabular} 
\end{table}

\begin{figure}[!htb]
        \centering
        \begin{subfigure}[b]{0.32\textwidth}
                \centering
                \includegraphics[height=3.7cm,width=5.6cm]{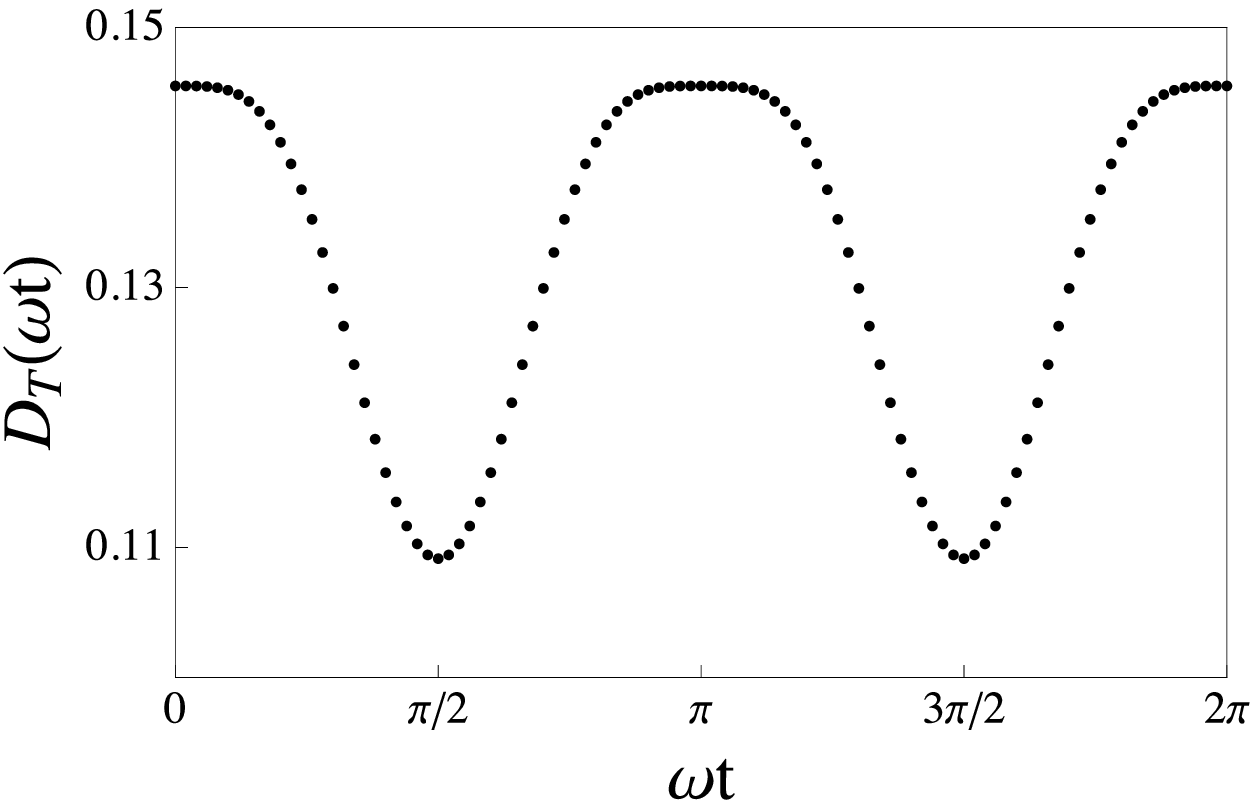}
                \caption{\small $D_T(\omega t)$ }           
        \end{subfigure} 
         \begin{subfigure}[b]{0.32\textwidth}
                \centering
                \includegraphics[height=3.7cm,width=5.6cm]{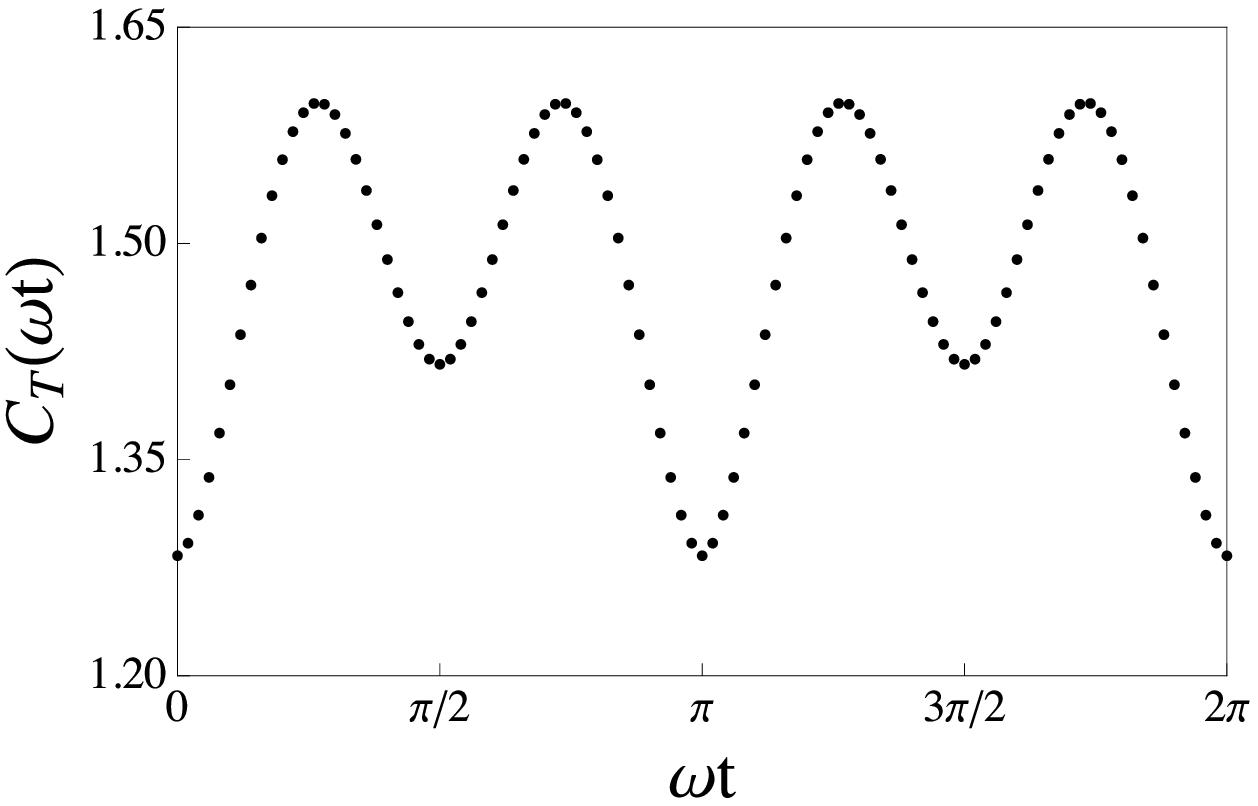}
                \caption{\small $C_T(\omega t)$}              
        \end{subfigure}
   \caption{\small Time evolution of disequilibrium and LMC complexity in $NH_3$}\label{fig:NH3-D-C}
\end{figure}

Finally, disequilibrium $D_T$ and LMC complexity $C_T$ versus time are plotted in Fig. \ref{fig:NH3-D-C}. As it is expected, $D_T$ is maximized when $S_T$ is minimized (Fig. \ref{fig:NH3-Shannon-Net-1}) and vice-versa. Complexity on the other hand, reaches its relative minimum values, when either $S_T$ or $D_T$ are taking their extreme values.

\subsection{ISWP}

\begin{figure}[!htb]
        \centering
        \begin{subfigure}[b]{0.32\textwidth}
                \centering
                \includegraphics[height=3.7cm,width=5.6cm]{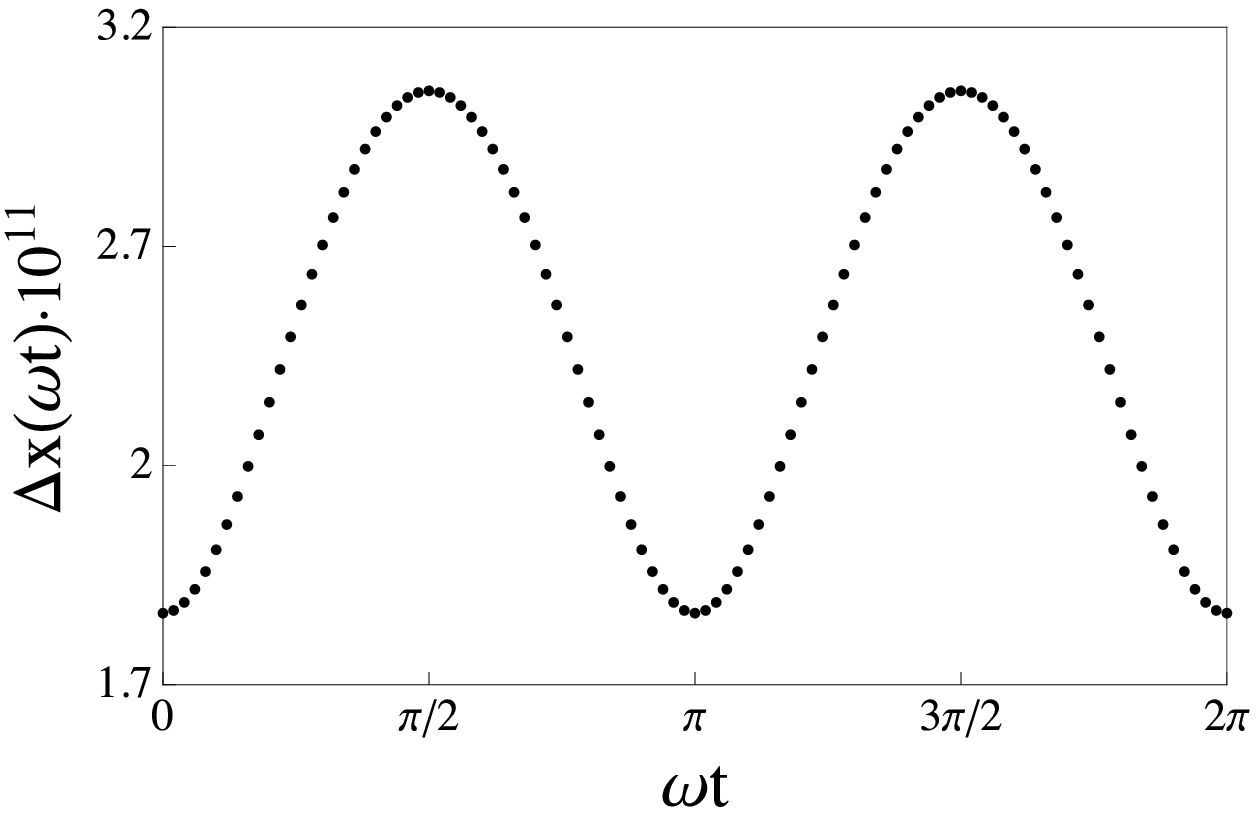}
                \caption{}
                \label{fig:ISWP-HPx}
        \end{subfigure}
        \begin{subfigure}[b]{0.32\textwidth}
                \centering
                \includegraphics[height=3.7cm,width=5.6cm]{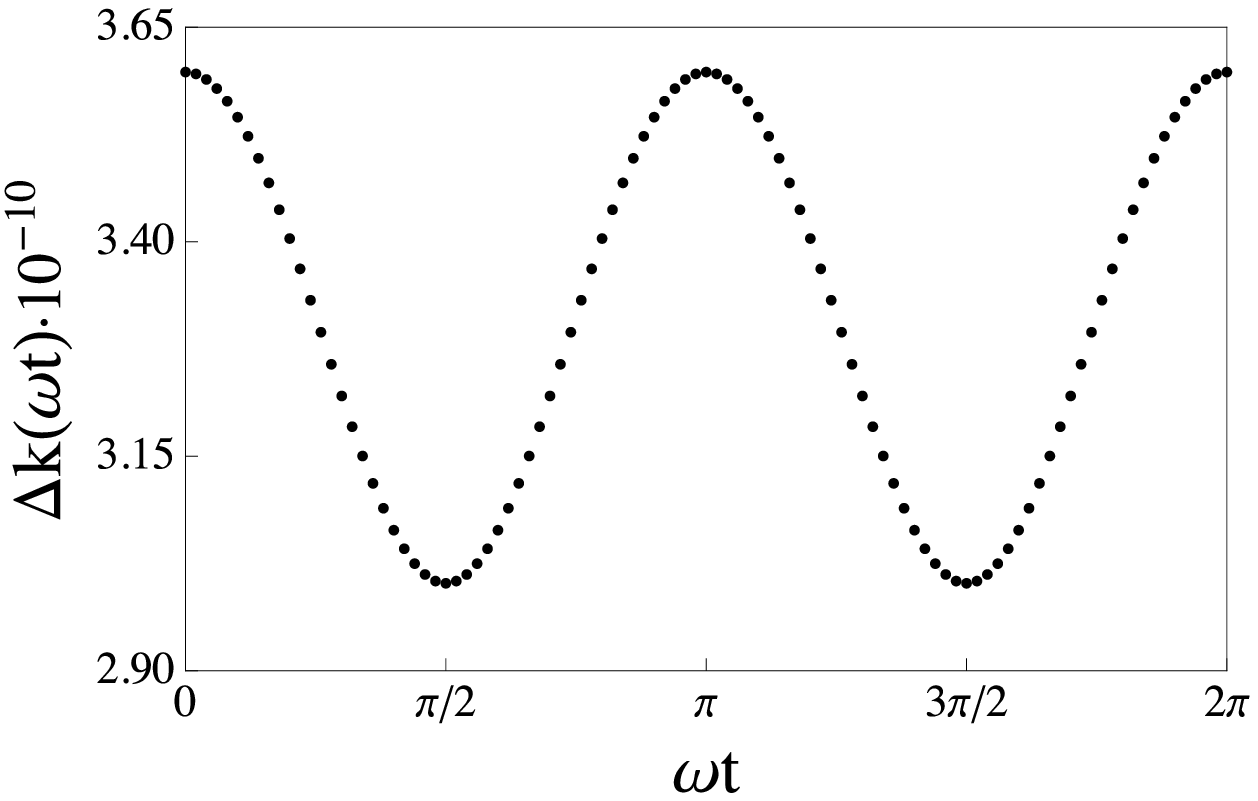}
                \caption{}
                \label{fig:ISWP-HPk}
        \end{subfigure}
        \begin{subfigure}[b]{0.32\textwidth}
                \centering
                \includegraphics[height=3.7cm,width=5.6cm]{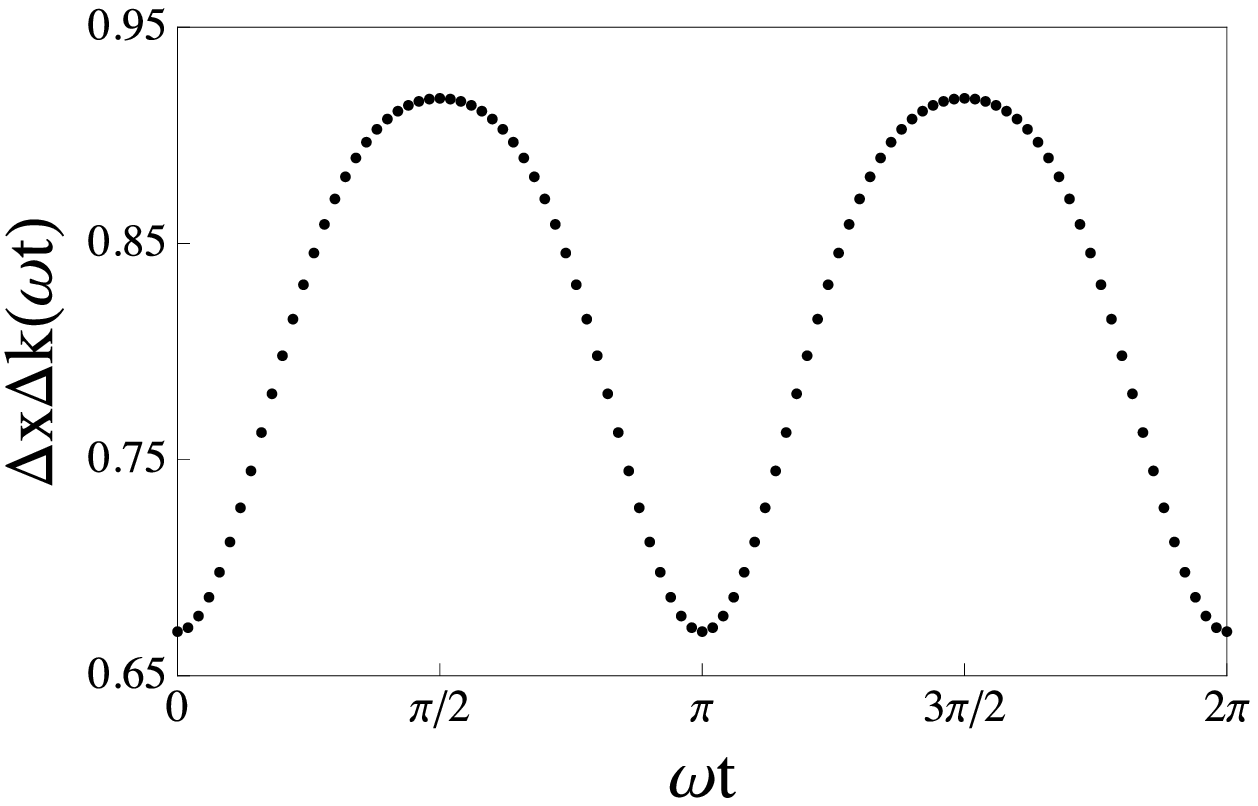}
                \caption{}
                \label{fig:ISWP-HP}
        \end{subfigure}
        \caption{\small Time evolution of Heisenberg relation in ISWP \\ a)in position space $\Delta x$, b) in momentum space $\Delta k$ and c)$\Delta x \Delta k$}\label{fig:ISWP-HP-ALL}
\end{figure}

We illustrate below the graphs of the same statistical measures for ISWP. Heisenberg uncertainty relation in Fig. \ref{fig:ISWP-HP-ALL} shows the same behavior as in DSWP in Fig. \ref{fig:NH3-HP-ALL}. 

\begin{figure}[!htb]
        \centering
        \begin{subfigure}[b]{0.32\textwidth}
                \centering
                \includegraphics[height=3.7cm,width=5.6cm]{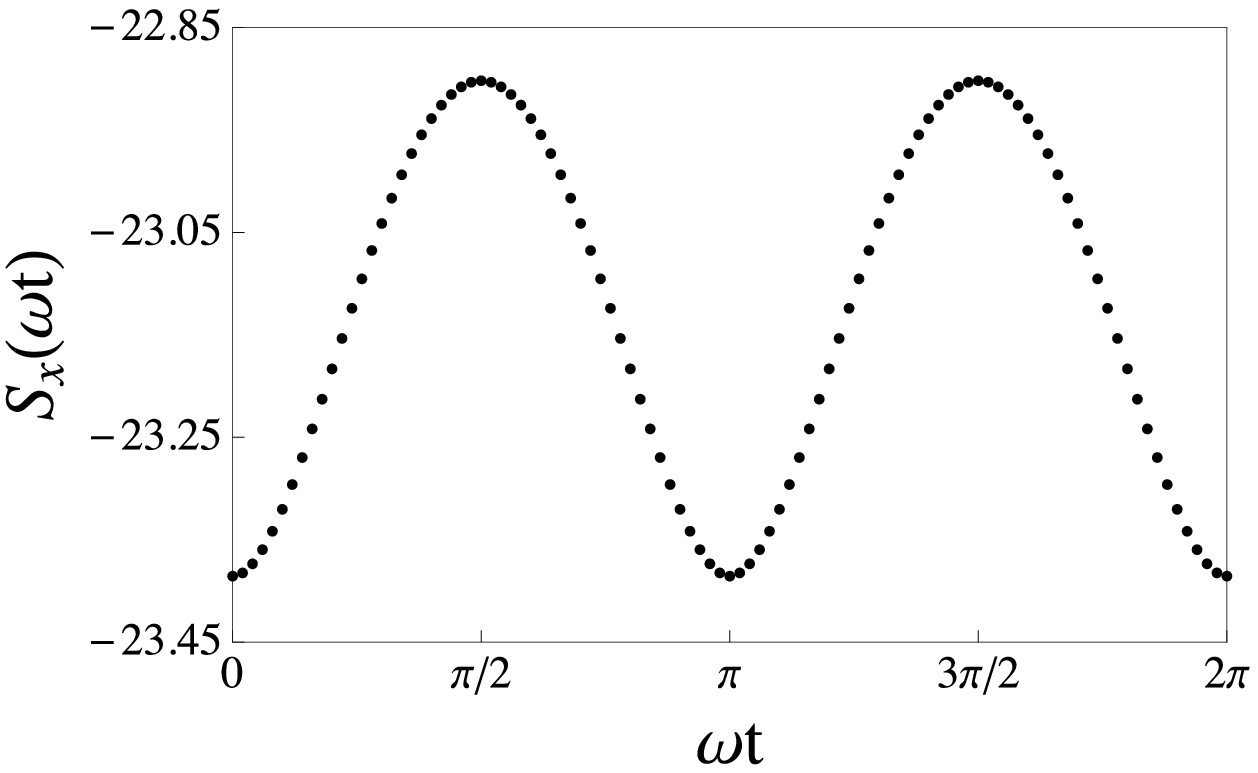}
                \caption{}
                \label{fig:ISWP-Shannon-Position}
        \end{subfigure}
        \begin{subfigure}[b]{0.32\textwidth}
                \centering
                \includegraphics[height=3.7cm,width=5.6cm]{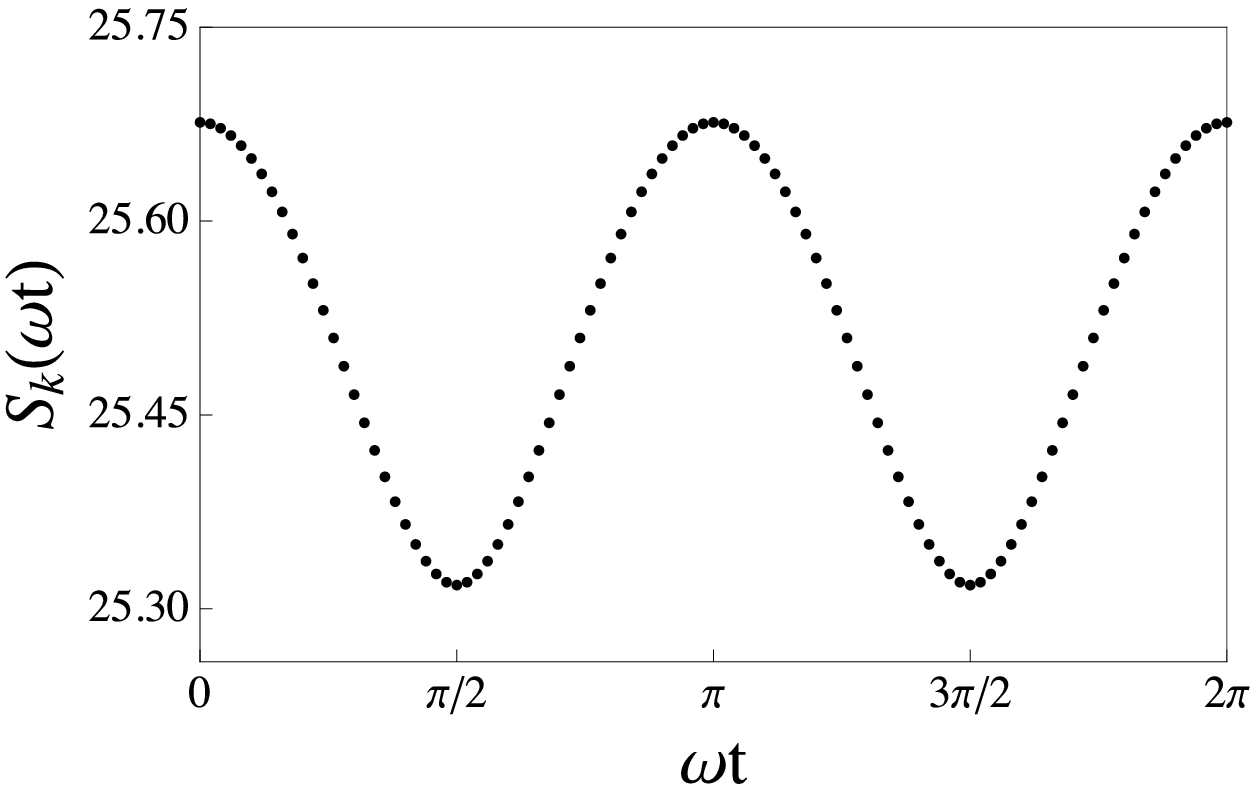}
                \caption{}
                \label{fig:ISWP-Shannon-Momentum}
        \end{subfigure}
        \begin{subfigure}[b]{0.32\textwidth}
                \centering
                \includegraphics[height=3.7cm,width=5.6cm]{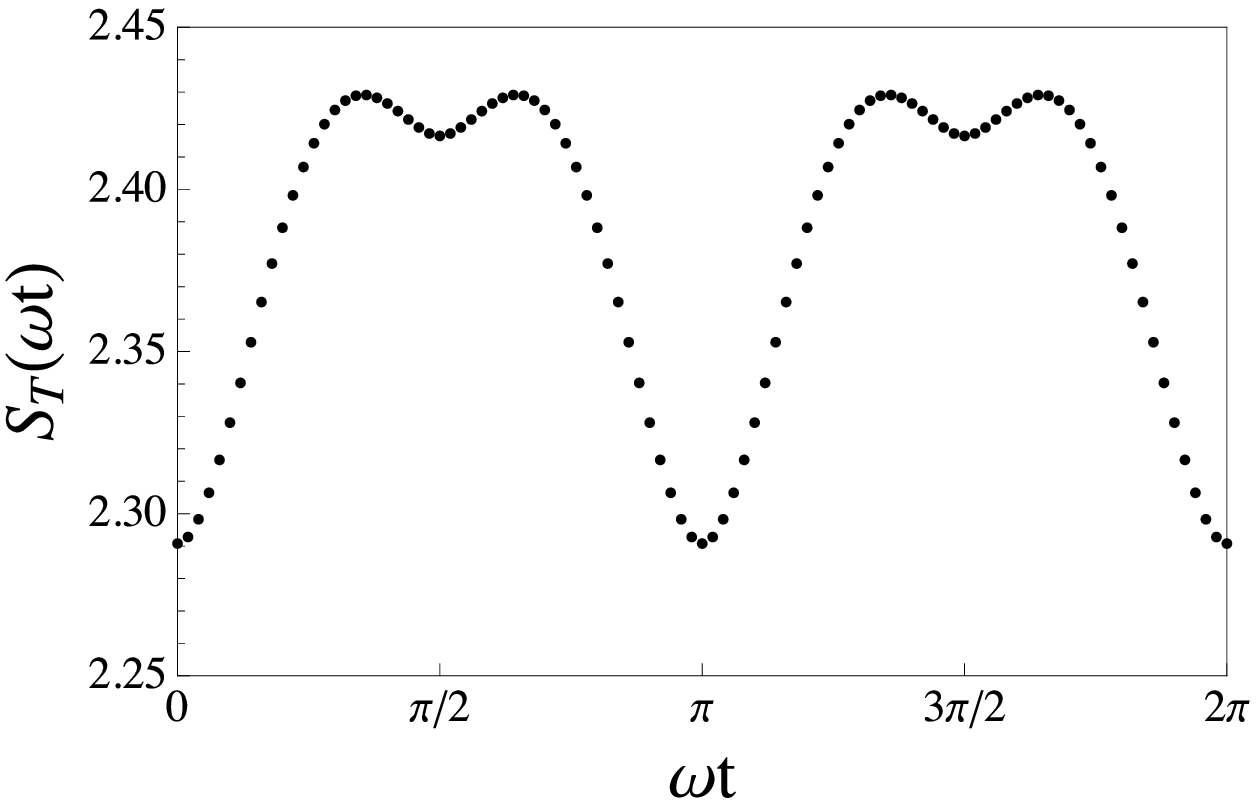}
                \caption{}
                \label{fig:ISWP-Shannon-Net}
        \end{subfigure}
        \caption{\small Time evolution of Shannon information entropy in ISWP \\ a)in position space $S_x$, b) in momentum space $S_k$ and c)$S_T=S_x+S_k$}\label{fig:ISWP-Shannon}
\end{figure}

On the other hand, while the results for both $S_x$ and $S_k$ are similar to those of the DSWP case (Figs. \ref{fig:ISWP-Shannon-Position} and \ref{fig:ISWP-Shannon-Momentum}), the net Shannon information entropy in Fig. \ref{fig:ISWP-Shannon-Net} presents an interesting fluctuation (local minimum value) when the particle is in the middle of the oscillation ($t=T/4, 3T/4$). This behavior is the result of the mutual overlap of the probability density curves in position space $\rho(x,t)$, which can be seen in Fig. \ref{fig:ISWP-Oscillation-Position}. 

\begin{figure}[!htb]
        \centering
        \begin{subfigure}[b]{0.32\textwidth}
                \centering
                \includegraphics[height=3.7cm,width=5.6cm]{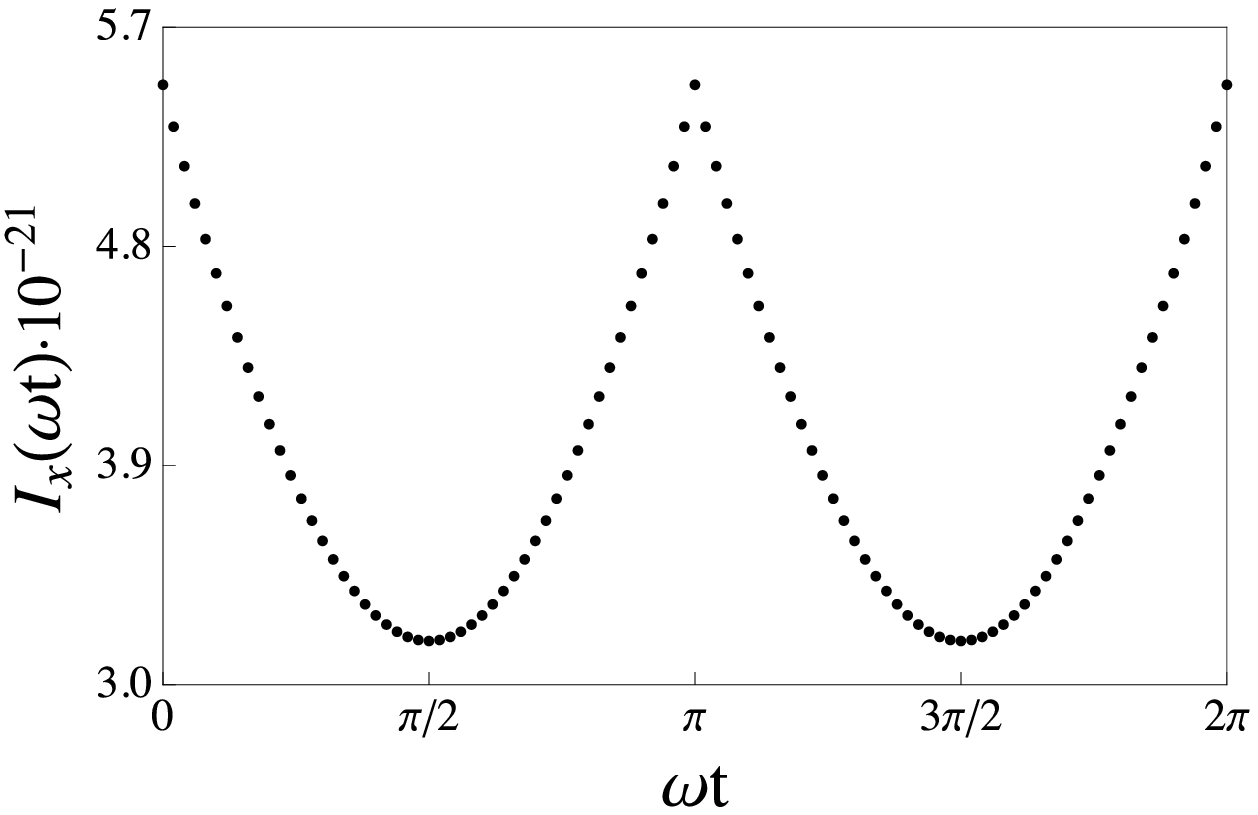}
                \caption{}
                \label{fig:position-space-fisher-ISWP}
        \end{subfigure}
        \begin{subfigure}[b]{0.32\textwidth}
                \centering
                \includegraphics[height=3.7cm,width=5.6cm]{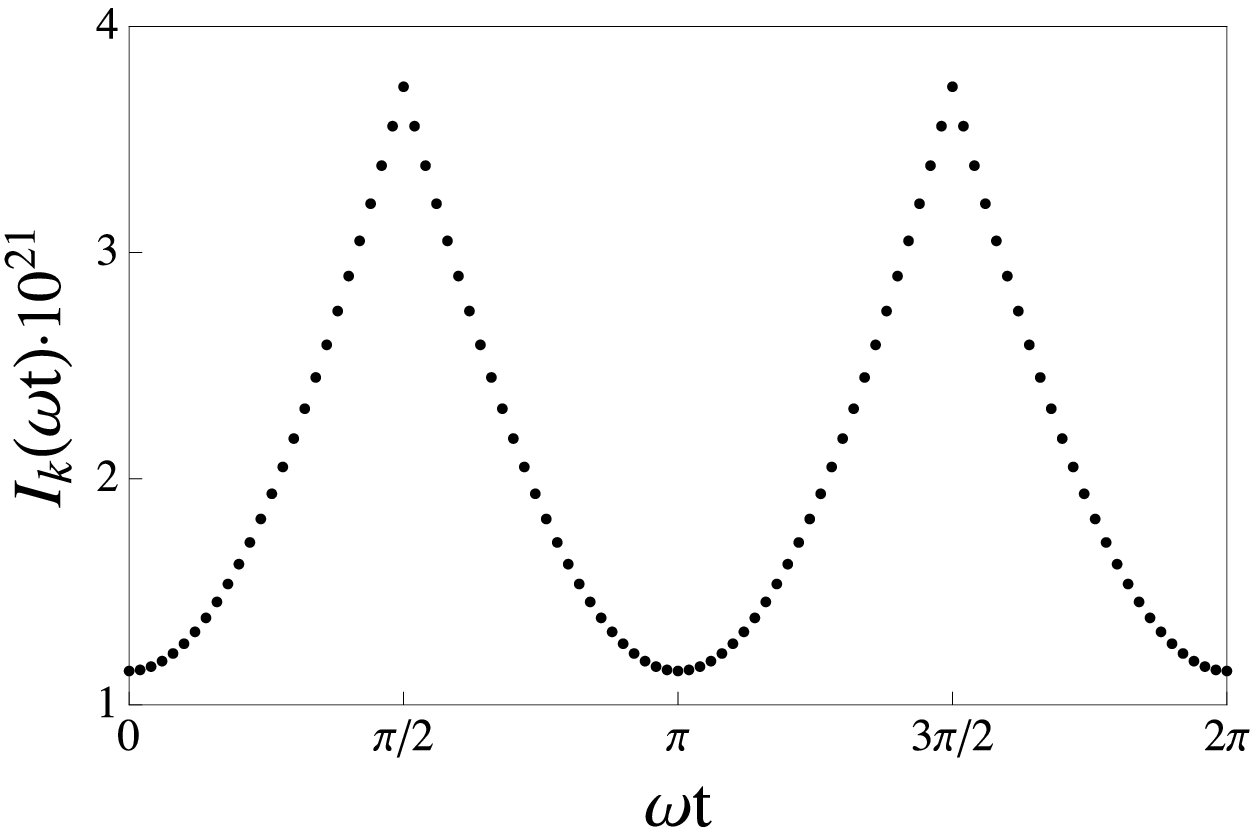}
                \caption{}
                \label{fig:momentum-space-fisher-ISWP}
        \end{subfigure}
        \begin{subfigure}[b]{0.32\textwidth}
                \centering
                \includegraphics[height=3.7cm,width=5.6cm]{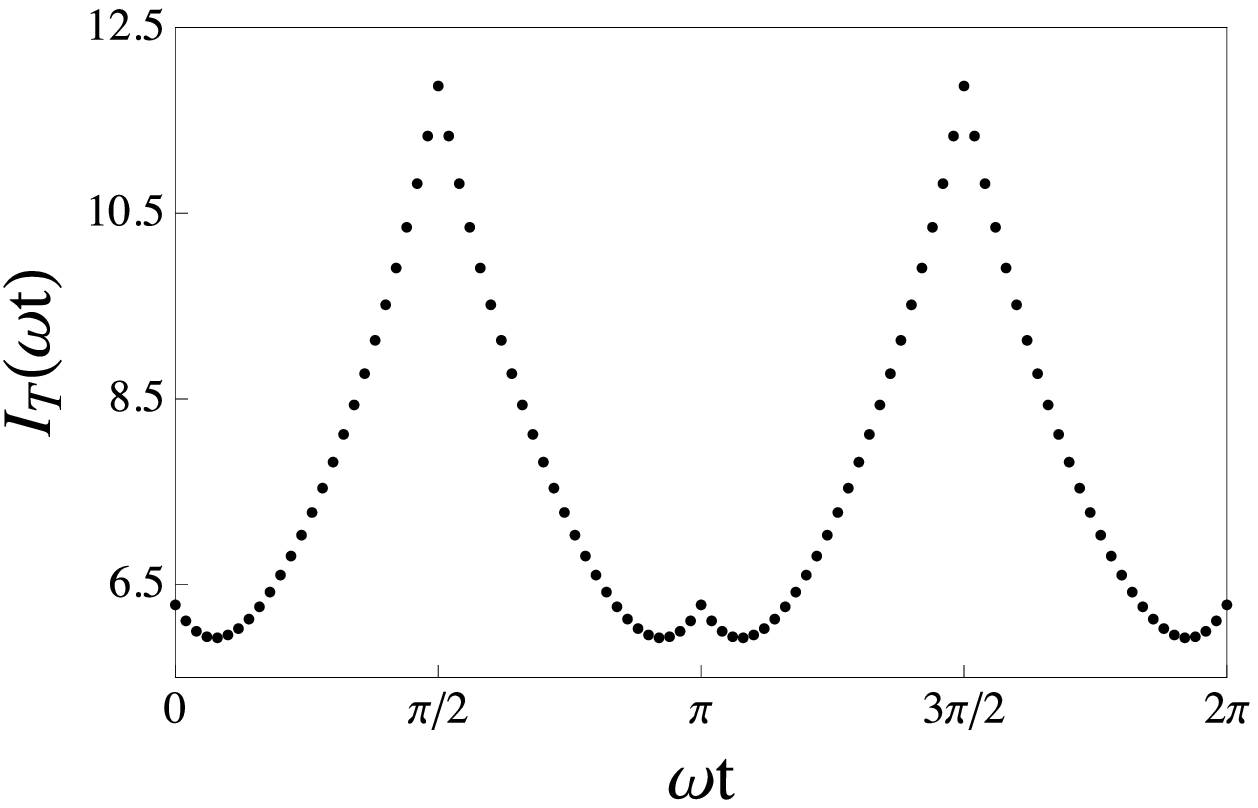}
                \caption{}
                \label{fig:final-fisher-ISWP}
        \end{subfigure}
        \caption{\small Time evolution of Fisher information in ISWP \\ a)in position space $I_x$, b) in momentum space $I_k$ and c)$I_T=I_x I_k$}\label{fig:ISWP-Fisher}
\end{figure}

Contrary to the net Shannon information entropy $S_T$, which reflects the features of the probability density in position space $\rho(x,t)$, the net Fisher information $I_T$ reflects the behavior of the probability density in momentum space $n(k,t)$. It is clear from Fig. \ref{fig:ISWP-Oscillation-Momentum} that curves of $n(k,t)$ overlap with each other when the particle is predominately in the left or in the right side of the well ($t=0,T/2,T$), which results in a local maximum value. The importance of Shannon and Fisher information entropies is now obvious, since these two measures manage to differentiate DSWP from ISWP case, providing a more sensitive analysis of the system. 

\begin{figure}[!htb]
        \centering
        \begin{subfigure}[b]{0.35\textwidth}
                \centering
                \includegraphics[height=3.7cm,width=5.6cm]{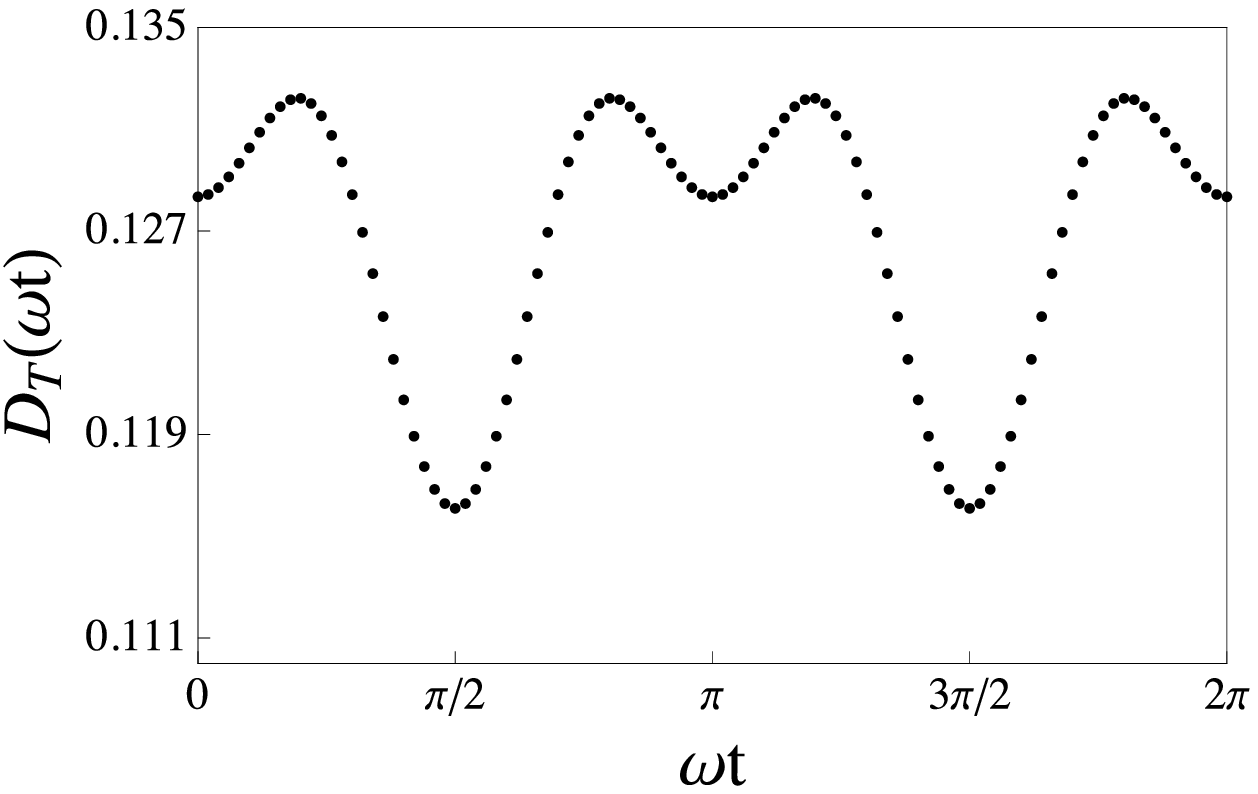}
                \caption{\small $D_T(\omega t)$ }  
        \end{subfigure} 
         \begin{subfigure}[b]{0.35\textwidth}
                \centering
                \includegraphics[height=3.7cm,width=5.6cm]{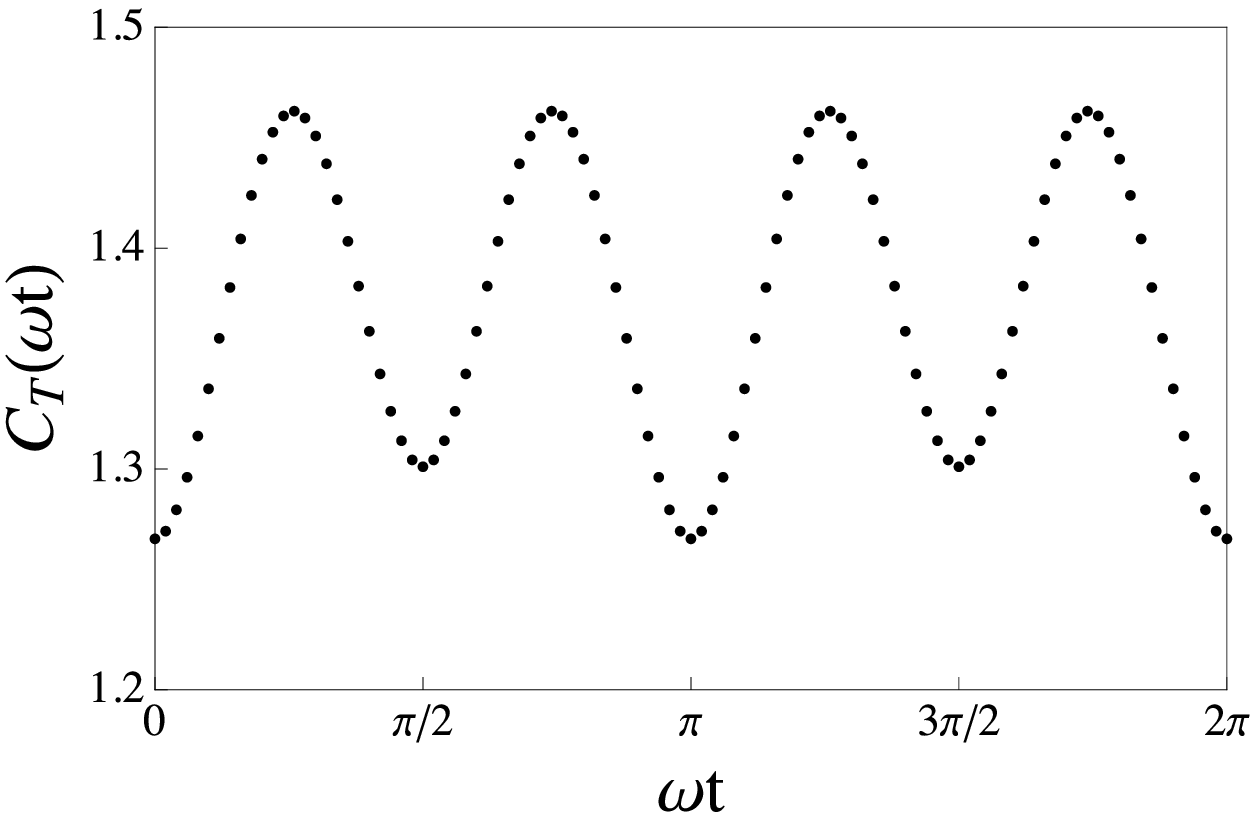}
                \caption{\small $C_T(\omega t)$}              
        \end{subfigure}
   \caption{\small Time evolution of disequilibrium and LMC complexity in ISWP}\label{fig:ISWP-D-C}
\end{figure}

Finally, in Fig. \ref{fig:ISWP-D-C} we plot disequilibrium and LMC complexity versus time.

\newpage

\section{Summary and Conclusions}\label{Summary and Conclusions}

We have found that when a particle tunnels through a barrier in a DSWP, the Heisenberg product of uncertainties and information entropy form a periodic function, which take their minimum values when the particle is located in one well, and their maximum when the particle is in the middle of the penetration region. Despite the absence of the tunneling effect in ISWP case, Heisenberg relation shows qualitatively the same behavior as in DSWP case. On the other hand, information entropies, disequilibrium and LMC complexity show slight but crucial qualitative differences, formulating a criterion, whether the tunneling effect is present or not in a quantum-mechanical system. 

Our results also verify the main differences between Shannon and Fisher information entropies. Fisher's information reflects the localization characteristics of the tunneling effect more sensitively than the Shannon information entropy, while the first one resembles to an exponential rise, and the latter to a logarithmical one. Moreover, Shannon's information entropy $S_T$ reflects the behavior of probability density in position space $\rho(x,t)$, while the Fisher information the behavior of probability density in momentum space $n(k,t)$. This can be observed from the results in both DSWP and ISWP cases. In this sense, Shannon information entropy is complementary to Fisher information.

In conclusion, it is worth mentioning that generally DSWP and ammonia molecule in particular, are of great interest in the field of quantum computation and information. Recent experimental and theoretical research \cite{Foot,Suzuki,Schroder} has shown that DSWP can be used to make quantum logic gates for ultracold atoms confined in optical lattices, while the ammonia molecule provides a nice vibrational system, which can be a promising candidate for achieving molecular quantum computation.


\begin{thebibliography}{1}

\bibitem{Cleeton}C. E. Cleeton and N. H. Williams, Phys. Rev. 45, 234 (1934).

\bibitem{Herzberg}G. Herzberg, Infrared and Raman Spectra of Polyatomic Molecules (Molecular Spectra and Molecular Structure, Volume 2), D. Van Nostrand, (1945) 

\bibitem{Basdevant}J-L Basdevant, J. Dalibard, Quantum Mechanics, Springer (2005).

\bibitem{Major}Fouad G. Major, The Quantum Beat Principles and Applications of Atomic Clocks, Springer, Second Edition (2007).

\bibitem{Peacock-Lopez}E. Peacock-Lopez The Chemical Educator Volume 11 Issue 6 (2006) pp 383-393.

\bibitem{Bialynicki-Birula}I. Bialynicki-Birula, J. Mycielski, Commun. Math. Phys. 44 (1975) 129.

\bibitem{Sears}S.B. Sears, Applications of information theory in chemical physics, Ph.D. thesis, University of North Carolina at Chapel Hill (1980).

\bibitem{KP}P. Karafiloglou, C.P. Panos, Chem. Phys. Lett. 389 (2004) 400.

\bibitem{Adami}C. Adami, Phys. Life Rev. 1 (2004) 3.

\bibitem{Shannon}C.E. Shannon, Bell System Technical Journal: 27:379-423 and 623-656 (1948).

\bibitem{Fisher}R.A. Fisher, Proc. Cambridge Philos. Soc. 22 (1925) 700.

\bibitem{GSCB}S.R. Gadre, S.B. Sears, S.J. Chakravorty, R.D. Bendale, Phys. Rev. A 32 (5) (1985) 2602.

\bibitem{Gadre}S.R. Gadre, Reviews of Modern Quantum Chemistry, A Celebration of the Contributions of Robert G. Parr, vol. 1, World Scientific, Singapore, 2002, Ch. Information theoretical approaches to quantum chemistry, pp. 108–147.

\bibitem{SPCM}K.D. Sen, C.P. Panos, K.C. Chatzisavvas, C.C. Moustakidis, Phys. Lett. A 364 (2007) 286.

\bibitem{Nalewajski1}R.F. Nalewajski, Information Theory of Molecular Systems, Elsevier Science Ltd, 2006.

\bibitem{Bonchev}D. Bonchev, Complexity in Chemistry, Taylor and Francis, 2003, Ch. Shannon's information and complexity, pp. 155–187.

\bibitem{LMC}R. L\' {o}pez-Ruiz, H.L. Mancini, X. Calbet, Phys. Lett. A 209 (1995) 321.

\bibitem{MS}H.E. Montgomery Jr., K.D. Sen, Phys. Let. A 372 (13) (2008) 2271.

\bibitem{CMP}K.C. Chatzisavvas, C.C. Moustakidis, C.P. Panos, J. Chem. Phys. 123 (2005) 174111.

\bibitem{SDL}J.S. Shiner, M. Davison, P.T. Landsberg, Phys. Rev. E 59 (2) (1999) 1459.

\bibitem{HSPSE}M. Hô, R.P. Sagar, J.M. Pérez-Jordá, V.H. Smith Jr., R.O. Esquivel, Chem. Phys.
Lett. 219 (1) (1994) 15.

\bibitem{HCSWGSE}M. Hô, B.J. Clark, V.H. Smith, D.F. Weaver, C. Gatti, R.P. Sagar, R.O. Esquivel, J. Chem. Phys. 112 (2000) 7572.

\bibitem{Manning}M.F. Manning, The Journal of Chemical Physics 3, 136 (1935).

\bibitem{Massen}S.E. Massen, Notes on Computational Quantum Physics, Thessaloniki (2005) (in Greek).

\bibitem{LSHR}R. L\' {o}pez-Ruiz, J. Sa\~{n}udo, Communication in Numerical Analysis, Volume 2012 (2012), 1-7.

\bibitem{Grif}David J. Griffiths, Introduction to Quantum Mechanics, Pearson; 2 edition (2003).

\bibitem{Cover}T. Cover, J. Thomas, Elements of Information Theory, Wiley-Interscience (1991).

\bibitem{MP}S.E. Massen, C.P. Panos, Phys. Lett. A 246 (1998) 530-533.

\bibitem{Frieden}B.R. Frieden, Science from Fisher Information, Cambridge Univ. Press, Cambridge, 2004.

\bibitem{Nalewajski2}R. F. Nalewajski, J. Math. Chem. (2013) 51:297-315.

\bibitem{Sanchez-Moreno}P. Sanchez-Moreno, A. R. Plastino, J. S. Dehesa, J. Phys. A: Math. Theor. Volume 44.

\bibitem{PNCT}C.P. Panos, N.S. Nikolaidis, K.Ch. Chatzisavvas, C.C. Tsouros, Phys. Lett. A 373 (2009) 2343-2350

\bibitem{Rao}C.R. Rao, Linear Statistical Interference and its Applications, Wiley, New
York, 1965.

\bibitem{Stam}A.J. Stam, Information and Control, Vol. 2, Issue 2, June 1959, Pages 101-112.

\bibitem{Catalan}R.G. Catalan, J. Garayand and R. Lopez-Ruiz, Phys. Rev. E 66 (2002) 011102.

\bibitem{Renyi}R\' {e}nyi , A. ``On Measures of Entropy and Information." Proc. Fourth Berkeley Symp. Math. Stat. and Probability, Vol. 1. Berkeley, CA: University of California Press, pp. 547-561, 1961.

\bibitem{Foot}C. J. Foot and M. D. Shotter, Am. J. Phys. 79, 762 (2011).

\bibitem{Suzuki}S. Suzuki, K. Mishima, K. Yamashita, Chem. Phys. Lett. 410 (2005) 358–364.

\bibitem{Schroder}M. Schroder, A. Brown, J. Chem. Phys. 131, 034101 (2009).














 


\end{thebibliography}
\end{document}